\documentclass[aps,prx,superscriptaddress,amsfonts,amsmath,amssymb,showpacs,floatfix,reprint]{revtex4-1}

\usepackage[english]{babel}
\usepackage{graphicx}					%Calls figure environment
\usepackage[pdftex]{epsfig}
\usepackage{ragged2e}
\usepackage[caption=false]{subfig}
\usepackage{amsmath,amssymb}
\usepackage{times}
\usepackage[colorlinks,linkcolor=blue,citecolor=blue,urlcolor=blue]{hyperref}
\usepackage[table,xcdraw]{xcolor}
\usepackage{epstopdf}
\usepackage{physics}
\usepackage[export]{adjustbox}
\usepackage{dsfont}
\usepackage{float}

\newcommand{\be}{\begin{equation}}
\newcommand{\ee}{\end{equation}}

\newcommand{\bee}{\begin{equation*}}
\newcommand{\eee}{\end{equation*}}

\begin{document}

\title{Minimal models of $\alpha$-Li$_2$IrO$_3$: On the range of the interactions, ground state properties, and magnetization processes.}
\author{Maria Laura Baez}
\affiliation{Dahlem Center for Complex Quantum Systems, Freie Universit\"{a}t Berlin, Germany}
\affiliation{Helmholtz-Zentrum f\"{u}r Materialien und Energie, Berlin, Germany} 

\date{\today}

\begin{abstract}

In recent years, a lot of effort has been devoted to the quest for experimental realizations of Kitaev interactions in spin systems. Recently, many materials have been synthesized which seem to realize extended Kitaev models, where Kitaev interactions are supplemented by Heisenberg and other bond dependent terms. The crystal and electronic structure of these materials renders the determination of a minimal model a non trivial pursuit. In this work, we will concentrate on one of these particular materials, $\alpha$-Li$_2$IrO$_3$, for which various minimal models have been proposed. Employing large scale Monte Carlo simulations we show how the number of prospective models can be reduced. We study in detail six models with different range of the interactions, and show how only two of those reproduce the most recent experimental results for this material. We obtain two possible minimal models, one of them with nearest neighbour interactions, while the other includes interactions up to third neighbours. Furthermore, we show that strong bond anisotropies and further neighbour interactions are crucial to stabilize the tilted counterotating spirals found in  $\alpha$-Li$_2$IrO$_3$. We further clarify the picture, and distinguish these two models by studying the magnetization processes. We predict the magnetization behaviour of these models, and propose future experimental directions.

\end{abstract}

\maketitle
%====================================================================
\section{Introduction}

The theoretical study of Kitaev materials is a flourishing part of the frustrated magnetism field \cite{Jackeli2009, Chaloupka-2010, song16, Lee-2016, Winter-2017}. Since Kitaev proposed his exactly solvable honeycomb model \cite{kitaev06}, which belongs to a wider range of quantum compass Hamiltonians \cite{Nussinov2015}, a great number of resources have been dedicated at studying this system. Kitaev's proposal kick-started a quest for experimental realizations of Kitaev spin liquids, and since then many materials have been synthesized that have been proposed to realize Kitaev interactions ($\mathrm{Na}_2\mathrm{IrO}_3$\cite{Singh2010}, $\mathrm{Li}_2\mathrm{IrO}_3$\cite{Singh2012,Takayama2015,Modic2014}, $\mathrm{Ru}\mathrm{Cl}_3$\cite{Plumb2014}, $\mathrm{Ba}_3\mathrm{IrTi}_2\mathrm{O}_9$\cite{Dey2012}, among others).

Until now, though, these materials not only present Kitaev interactions but are also accompanied by Heisenberg and other bond dependent exchanges\cite{Jackeli2009, Chaloupka-2010, Rau2014, Winter-2017}, the so called {\it extended Kitaev models}. These perturbing interactions have a strong effect on the ground state of the these materials. Furthermore, even though Kitaev interactions give rise to spin liquid states, the perturbations are so strong that the synthesized materials exhibit classical magnetic order.

While the quest for materials which {\it only} realize Kitaev interactions is still ongoing\cite{nasu16, kitagawa18}, the theoretical understanding of extended Kitaev models remains challenging. The complexity of their crystal structure and orbital hybridization \cite{Winter-2017} has provided significant challenges to their modelling. So far, we know that the minimal models for these materials can be built as modifications of the same basic Hamiltonian\cite{Jackeli2009, Chaloupka-2010, Rau2014, Winter-2017}, but where different variations of the model seem to be realized in different materials \cite{Singh2010, Singh2012, Takayama2015, Modic2014, Plumb2014, Winter-2017},  as for example in $\mathrm{Na}_2\mathrm{IrO}_3$\cite{Chaloupka-2010, Singh2012, choi12, yamaji14, hwan15, Winter2016} and $\alpha-\mathrm{Li}_2\mathrm{IrO}_3$ \cite{Singh2012, Biffin-2014, Reuther-2014, Kimchi-2015, Williams-2016, Winter-2017} . In particular, different materials present different bond anisotropies on the interactions and further neighbour exchanges. Since different variations of the same model can reproduce experimental results, the question that arises is related to how can one distinguish between these models further, and which experiments can be performed to asses their validity. In this work, we will attempt to answer this question in the context of $\alpha$-$\mathrm{Li}_2\mathrm{IrO}_3$.

$\alpha-\mathrm{Li}_2\mathrm{IrO}_3$ (isostructural to $\mathrm{Na}_2\mathrm{IrO}_3$) possesses a layered crystal structure where the Ir$^{4+}$ ions, surrounded by an octahedral cage of oxygens, form a honeycomb lattice \cite{Freund2016}. This lattice resides in the $[111]$ Cartesian plane.  Both compounds, $\mathrm{Na}_2\mathrm{IrO}_3$ and $\alpha-\mathrm{Li}_2\mathrm{IrO}_3$, present long range order, showing anomalies in the specific heat and in the magnetic susceptibility at a critical temperature $T_c \sim 15K$, with a Curie-Weiss temperature of $\Theta = -125(6)K$ and $\Theta = -33(3)K$ respectively \cite{Singh2012}.
$\mathrm{Na}_2\mathrm{IrO}_3$ presents a zig-zag ordered state and it would be expected that, given the similarities in the thermodynamic anomalies, $\alpha-\mathrm{Li}_2\mathrm{IrO}_3$ would also present a zig-zag order. While the thermodynamics show similarities in the behavior of the Na and Li compound, recent studies performed on single crystals and powder samples of $\alpha-\mathrm{Li}_2\mathrm{IrO}_3$ have shown that the magnetic order present in this material is not of zig-zag type, but of spiral nature. Magnetic resonant X-ray diffraction (MRXD) together with magnetic powder neutron diffraction determined a magnetic structure composed of counterrotating incommensurate coplanar spin spirals \cite{Williams-2016}, with a propagation wavevector $\mathbf{q}=(0.32(1),0,0)$. At the same time, the authors of Ref. \cite{Williams-2016} were able to determine that the plane of rotation of the spirals is uniform between the different sublattices and tilted with respect to the lattice plane by $80^o$. In their studies they see that the MRXD, at a temperature of $5K$, presents satellite peaks at positions $\mathbf{\tau} \pm \mathbf{q}$, where $\mathbf{\tau}$ are the positions of allowed structural reflections $\mathbf{\tau}=(h,k,l)$, with $h+k=$even.

Similarities have been found between the ground state of $\alpha-\mathrm{Li}_2\mathrm{IrO}_3$ and the two structural polytypes, $\beta-\mathrm{Li}_2\mathrm{IrO}_3$ and $\gamma-\mathrm{Li}_2\mathrm{IrO}_3$, which correspond to hyper-honeycomb and stripy-honeycomb magnetic lattices respectively. All three of these polytypes are members of the ``harmonic honeycomb'' structural series \cite{Modic2014}, and the similarities in their magnetic ordering have lead to proposals of universality between the members of the family of harmonic honeycomb Iridates \cite{Kimchi-2015}. 

It has been shown previously that the magnetic ordering of the $\beta$ and $\gamma$ structures are well described by a dominant ferromagnetic Kitaev interaction, combined with other smaller exchange terms of the form of Heisenberg and bond dependent terms, \cite{Biffin-2014, Lee-2015, Kimchi-2015, Lee-2016}.

Given the similarities between the different polytypes, it is expected that the model which correctly describes the magnetic structure of $\alpha$--$\mathrm{Li}_2\mathrm{IrO}_3$ will be composed of dominant Kitaev interactions, supplemented by weaker Heisenberg and bond dependent terms. While in principle this proposal is in agreement with what is known for the  $\beta$ and $\gamma$ polytypes, the particularities of the minimal model have not yet been defined. Many models have been proposed which reproduce some characteristics of the $\alpha$ polytype, but which have been tested only on toy models, or by a Luttinger-Tisza (LT) approximation, which are not ensured to succeed at the detection of incommensurate states. 

Regarding $\alpha$--$\mathrm{Li}_2\mathrm{IrO}_3$, the range of its interactions is still an open question. There have been proposals where this material is modeled as a nearest neighbour \cite{Singh2012, Kimchi-2015, Williams-2016}, second neighbour \cite{Reuther-2014}, or third neighbour \cite{Winter-2017} extended Kitaev model, and in this work we perform a comparative analysis between all those models, via Monte Carlo simulations. We aim at exploring the possible ground states of the different models via a method which is not biased towards any type of magnetic order, and which can deal with the incommensurate nature of the expected ground state appropriately. We find that among the six proposed models up to date, only two, a nearest and a third neighbour model, can reproduce the experimental features correctly. The nearest neighbour model we study is a variation of the model proposed by I. Kimchi {\it et. al.} \cite{Kimchi-2015, Williams-2016}. In the model of Ref.\cite{Kimchi-2015}, bond dependent interaction are allowed by the crystallographic symmetry, and was studied by the authors in a toy model consisting of 1D zig-zag chains. We will extend this model to include further bond dependent interactions, and we will study it employing large scale Monte Carlo simulations.  On the other hand, we also find that the third neighbour model proposed by Winter {\it et. al.} via DFT calculations\cite{Winter-2017} can also accurately reproduce the experimental results, but provided that bond anisotropies are included.

The discrepancy between the range of the interactions of these two models raises the question of how we can further distinguish them to ascertain which one corresponds to $\alpha$--$\mathrm{Li}_2\mathrm{IrO}_3$. We propose to answer this question studying the magnetization behaviour of both models.  Our aim is that the differences in the magnetization processes with different  applied field directions can provide an efficient experimental route to probe which of the proposed models is more feasible.  In the presence of a magnetic field, we find that both models present different magnetization processes depending on the direction in which the field is applied. This indicates an experimentally realizable way of determining if one of the proposed models is correct, by studying the low temperature magnetization of single  crystals of $\alpha$--$\mathrm{Li}_2\mathrm{IrO}_3$.

The paper is structured as follows: in section \ref{MandM} we show the fundamental characteristics of the studied models, and briefly describe the numerical method employed. In section \ref{Gp} we show the ground state properties for those models which reproduce the experimental results, and in section \ref{Mp} we study in detail the magnetization processes for those same models. Finally in section \ref{DandC} we discuss our results in the context of the available experimental evidence, and propose further theoretical and experimental studies which can help determine which minimal model corresponds to  $\alpha$--$\mathrm{Li}_2\mathrm{IrO}_3$.

\section{Models and method} 
\label{MandM}

\subsection{Models} 
We classify the different models according to the range of their interactions, into nearest, second, and third neighbour models. While here we mention the common characteristics of all of them, in the main text we will only show the results for those models which reproduce the experimental features of $\alpha-\mathrm{Li}_2\mathrm{IrO}_3$. In Appendices \ref{I_c}, \ref{ITM}, \ref{2N}, and \ref{2NI_c}  we show the ground state properties for the rest of the models.

\begin{figure}[h!]
\centering
\includegraphics[scale=0.2]{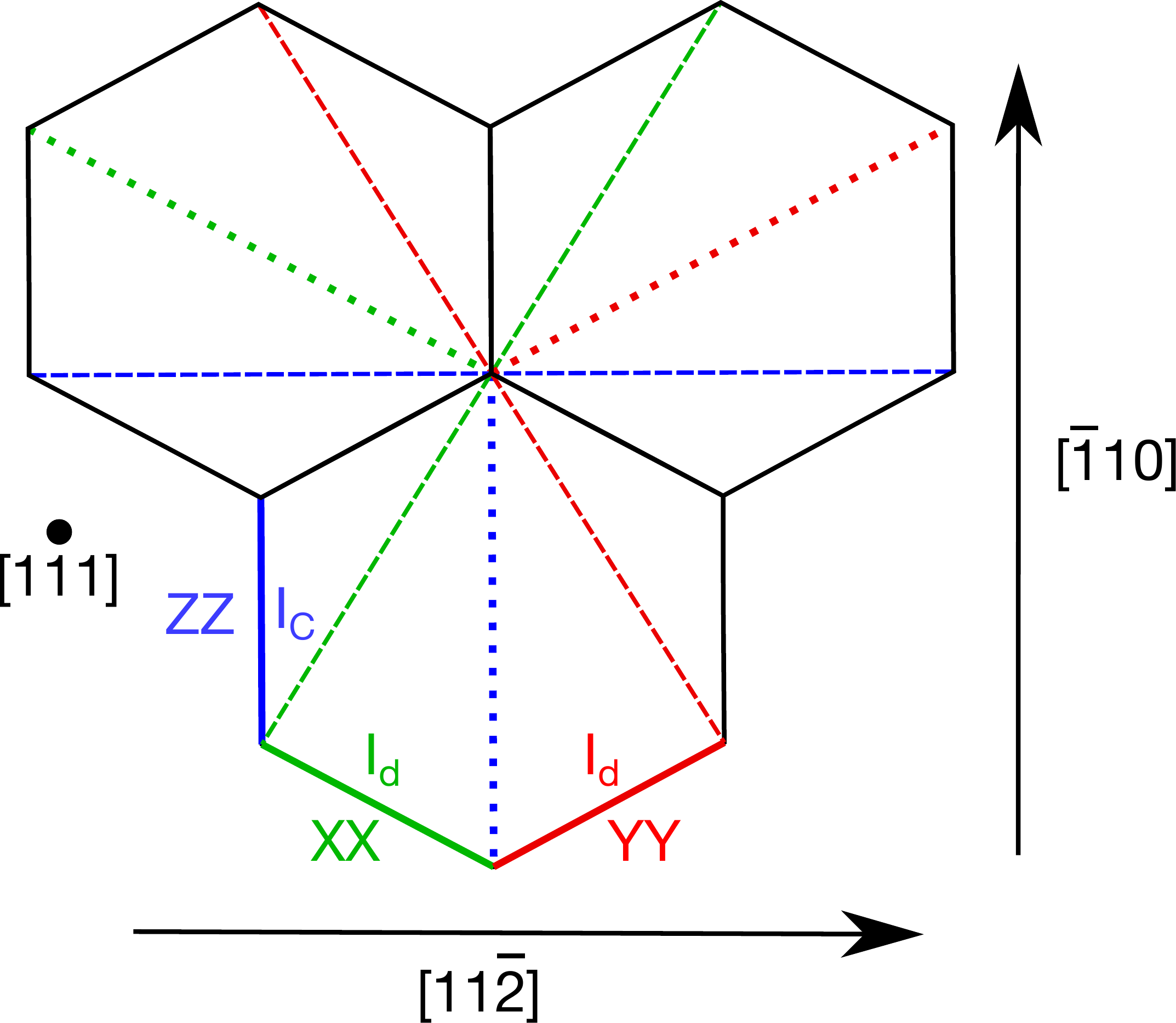}
\caption{(a) Colour coded honeycomb lattice. Full lines correspond to nearest neighbour interactions, dashed and dotted to second and third neighbours respectively. We colour coded the Kitaev exchanges $S_i^xS_j^x$ (green), $S_i^yS_j^y$ (red), and $S_i^zS_j^z$ (blue). We indicate the $I_c$ and $I_d$ terms for nearest neighbour interactions. The lattice plane is the $[111]$ plane, and the magnetic spirals propagate in the $[11\bar{2}]$ direction, perpendicularly to the $zz$-bonds. The $[\bar{1}10]$ direction is parallel to the $zz$-bonds. }
\label{lattice}
\end{figure}

The models treated here are variations of the extended Kitaev model 
\be
\mathcal{H} = J\sum_{ij}\mathbf{S}_i\mathbf{S}_j+ \sum_{ij \in \gamma-\mathrm{bonds}} \left( K S^{\gamma}_iS^{\gamma}_j + \Gamma (S^{\alpha}_iS^{\beta}_j+S^{\beta}_iS^{\alpha}_j)\right)\,,
\label{general_iridate}
\ee
where $\gamma = \{x,y,z\}$, and where $\alpha$ and $\beta$ indicate the two spin components perpendicular to $\gamma$. $J$ represents a Heisenberg coupling, while the bond dependent terms contain Kitaev interactions coupled by $K$. The $\Gamma$ exchange couples two orthogonal spin components, $\alpha$ and $\beta$, along the bond with Kitaev interactions in the $\gamma$ spin component.  A diagram of the lattice and the Kitaev interactions is shown in Fig.~\ref{lattice}.

The nearest ($\mathcal{H}(J_1, K_1, I_c, I_d)$) and second neighbor ($\mathcal{H}(J_{1,2}, K_{1,2}, I_c, I_d)$) models we study are an anisotropic version of Eq.\ref{general_iridate}, based on the model proposed in Refs.~ \cite{Kimchi-2015,Williams-2016}:
\begin{align}
 \mathcal{H}(J_n, K_n, I_c, I_d)&=\sum_n\big[I_c\sum_{<ij>}S^{r_{ij}}_iS^{r_{ij}}_j+I_d\sum_{<ij>}S^{r_{ij}}_iS^{r_{ij}}_j\notag\\
 & +J_n\sum_{<ij>_n}\mathbf{S}_i\cdot\mathbf{S}_j +K_n\sum_{<ij>_n}\sum_{\gamma}S^{\gamma}_i S^{\gamma}_j   \big]
\label{Hamiltonian}
\end{align}
where $<ij>_n$ denote a sum over $n$-th neighbours, and $\mathbf{S}_i$ = $(S^x_i, S^y_i, S^z_i )$ denotes the classical spin acting on site $i$. The terms containing the couplings $I_c$ and $I_d$ are Ising terms that couple the spins components parallel to the bond orientation, i.e $\mathbf{S}^{r_{ij}} = \mathbf{S}\cdot\mathbf{\hat{r}}_{ij}$, where $\mathbf{\hat{r}}_{ij}$ is the unit vector connecting the spins at sites $i$ and $j$. Please note that the Kitaev model has a particular symmetry in the bond isotropic case, where a $60^o$ rotation in real and spin space leave the ground state invariant, and this symmetry is preserved in the model of Eq.~\ref{Hamiltonian} when $I_c = I_d$. In the real material, the octahedral cage enclosing the Ir$^{4+}$ atoms is not perfect, presenting deformations. These deformations induce a bond anisotropy on the interactions, where the couplings of the Ising terms is not the same on the $zz$-, $xx$- and $yy$-bonds. For this, we choose $I_c$ to be active only on the {\it zz}-bonds, while $I_d$ acts on the rest of them (zig-zag bonds). 

We studied two nearest neighbour models, the model given by  $\mathcal{H}(J_1, K_1, I_c \neq 0, I_d=0)$ contains dominant ferromagnetic Kitaev interactions $K_1 < 0$ as well as small antiferromagnetic Heisenberg terms  $J_1 > 0$, the Ising term is ferromagnetic as well, $I_c < 0$. The second nearest neighbour model also includes terms with $I_d < 0$, $\mathcal{H}(J_1, K_1, I_c \neq 0, I_d\neq 0)$. For the second neighbour models we studied an isotropic model, $\mathcal{H}(J_{1,2}, K_{1,2}, I_c = 0, I_d=0)$, with nearest and second neighbour Heisenberg interactions, $J_1 > 0$ and $J_2 < 0$ respectively, as well as nearest and second neighbour Kitaev exchanges, $K_1 < 0$ and $K_2 > 0$. Finally we study an anisotropic second neighbour model, $\mathcal{H}(J_{1,2}, K_{1,2}, I_c \neq 0, I_d = 0)$, including terms with $I_c < 0$.

For the third neighbor case, $\mathcal{H}(J_{1,3}, K_{1,2}, \Gamma_{1,2})$, we implemented the model proposed by Winter {\it et.al}. based on DFT calculations \cite{Winter2016}. In their work, they propose a model with Heisenberg, Kitaev, and other bond dependent interactions, some of them ranging up to third neighbours. The effective Hamiltonian obtained results in
\begin{align}
 &\mathcal{H}(J_{1,3}, K_{1,2}, \Gamma_{1,2})=J_1\sum_{<ij>_1}\mathbf{S}_i\cdot\mathbf{S}_j+K_1\sum_{<ij>_1}S^{\gamma}_i S^{\gamma}_j \notag\\
  &+ \Gamma_1 \sum_{<ij>_1}(S^{\alpha}_i S^{\beta}_j+S^{\beta}_i S^{\alpha}_j) +K_2\sum_{<ij>_2}S^{\gamma}_i S^{\gamma}_j \notag\\
  & + \Gamma_2 \sum_{<ij>_2}(S^{\alpha}_i S^{\beta}_j+S^{\beta}_i S^{\alpha}_j) +J_3\sum_{<ij>_3}\mathbf{S}_i\cdot\mathbf{S}_j\,,
\label{Hamiltonian_W2}
\end{align}
where $X_n$, with $X = J$, $K$, or $\Gamma$, represent the exchange coupling for an interaction between $n$th-neighbours, $\sum_{<ij>_n}$ represents a sum over $n$th-neighbours, and $\{\alpha, \beta, \gamma\} =\{x,y,z\}  $ indicate the spin component. A particularity of this model is that the Kitaev and $\Gamma$ exchanges are equal in magnitude but opposite in sign, $|K_n| = -|\Gamma_n|$. 

Ref.~\cite{Winter2016} reports strong bond anisotropies in the model, an observation which coincides with previous proposals for this material \cite{Kimchi-2015}. To gain an understanding of such complex Hamiltonian we first studied the bond isotropic case $\mathcal{H}_I(J_{1,3}, K_{1,2}, \Gamma_{1,2})$, to afterwards introduce anisotropies, giving rise to the anisotropic third neighbour model $\mathcal{H}_A(J_{1,3}, K_{1,2}, \Gamma_{1,2})$. The following bond anisotropies based on Ref.~\cite{Winter2016} are introduced,
\begin{align}
 J_1^{XY} = J_1 -\delta \quad  J_1^{Z} = J_1 +\delta \notag \\
 K_1^{XY} = K_1 -\delta \quad  K_1^{Z} = K_1 +\delta \notag \\
 \Gamma_1^{XY} = \Gamma_1 -\delta \quad  \Gamma_1^{Z} = \Gamma_1 +\delta \notag \\
 \Gamma_2^{XY} = \Gamma_1 -\delta \quad  \Gamma_1^{Z} = \Gamma_1 +\delta \,,
 \label{aniso}
\end{align}
where the quantities $J_1$, $K_1$, $\Gamma_1$, and $\Gamma_2$ are understood as bond averaged interactions, and the superscript indicate on which bond these interactions act.  

In Table~\ref{Winter_bonds} we show the bond average value of the different couplings together with the anisotropic component $\delta$ that reproduce the experimental results.
\begin{table}[h!]
    \begin{tabular}{ | l | c | c | c | c |}
      \hline
	& $\delta_{W}$ & Bond average W & Anisotropic interactions \\ \hline
      $J_1$ & 0.14 & 0.2 & $J_1^{XY} = 0.06 \quad J_1^Z = 0.34$ \\ \hline
      $K_1$ & 0.34 & -1 & $K_1^{XY} = -1.34 \quad K_1^Z = -0.66$ \\ \hline
      $\Gamma_1$ & 0.195 & 1 & $\Gamma_1^{XY} = 0.805 \quad \Gamma_1^Z = 1.195$ \\ \hline
      $K_2$ & 0 & -0.275 & $K_2^{XY} = -0.275 \quad K_2^Z = -0.275$  \\ \hline
      $\Gamma_2$ & -0.06 & 0.275 & $\Gamma_2^{XY} = 0.335 \quad \Gamma_2^Z = 0.215$ \\ \hline
      $J_3$ & 0 & 0.3 & $J_3^{XY} = 0.3 \quad J_3^Z = 0.3$ \\ \hline
      \end{tabular}
  \caption{Values that reproduce the experimental results with the corresponding bond anisotropies for the anisotropic third neighbour model. All values are given in terms of $|K_1|$.}
  \label{Winter_bonds}
\end{table}

\subsection{Method}
\label{method}
The numerical solution consist on Monte Carlo simulations for classical Heisenberg spins implementing a Metropolis-Hastings algorithm. Even though we want to study ground state properties of the bulk, some of these ground states are incommensurate phases which, to the effect of the algorithm, means that we implement free edge boundary conditions (FEBs), where the spins on the edges are exposed to a putative vacuum. The presence of FEBs carries some added effects, and as such the system will exhibit edge modes that are, in principle, not relevant to the study of the bulk physics. We show in Appendix \ref{sec:kit-heis} that the edge modes do not affect the bulk properties of the data, provided the lattice sizes are sufficiently big. We achieve this via a benchmark of our code implementing FEBs against the results of Ref.~\cite{Price-2013} for the Kitaev-Heisenberg model ($J_1 \neq 0$, $K_1 \neq 0$) with periodic boundary conditions. To minimize the finite size effects and further enhance equilibration we implement an iterative minimization and parallel tempering algorithms respectively, with system sizes ranging from 2400 up to 5400 sites. 

\begin{figure}[h!]
\centering
\includegraphics[scale=0.3,angle=0]{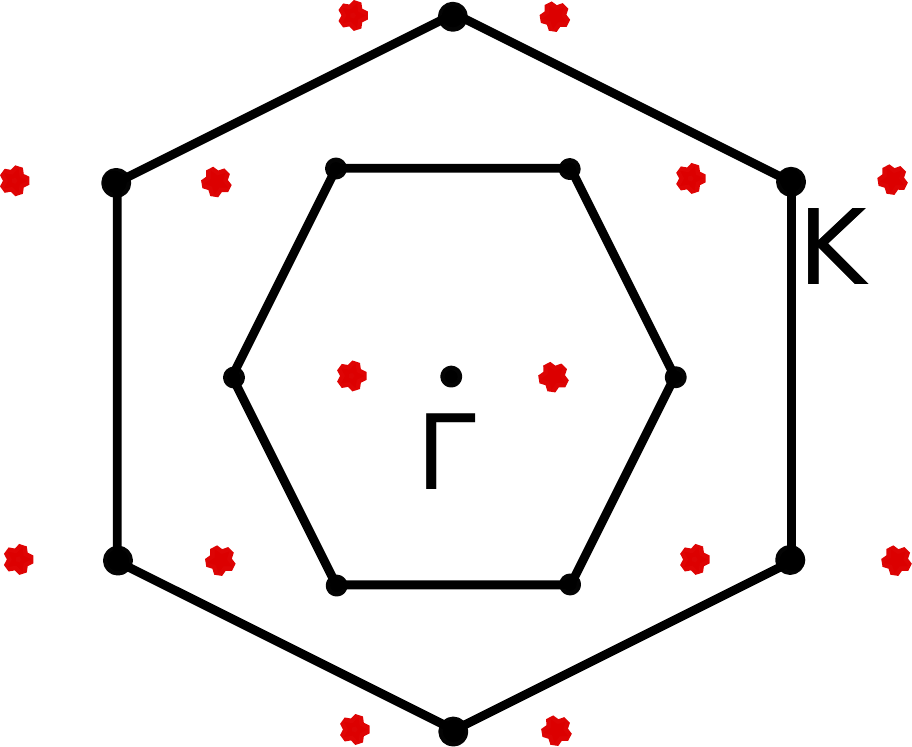}
\caption{Reciprocal space diagram of the honeycomb lattice showing in red the positions of the magnetic Bragg peaks corresponding to the ground state of $\alpha$--$\mathrm{Li}_2\mathrm{IrO}_3$. The black dots correspond to the allowed structural Bragg peaks $\tau$, while the red stars correspond to the satellite peaks $\tau + \mathbf{q}$. The inner hexagon is the first Brillouin zone, while the outer hexagon is the extended Brillouin zone.}
\label{sketch_S(q)}
\end{figure}

To identify the different states in the phase diagram we will rely on the study of the Fourier transform of the spin-spin correlation function
\be 
C_{ij} = \langle \mathbf{S}_i\cdot\mathbf{S}_j \rangle - \langle \mathbf{S}_i \rangle\langle \mathbf{S}_j \rangle\,,
\label{corrfunc}
\ee
where the average is taken over different Monte Carlo sweeps. For the ground state of $\alpha$--$\mathrm{Li}_2\mathrm{IrO}_3$ we expect to find peaks at positions indicated by the red stars in Fig.~\ref{sketch_S(q)} in accordance with the results of Ref.\cite{Williams-2016}. We can determine the propagation wavevector of the magnetic spirals from the maxima of the Fourier transform of Eq.~\ref{corrfunc}, as they will be located at the points $ \tau \pm \mathbf{q}$, with $\mathbf{q} = (\pm q, 0)$ the propagation wavevector. At the same time, and since Monte Carlo produces the spin pattern of the state, we confirm our results by calculating the relative angle between nearest neighbour spins in one spiral. 

To study the magnetization processes we will be interested in the ferromagnetic order parameter, 
\begin{equation}
|\mathbf{OP}_{FM}| = \frac{1}{N}|\sum_i\langle\mathbf{S}_{i}\rangle|\,,
\end{equation}
where the average is taken over different Monte Carlo sweeps, and the magnetization in the direction of an externally applied field,
\begin{equation}
OP_{M\cdot H} =\frac{1}{N}(OP_{FMx}H_{x}+OP_{FMy}H_{y}+OP_{FMz}H_{z})\,,
\end{equation}
where N is the number of sites, $\mathbf{S}_{i}$ represent the spin at site $i$, and $OP_{FMi}$ is the $i$-th component of the ferromagnetic order parameter. To probe the existence of a ferromagnetic order with a particular polarization the ferromagnetic order parameter has to be projected onto this direction. As we will include magnetic fields, we choose to study the projection of  $\mathbf{OP}_{FM}$ in the direction of the applied field. To estimate critical fields we will also employ the associated response function of the ferromagnetic order parameter, the magnetic susceptibility
\be
\chi = \frac{d \mathbf{FM}_{OP}}{d H} = \frac{1}{N}\frac{1}{T}\sum_i\left(\langle\mathbf{S}_{i}^2\rangle-\langle\mathbf{S}_{i}\rangle^2\right)
\ee
where T is the temperature of the simulation.

When everything is taken into consideration, our simulations consist of system sizes ranging from 2400 to 5400 Heisenberg spins, with a temperature consistently set at $T = 0.001$ in units of the dominant coupling of the model. We employ $2$x$10^6$ Monte Carlo sweeps, of which $10^5$ are used as equilibration steps, and the rest are employed to calculate observables. Furthermore, we average over 10 independent runs per data point. We interlace these sweeps with parallel tempering swaps, for 32 replicas, every 100 sweeps. Once the low temperature state was equilibrated we employ an iterative minimization algorithm with a threshold $10^{-20}$, to minimize the energy further. 

\section{Ground state properties}
\label{Gp}

\subsection{Nearest neighbour model, $\mathcal{H}(J_1, K_1, I_c \neq 0, I_d\neq 0)$}
In the following, we will show the results for the ground state properties of the model described by $\mathcal{H}(J_1, K_1, I_c \neq 0, I_d\neq 0)$ (Eq.~\ref{Hamiltonian}). The limit $I_d = 0$ corresponds to the $\mathcal{H}(J_1, K_1, I_c \neq 0, I_d = 0)$ model  (Eq.~\ref{Hamiltonian}), which we show in Appendix \ref{I_c}. 
\subsubsection{Phase diagram}
\begin{figure*}
\includegraphics[scale=1]{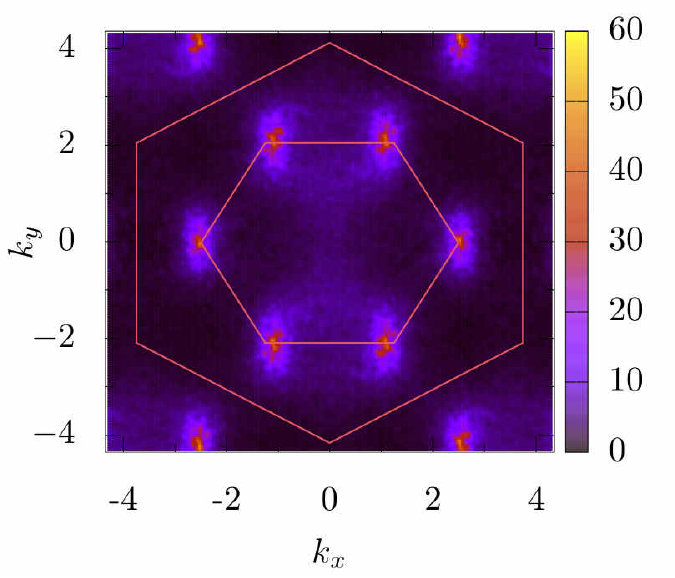}
\includegraphics[scale=0.4]{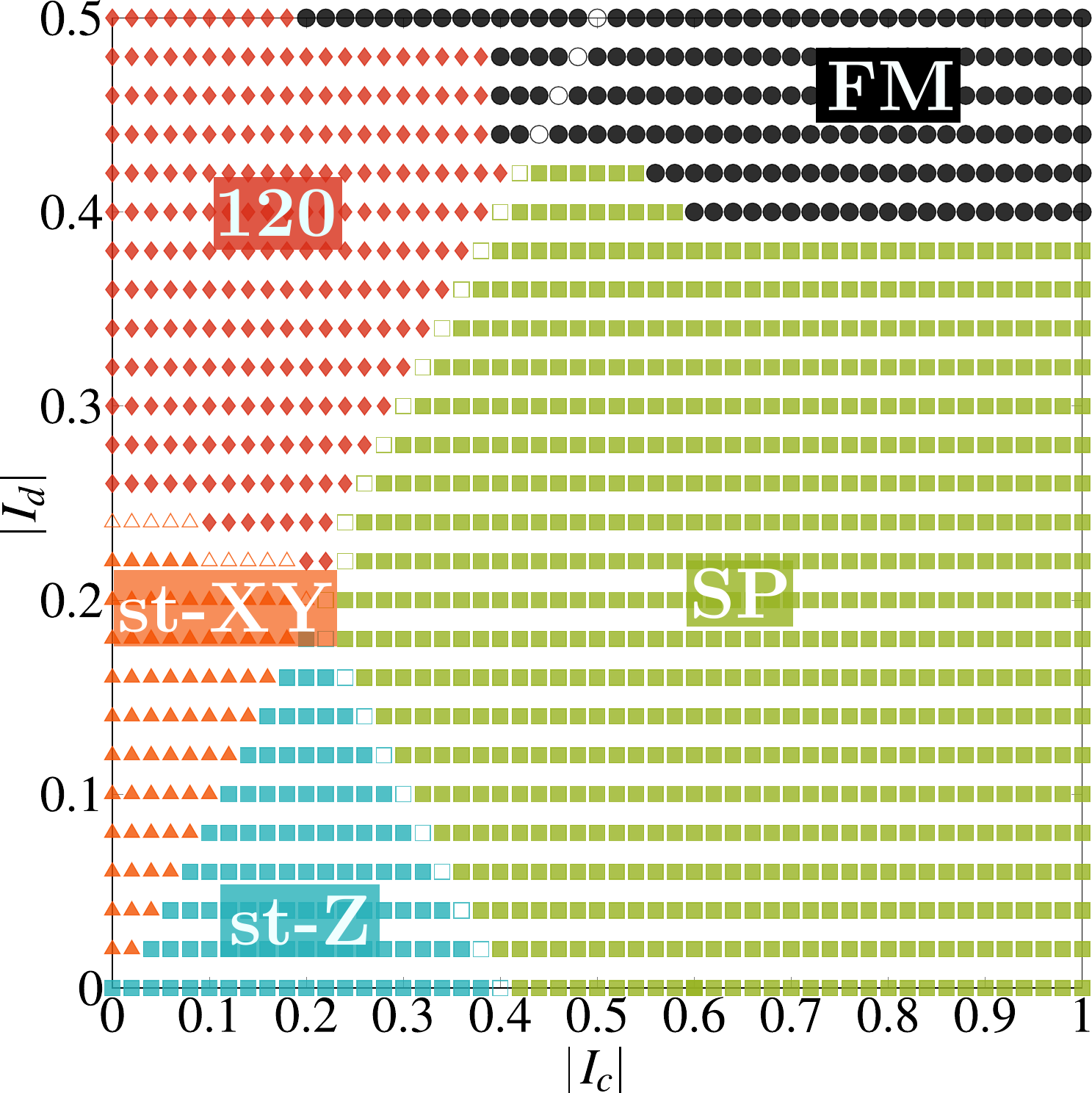}
\caption{{\it Left}: Fourier transform of the correlation function for the $120^o$ order present in the $I_c=0$ regime, for $I_d = -0.5$. The maxima are located close to the corners of the first Brillouin zone, indicating the tendency of the system to order according to a slightly distorted $120^o$ order.  {\it Right}: Phase diagram for the nearest neighbour model $\mathcal{H}(J_1, K_1, I_c \neq 0, I_d\neq 0)$. Blue squares represent the st-Z order, orange triangles are st-XY, red diamonds the $120^o$, green squares the spiral (SP), and black dots correspond to ferromagnetic order (FM). The open marks show the zone boundaries between the st-XY, st-Z, $120^o$, and incommensurate phases. The line of open circles in the ferromagnetic phase indicates the line along which vortex-like defects appear.}
\label{120-phdIcId}
\end{figure*} 

We study the model described in Eq.~\ref{Hamiltonian} employing as a starting point the suggested values of the exchange couplings given in Ref.~\cite{Williams-2016}. We will set fixed values for the Heisenberg and Kitaev interactions $J/|K|=0.2$, $K/|K|=-1$. To analyse the effect of bond dependent interactions we allow the couplings $I_c/|K|$ and $I_d/|K|$ to move in the range $\{-1, ... 0\}$.  Studying both the real space configuration of the spins, as well as the correlation function in reciprocal space we are able to map the phase diagram for finite $I_c$ and $I_d$. We show this phase diagram in Fig.~\ref{120-phdIcId}(right), where we observe a rich behaviour, with different commensurate-incommensurate transitions. The  green squares corresponds to the regions of the phase diagram where an incommensurate spin spiral state is found. 

The regime $I_c = 0$ and for moderate values of $I_d$ ($I_d > -0.25$) we observe a degenerate stripy phase. This state arise from the presence of magnetic domains exhibiting stripy phases polarized in the X (st-X phase) and Y (st-Y phase) directions. We denote this state by st-XY (orange triangles). For smaller $I_d$ we obtain a distorted $120^o$ order (red diamonds). This latter state was previously studied via a soft spin approximation\cite{Williams-2016}, and identified as an incommensurate spin spiral propagating in the vertical direction of Fig.~\ref{lattice}. The Fourier transform of the correlation function for this phase presents maxima close to the first Brillouin zone, Fig.~\ref{120-phdIcId}(left), while the spin pattern in real space shows that the state is that of a distorted $120^o$. For small enough $I_d$, the distortion in the $120^o$ is more pronounced, and we expect that in the limit $I_d \rightarrow -\infty$ we recover the perfect $120^o$ order. 

\begin{figure}[h!]
\centering
\includegraphics[scale=0.15]{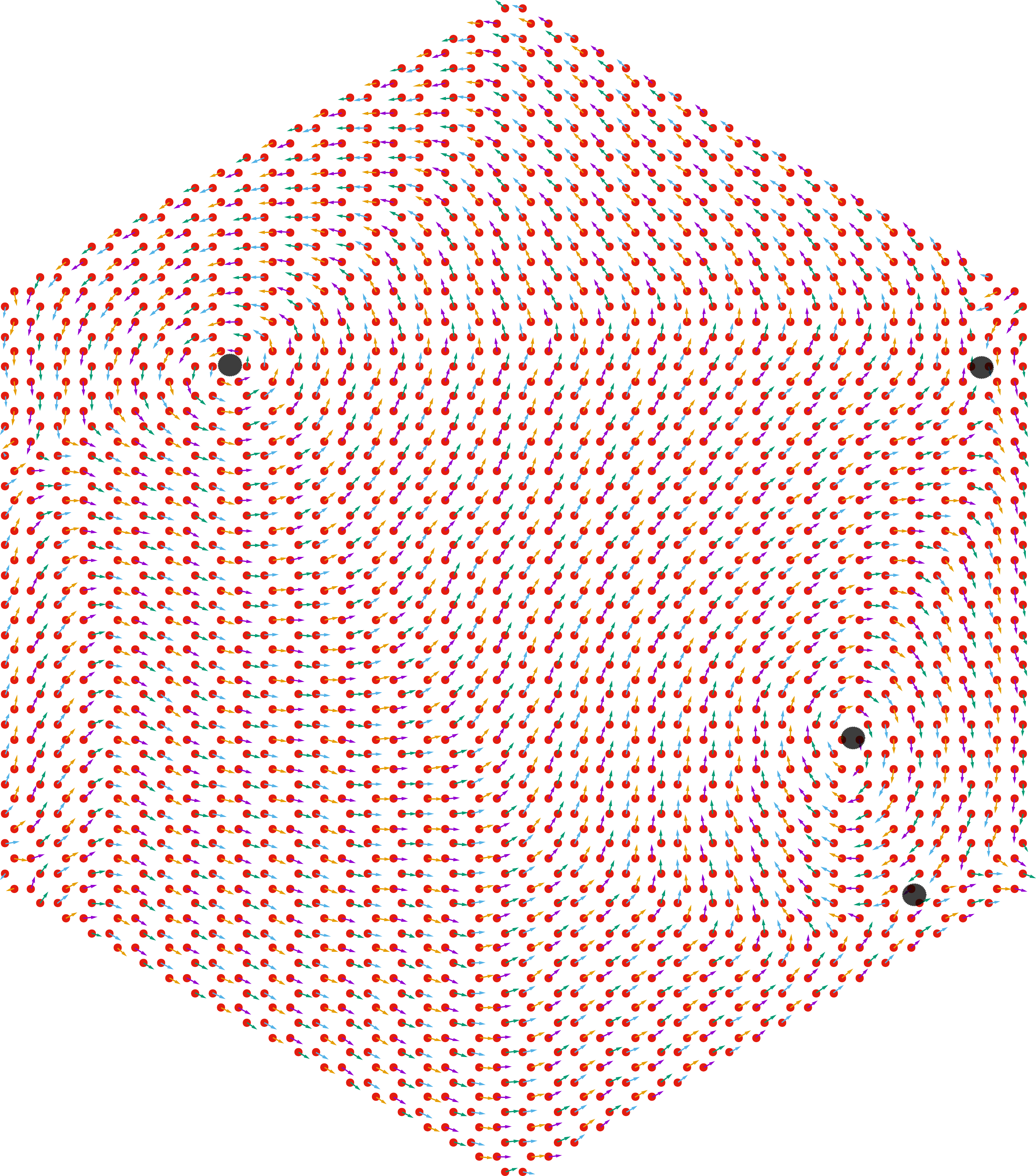}
\caption{Real space configuration of the ferromagnetic order for the model shown in Eq.~\ref{Hamiltonian} with parameters $I_c = I_d = - 0.5$. We show a snapshot of the spin configuration from the $ [111]$ direction. The spins are oriented in the lattice plane. We mark the vortex fluctuations by black dots. }\label{ferro}
\end{figure}

The degeneracy shown in the st-XY phase is expected since the Heisenberg-Kitaev model stabilizes a triple degenerated stripy phase \cite{Price-2013}. The inclusion of a small bond dependent interaction strengthening one particular bond, breaks the degeneracy of the stripy phase.   When both $I_c$ and $I_d$ are non-zero, we see a clear separation of phases through the line $I_c = I_d$. When $|I_c| > |I_d|$ a counterrotating spiral state dominates the phase diagram. This phase reproduces the experimental results for $\alpha-\mathrm{Li}_2\mathrm{IrO}_3$ and will be studied in detail in the next section.

\begin{figure*}
\begin{minipage}{0.6\textwidth}
\includegraphics[scale=0.4]{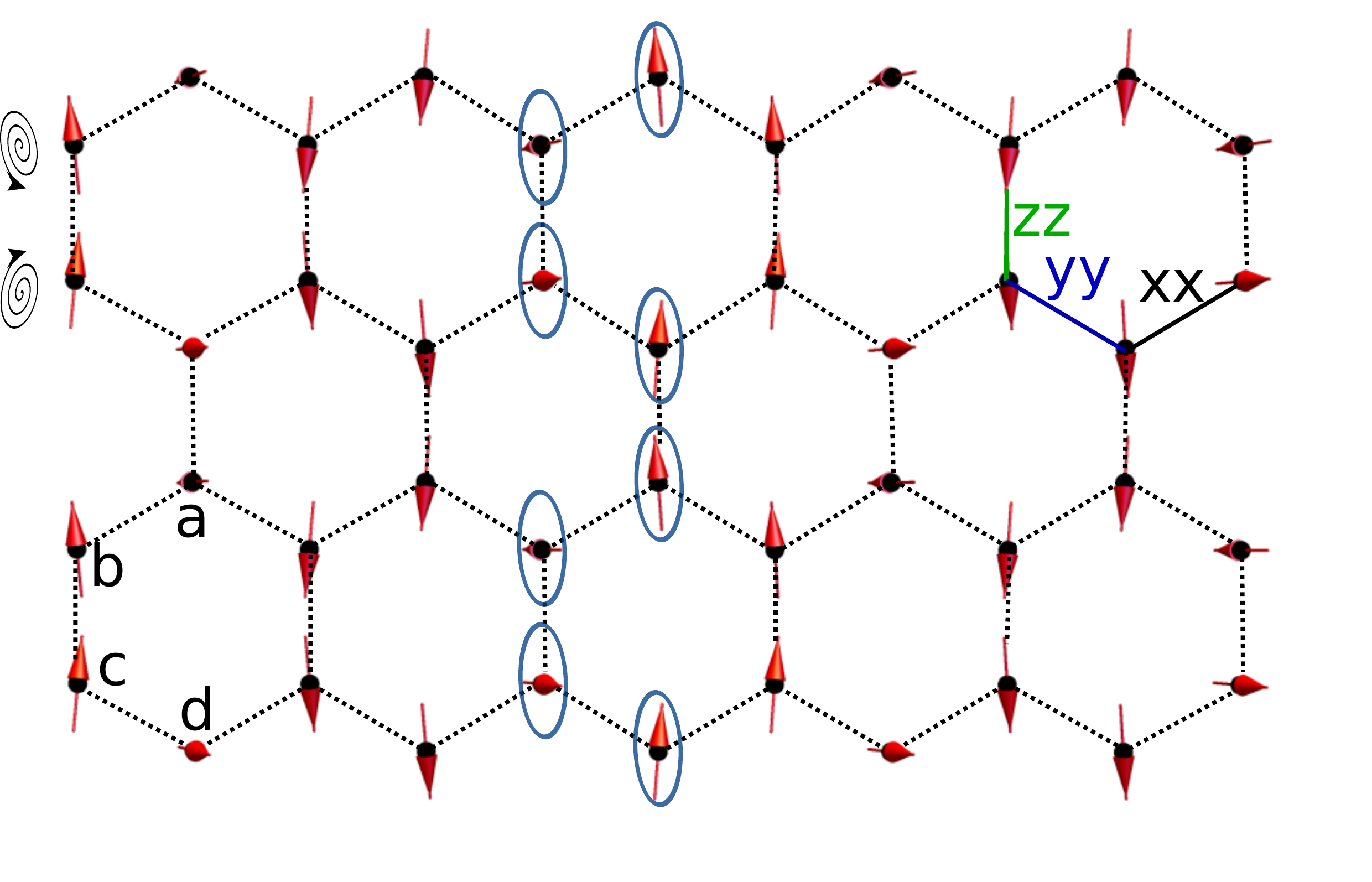}
\end{minipage}
\begin{minipage}{0.3\textwidth}
\includegraphics[scale=0.25]{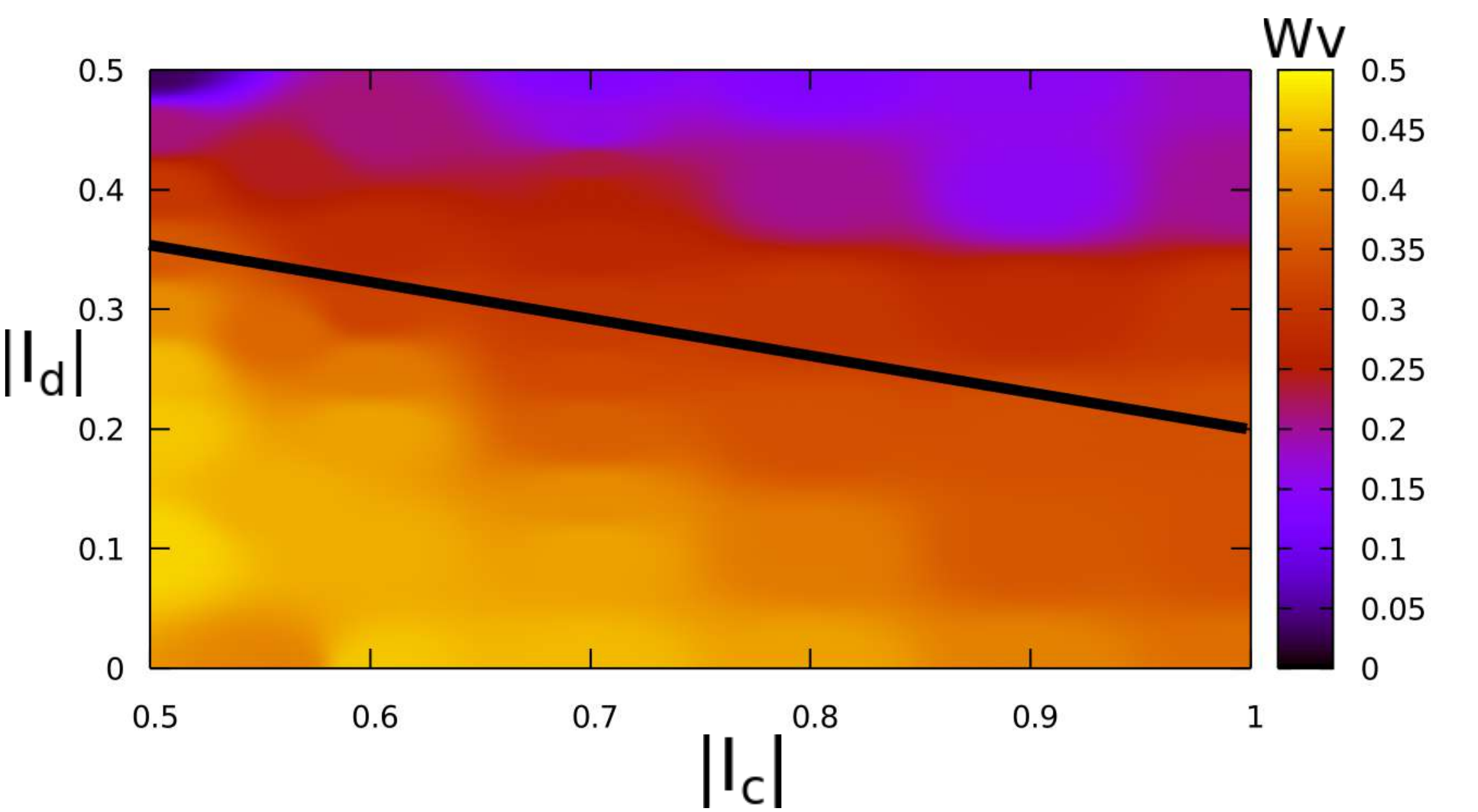}
\includegraphics[scale=0.25]{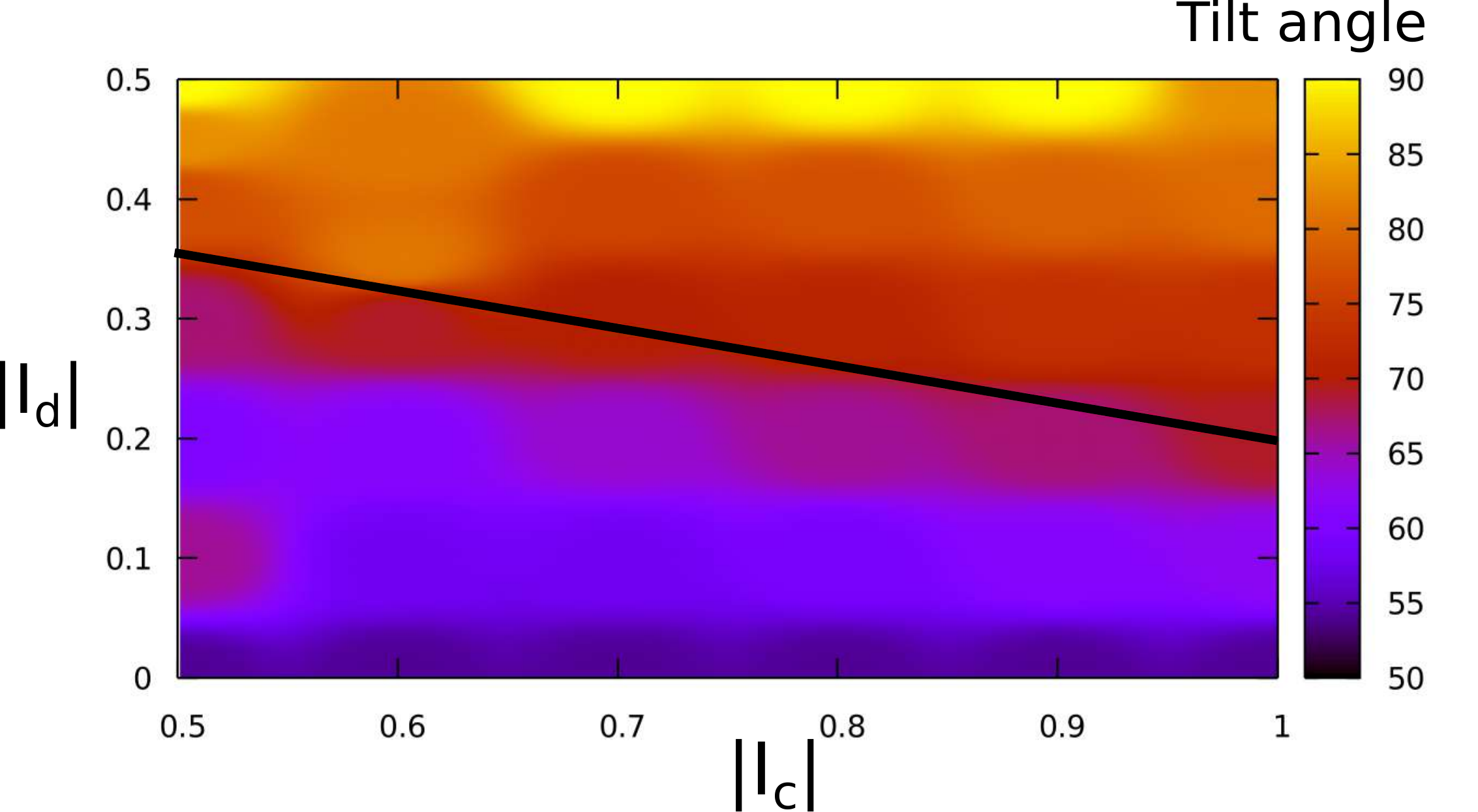}
\end{minipage}
\caption{{\it Left}: Spiral structure obtained from Monte Carlo simulations for the nearest and third neighbour models. We show here the results for the nearest neighbour model with coupling strengths $I_c = -0.5$ and $I_d = -0.35$. The results for the third neighbour model are the same as the ones for the nearest neighbour model. The wavevector and tilt angle coincide with the experimental results $q = 0.31(1)$ and $\theta \sim 80^o$. The circular drawings on the upper right indicate the rotation direction of two neighbouring spirals. Blue circles show the plane of rotation, with the plane tilt. By a, b, c, and d we indicate the four sublattices where spins of each sublattice form one spiral along the slab. Finally, and for the sake of completeness ,we indicate the bonds that correspond to the Kitaev interaction as defined in the model. {\it Right}:  Heat map of the wavevector (top) and tilt of the rotation plane's angle (bottom) for the spiral phase in the nearest neighbour model, for the regime $I_c/K > 0.5$. The black like represents the line of wavevector $q=0.32$, in units of $2\pi$.}
\label{spiral}
\end{figure*}

Finally, we note that, at high enough values of $|I_c|$ and $|I_d|$ the model adopts a ferromagnetic order with polarization in the lattice plane. This order lives on both sides of the $I_c = I_d$ line, presenting a ferromagnetic order with a net magnetization in the direction of the $zz$-bonds when $|I_c| > |I_d|$, and perpendicular to it in the  $|I_c| < |I_d|$ case. The line in the phase diagram separating both polarizations is special given that the low energy fluctuations in this regime are of a different nature. In this ferromagnetic state, and over the line $I_c=I_d$, the low energy fluctuations are vortex-like, and appear in pairs of vortex-antivortex fluctuations. An example of a spin pattern presenting this behaviour is shown, for a calculation over 2400 sites, in Fig.~\ref{ferro}. 

These vortex fluctuations where studied in systems ranging from 24 to 5400 lattice sites, and they consistently appear in all the studied system sizes over this particular line in the phase diagram, which rules out this behaviour as a finite size effect.  A detailed study of these vortices is beyond the scope of this paper, and is left for future study.

\subsubsection{Spiral properties}

A big part of the phase diagram on Fig.~\ref{120-phdIcId} is dominated by an incommensurate phase (green squares). This state represents an incommensurate counterrotating spiral which propagates in the horizontal direction according to Fig.~\ref{lattice} (the direction perpendicular to the {\it zz}-bonds). In the regime $I_d = 0$ the wavevector varies between $0.5$ and $0.4$ (in units of $2\pi$) and the plane of rotation is tilted with respect to the lattice plane by $54^o$, i.e, the rotation plane is oriented parallel to the XY-Cartesian plane. 

In the regime where both $I_c \neq 0$ and $I_d \neq 0$ the commensurate phases survive for values of $I_d$ down to $-0.4$ and for values of $I_c$ such that $|I_c| > |I_d| - 0.4$ for $|I_d| < 0.2$, and $|I_c| > |I_d|$ for $ 0.2 < |I_d| < 0.4$.  In this regime, some properties of the spiral phase found for the $I_d = 0$ case are modified. As $I_c$ and $I_d$ are varied, the wavevector varies between $0.5$ and $0$ (wavevector $0$ correspond to the onset of ferromagnetic order). The rotation plane's tilt now also varies along the phase diagram, between $\sim 50^o$ (consistent with the spin spiral known to appear at $I_d = 0$) and $\sim 90^o$. We show in Fig.~\ref{spiral} a real space pattern of the spin spiral, at a point in the phase diagram which reproduces the experimental results.

In Fig.\ref{spiral}(right)we show two heat maps, one for the wavevector and another for the tilt angle of the rotation plane, respectively. We can see the variation of these quantities in the phase diagram. In the heat map corresponding to the wavevector (Fig.~\ref{spiral}(top right)) we indicate with a black line the zone of wavevector $q=0.32$ (in units of $2\pi$) and superimpose this line over the tilt angle heat map (Fig.~\ref{spiral}(bottom right)). Please note that this mark is a guide to the eye, it does not arise from a fit to the data.

\begin{figure}[h!]
\centering
\includegraphics[scale=1]{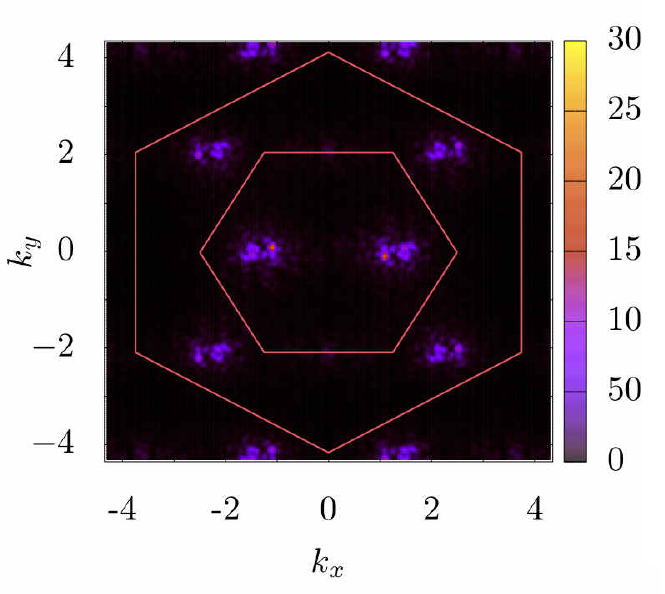}
\caption {Correlation function for the incommensurate spin spiral state at $I_c=-0.5$ and $I_d = -0.35$. We observe maxima as satellite points of the $\Gamma$ point, and secondary maxima as satellites of the $K$ points.}\label{sp2}
\end{figure}

Further confirming the existence of counterrotating spin spirals reproducing the experimental results, we show the Fourier transform of the correlation function in Fig.~\ref{sp2}. This figure was calculated for the parameters $I_c = -0.5$ and $I_d = -0.35$, which reproduce the wavevector and rotation plane tilt found in experiments.
The correlation function presents maxima at the expected positions. They appear as satellites of the $\Gamma$ point, with positions $\tau + \mathbf{q}$, where $\tau$ indicates the location of the $\Gamma$ points and $\mathbf{q} = (\pm 0.32, 0)$ in units of $2\pi$, which coincide with the experimental results mentioned in the introduction.

\subsection{ Anisotropic third neighbour model, $\mathcal{H}_A(J_{1,3}, K_{1,2}, \Gamma_{1,2})$ }

\begin{table}[h]
  \begin{center}
    \begin{tabular}{ | l | c | c | c |}
      \hline
	Bond & $J_n$ & $K_n$ & $\Gamma_n$ \\ \hline
	$X_1$, $Y_1$ & $\mathbf{-1.0}$ & $\mathbf{-13.0}$ & $\mathbf{+6.6}$ \\ \hline
	 $Z_1$ & $\mathbf{-4.6}$ & $\mathbf{-4.2}$ & $\mathbf{+11.6}$ \\ \hline
	$X_2$, $Y_2$ & $+0.9$ & $\mathbf{-2.9}$ & $\mathbf{+3.0}$ \\ \hline
	 $Z_2$ & $-0.9$ & $+0.1$ & $\mathbf{+1.5}$ \\ \hline
	$X_3$, $Y_3$ & $\mathbf{+4.7}$ & $-0.2$ & $0$ \\ \hline
	 $Z_3$ & $\mathbf{+4.4}$ & $+0.4$ & $-0.1$ \\ \hline
            \end{tabular}
  \caption{Values of the anisotropic interactions as obtained in Ref.~\cite{Winter2016}. All interactions are given in meV. For the study of the anisotropic model we have selected those interactions which are greater than one (bold).}
  \label{Winter_ani}
  \end{center}
\end{table}

The third neighbour model, $\mathcal{H}(J_{1,3}, K_{1,2}, \Gamma_{1,2})$, was proposed by Winter {\it et. al.} in Ref.~\cite{Winter2016} where it was determined (via a combination of DFT and exact diagonalization on a small cluster of hexagons) that this model presents large anisotropies in various parameters. The isotropic model, $\mathcal{H}_I(J_{1,3}, K_{1,2}, \Gamma_{1,2})$, presents incommensurate spin spirals in the ground state (results for the isotropic model are shown in Appendix.~\ref{ITM}). However, since the model is bond isotropic, the spin spirals can propagate in three symmetry allowed directions. The powder MRXD measurements performed on $\alpha$-Li$_2$IrO$_3$ indicate that the material does not seem to show any type of degeneracy of the ground state.  While this degeneracy could be broken by order by disorder effects, the fact that the model proposed by Winter {\it et. al.} \cite{Winter2016} exhibits strong bond anisotropies seem to indicate that these anisotropies need to be included in the model to break the degeneracy. In the following we will treat the model shown in Eq.~\ref{Hamiltonian_W2} including these anisotropies and study what their effect is on the phase diagram and the spiral properties of the system.

The values of the exchange parameters, as obtained in Ref.\cite{Winter2016}, are shown in table \ref{Winter_ani}. Studying this table it becomes clear that an analysis considering that the third neighbour model is bond isotropic is an excessive simplification. We will introduce the anisotropies in the model as shown in Eq.~\ref{Hamiltonian_W2} in the following way: for a given coupling $\alpha$ we have a bond anisotropy which differentiates the X and Y bond ($\alpha_{XY}$) from the Z bond ($\alpha_Z$).  We will define $\alpha_m$ as the bond average of the exchange coupling, $\alpha_{XY} +\alpha _{z}/2$, and $\delta$ as the anisotropy constant (with an appropriate sign) such that (for a direct comparison please look at Eqs.~\ref{aniso})
\be
\alpha_{XY} = \alpha_m - \delta \quad \alpha_{Z} = \alpha_m + \delta
\ee
This way, calculating the bond average from Table~\ref{Winter_ani} we can determine what the anisotropy for each exchange is. Please note that after this process is performed all couplings and anisotropy constants are expressed in terms of $|K^{XY}_1| = 13$. The values obtained for the anisotropy constants are given in Table.~\ref{wint_ani_c}.
\begin{table}[h!]
\begin{center}
  \begin{tabular}{ | l | c |}
     \hline
	& $\delta$ \\ \hline
      $J_1$ & 0.14 \\ \hline
      $K_1$ & 0.34 \\ \hline
      $\Gamma_1$ & 0.195 \\ \hline
      $K_2$ & 0  \\ \hline
      $\Gamma_2$ & -0.06 \\ \hline
      $J_3$ & 0 \\ \hline
      \end{tabular}
  \caption{Values of the anisotropy constants for the anisotropic third neighbour model, extracted from Ref.~\cite{Winter2016} following the process described in the text. All values are given in terms of $|K_1|$.}
  \label{wint_ani_c}
\end{center}
\end{table}

When we study the bond isotropic model with the exchange constants obtained from Table.~\ref{Winter_ani} we obtain a zig-zag order, which agrees with the results from Ref.~\cite{Winter2016}. In the bond anisotropic case, where the interaction couplings reduce to those shown in Table.~\ref{Winter_ani}, we again obtain a zig-zag state \footnote{Please note that in the phase diagrams shown in this paper no zig-zag order is shown, as we have concentrated in mapping a part of the diagram which exhibits incommensurate spirals. If we were to map the full phase diagram we would see (and we have confirmed this via numerical simulations) that, as in Ref.~\cite{Winter2016} for $J_3 \gtrsim 0.4$ a zig-zag phase is present.}. By changing the bond averages but maintaining the anisotropic parameters $\delta$ constant, we can map a phase diagram including anisotropies. Changing the bond averages is not a radical idea, since the values obtained by Winter {\it et. al.} \cite{Winter2016} have been obtained via exact diagonalization on small clusters. This, combined with the uncertainty in the crystal structure which has been resolved until this point indicates that, while the nature of the interactions might not change, their coupling strength could. 

\subsubsection{Phase diagram}
\label{W_ani_pHd}

\begin{figure}[h!]
\includegraphics[scale=0.45]{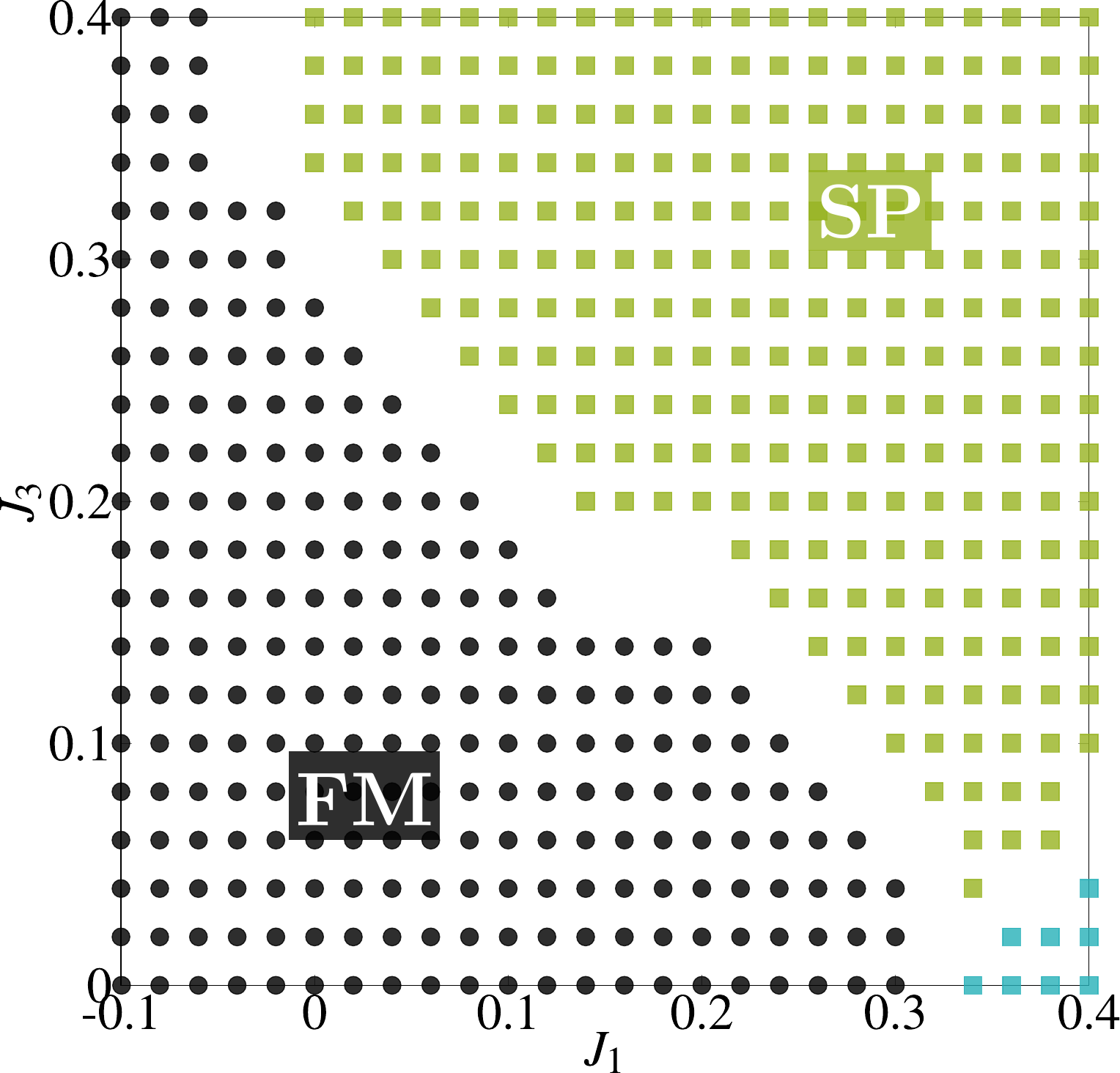}
\caption{Phase diagram for the anisotropic third neighbour model shown in Eq.~\ref{Hamiltonian_W2}. Black dots correspond to ferromagnetic order (FM), light blue squares to stripy order (st-Z). The incommensurate states are represented by green squares (SP). }\label{phd_W_aniso}
\end{figure}

The phase diagram of the anisotropic third neighbour model is shown in Fig.~\ref{phd_W_aniso} top. We map the phase diagram for the following  coupling strengths: a dominant nearest neighbour Kitaev coupling $K_1 = -1$ supported by (here and in the following, all exchange couplings are given in units of $|K_1|$) $K_2 = -0.275$, $\Gamma_1 = 1$, $\Gamma_2 = 0.275$, $J_1 \in (-0.1,..., 0.4)$, and $J_3 \in (0,..., 0.4)$. The phase diagram presents two dominant phases, a ferromagnetic state (black dots) and an incommensurate state (green squares). At the bottom right corner a small stripy phase is observed (blue squares). A comparison with the phase diagram for the isotropic model (Fig.~\ref{phd_W_iso}) indicates that both the ferromagnetic and the spin spiral states are displacing the rest of the phases.

The ferromagnetic phase present in this model exhibits an in plane net magnetization, in the direction parallel to the $zz$-bonds. Domain walls separate two domains exhibiting the two possible orientations of the polarization, where these domain walls are realized by spins aligned antiferromagnetically. As $J_1$ and $J_3$ are increased the domains multiply, until the system enters an incommensurate phase.

\begin{figure}[h!]
\includegraphics[scale=0.13]{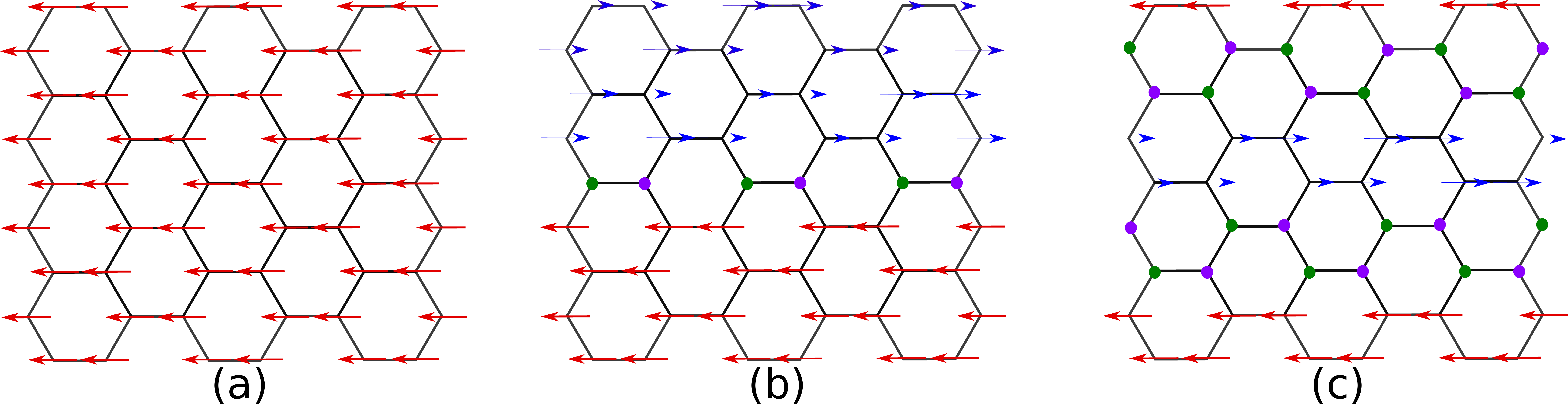}
\caption{Sketch depicting the commensurate/incommensurate transition in the anisotropic third neighbour model projected on the [111] plane for $J_3 = 0.2$. {\it (a)} Ferromagnetic state at $J_1 = -0.1$. {\it (b)} When $J_1$ is increased (here we show $J_1 = 0$) ferromagnetic domains are separated by antiferromagnetic domain walls (green dots depict spins pointing outside the page, and purple are spins pointing inside the page). {\it (c)} at a critical value of $J_1 \sim 0.15$ the system enters the spiral phase which can be depicted in the [111] plane as alternating ferromagnetic/antiferromagnetic domains.  }\label{phd_W_aniso2}
\end{figure}

This commensurate-incommensurate transition can be seen as the generation of magnetic domains overpowering this ferromagnetic state. In Fig.~\ref{phd_W_aniso2} we show a sketch of the projected spin pattern on the [111] plane (the lattice plane) for a cut through a fixed value of $J_3$, where the system transitions from a ferromagnetic to a spiral state. To simplify the argument we have assumed for this discussion that the angles between the spins in the resulting spirals are $45^o$, and there is no tilt angle in the rotation plane \footnote{The reason for choosing this particular example is that of convenience, given that at this angle and plane tilt, the spin spiral can be seen as alternating ferromagnetic/antiferromagnetic domains of size two. For another wavevector there would also be ferromagnetic and antiferromagnetic domains but the size of these regions would not be the same.}. For this particular example, at small values of $J_1$ the state presents no domain walls (Fig.~\ref{phd_W_aniso2} (a)). As $J_1$ is increased domain walls start to span the length of the system, separating big domains of ferromagnetic order (Fig.~\ref{phd_W_aniso2} (b)) with opposing polarization vectors. For even bigger $J_1$ the system now contains ferromagnetic and antiferromagnetic domains spanning two sites in the vertical direction, each. This is seen in Fig.~\ref{phd_W_aniso2} (c), which corresponds to spin spirals of a wavevector such that the angle between spins is $45^o$. 

\subsubsection{Spiral properties}

\begin{figure}[h!]
\centering
\includegraphics[scale=0.95]{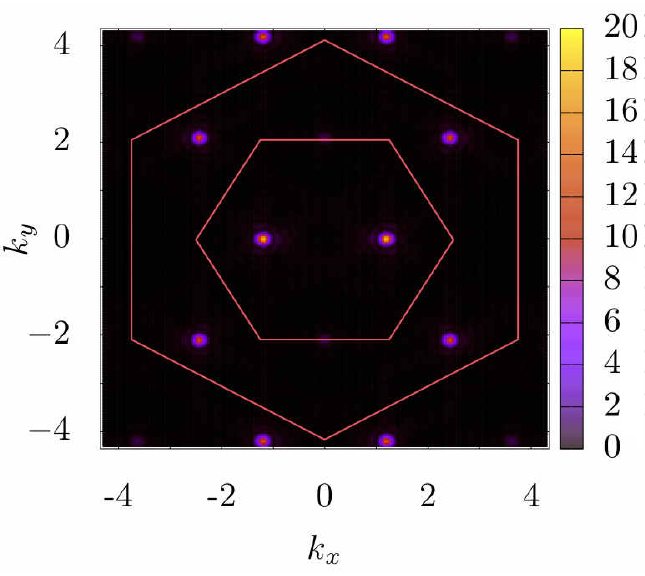}
\caption{Fourier transform of the correlation function for the spiral phase ($J_1 = 0.3$, $J_3 = 0.3$). This state reproduces the experimental features: wavevector $(q,0)$ with $q=0.32$, and tilt of the plane of rotation of $80^o$.}
\label{Sq_W_aniso}
\end{figure}

As can be seen from Fig.~\ref{phd_W_aniso}, a big part of the phase diagram is dominated by an incommensurate state.  In Fig.~\ref{Sq_W_aniso} we show the Fourier transform of the spin pattern obtained from this model (the real space spin pattern is shown in Fig.~\ref{spiral}(left)). From this we observe that the maxima are located inside the Brillouin zone as satellites of the $\Gamma$ point, and secondary maxima appear as satellite points of the $K$ points. From the Fourier transform we can further extract a wavevector $(q,0)$, $q=0.32$ in units of $2\pi$, which coincides with the analysis performed on the real space spin pattern which we show below. Overall, this is consistent with the MRXD results.

The incommensurate state are spin spirals of the same nature as those found for the nearest neighbour model studied in the previous section. In essence, the state is such that planar spirals propagate in the direction perpendicular to the $zz$-bonds. Furthermore, the spirals counterotate, with those formed by spins on sublattice $a$ and $c$ rotating with opposite chirality to those formed by the spins of sublattices $b$ and $d$. A real space pattern of the spirals can be observed in Fig.~\ref{spiral}(left), where we show a spiral which coincides with the experimental results for $\alpha$-Li$_2$IrO$_3$, exhibiting a wavevector $(q,0)$ where $q=0.32$ in units of $2\pi$, and a tilt of the rotation plane of $\sim 80^0$.

Comparing this spiral with that found for the  bond isotropic model, it is not surprising to notice that the effect of the anisotropies was that of destroying the degeneracy of the spiral phase. Recall that in the bond isotropic model our spirals were planar spirals, but also degenerate, this degeneracy arising from the fact that the Hamiltonian retains the discrete Kitaev symmetry. Thus the spirals were free to propagate in three possible directions. 

\begin{figure}[h!]
\includegraphics[scale=0.27]{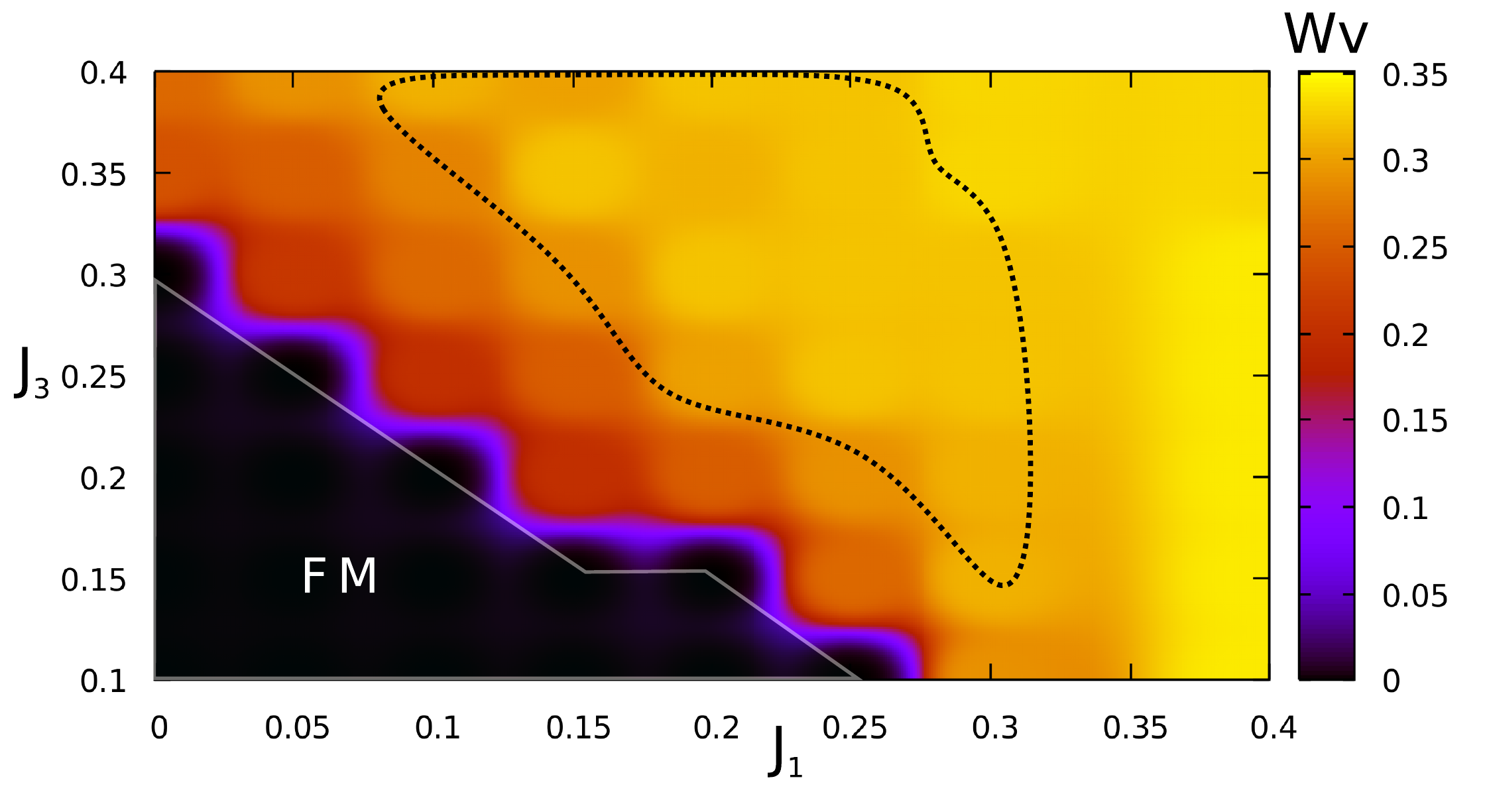}
\includegraphics[scale=0.27]{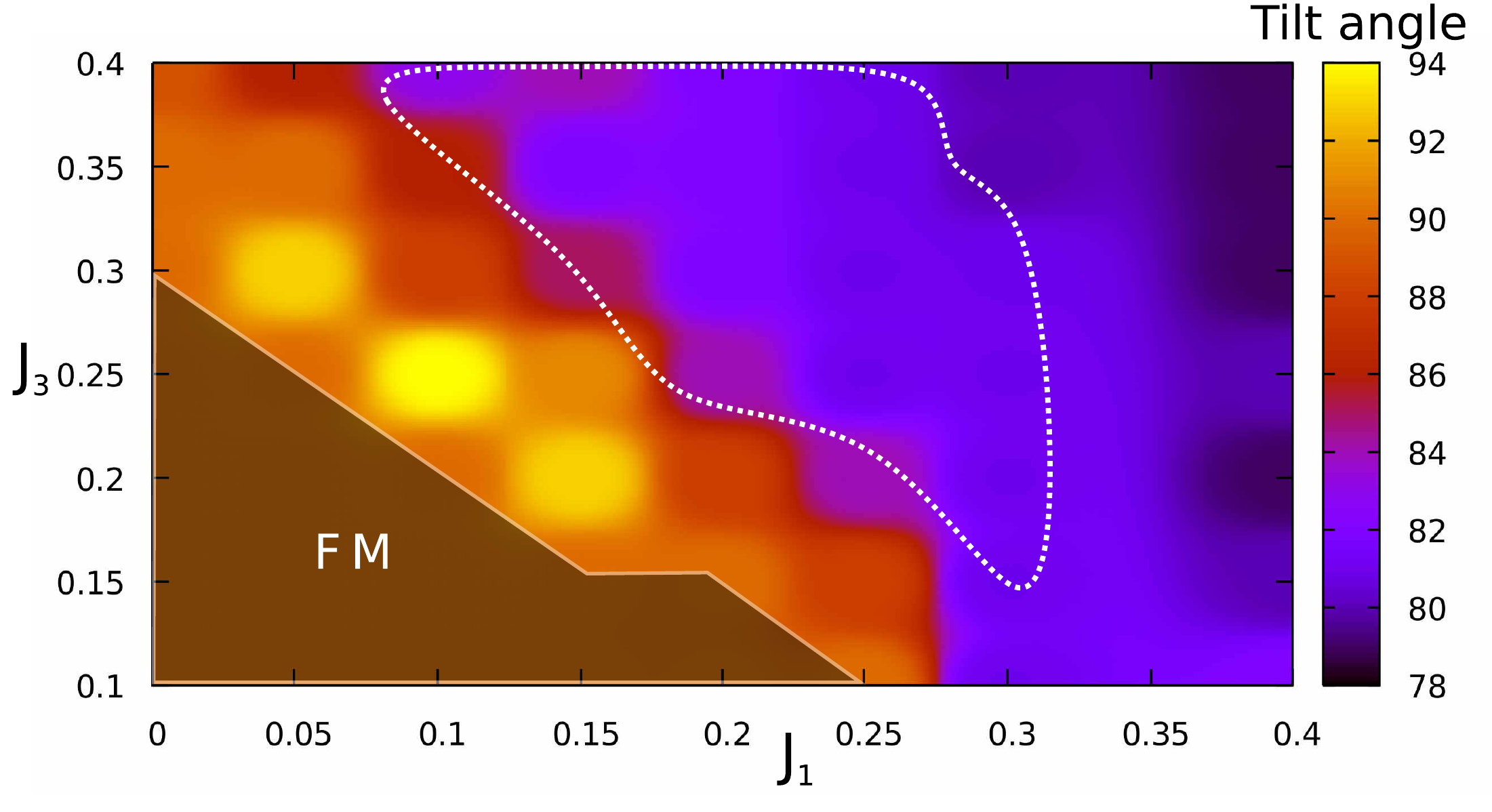}
\caption{Heat map of the wavevector (top) and rotation plane's tilt (bottom) for the spiral phase in the anisotropic third . The dotted line indicates the region of the wavevector map where the wavevector coincides with the experimental value $q = 0.32(1)$. Superimposing this dotted line on the tilt angle heat map we show there is a broad region of the phase diagram where the tilt of the rotation angle is $80^o$, as was seen in experiments. Please note that these heat maps and the dotted lines are a guide to the eye, where an interpolation has been performed between the points obtained from our simulations. Nonetheless our simulations clearly show an extended region where the experimental values are recovered. }\label{spiral2}
\end{figure}

Throughout the phase diagram, we notice that the wavevector and tilt angle change. In Fig.~\ref{spiral2} we show a heat map of the variation of the wavevector (Fig.~\ref{spiral2}(top)) and of the tilt angle (Fig.~\ref{spiral2}(bottom)) as the exchanges $J_1$ and $J_3$ are modified.

\section{Magnetization processes}
\label{Mp}

Up to this point, we have performed an analysis of the possible minimal models for $\alpha$-Li$_2$IrO$_3$. All these models where proposed either on the basis of symmetry allowed interactions \cite{Singh2012, Rau2016, Rau2014, Khaliullin2005, Jackeli2009}, or on DFT studies \cite{Winter2016, Winter-2017}. Comparing the results for the models shown previously, and those shown in the appendices, we have determined that only two models reproduce the experimental features of the material. While all models present incommensurate spin spiral phases, only the nearest neighbour ($\mathcal{H}(J_1, K_1, I_c \neq 0, I_d\neq 0)$) and anisotropic third neighbour models reproduce not only the counterotating nature and wavevector of the spirals, but also the tilt of the rotation plane. 

Now we want to put forward a prediction regarding the magnetization behaviour of these models in a way which can be verifiable experimentally. The fact that the models which reproduce the experimental results have different range of interactions mean that more studies are needed to reduce the number of possible models further. For this we have chosen to study the magnetization processes of the different models by applying an external magnetic field in different directions. Given the different interactions and bond anisotropy of both models, the behaviour will be different for both models, depending on the direction of the applied field. As we will see, the magnetization processes of the nearest neighbour and anisotropic third neighbour models are radically different, which indicates an experimentally feasible way of probing whether one of this models is in fact a good representation of $\alpha$-Li$_2$IrO$_3$.

\subsection{Nearest neighbour model, $\mathcal{H}(J_1, K_1, I_c \neq 0, I_d\neq 0)$}
We will study the magnetization processes of the model given by Eq.~\ref{Hamiltonian}, for magnetic fields applied in three different  global directions: $H \parallel [111]$, $H \parallel [\bar{1}10]$, and $H \parallel [11\bar{2}]$. Of these three directions, $[111]$ corresponds to the direction perpendicular to the lattice plane, while $[\bar{1}10]$  ($[11\bar{2}]$) is the direction parallel (perpendicular) to the $zz$- bonds (see Fig.~\ref{lattice}). 

\begin{figure}[h!]
\includegraphics[scale=0.8]{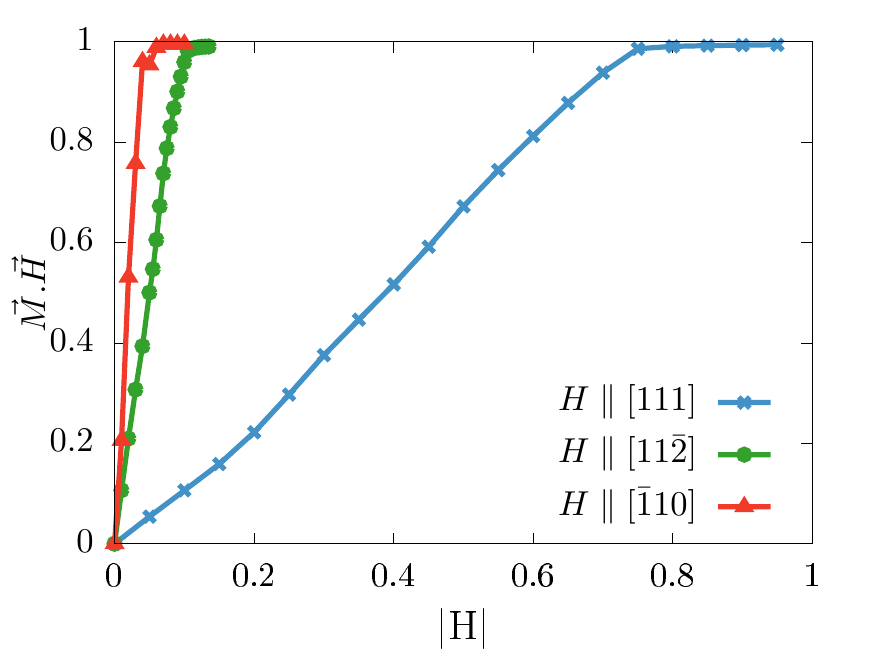}
\caption{ Magnetization for the model shown in Eq.~\ref{Hamiltonian} in the direction of the field as a function of field intensity. We show the results for the three field directions, $H \parallel [\bar{1}10]$ (red curve), $H \parallel [11\bar{2}]$ (green curve) and $H \parallel [111]$.  An easy plane anisotropy is observed where the critical field for the external field in the lattice plane saturating much faster than for the out of plane field.}\label{easy_axis}
\end{figure}

We will concentrate on the point of the phase diagram that reproduces the experimental results: $I_c = -0.5$ and $I_d = -0.35$ (in units of the Kitaev coupling $|K|$). Fig.~\ref{easy_axis} shows the magnetization curves for the different directions of the field. While the magnetization process is consistently the same for all field directions, ferromagnetic domains are created, which increase in size until they dominate the system, we observe different critical fields for the different field directions. For field directions in the lattice plane ($H \parallel [\bar{1}10]$, $H \parallel [11\bar{2}]$) the critical field is much lower than for the out of plane direction ($H \parallel [111]$). This indicates the existence of an easy plane anisotropy. 

The origin of this easy plane can be seen from the position of the experimentally relevant point in the phase diagram, as this point is close to the phase boundary between the spiral and the ferromagnetic phase. We have seen before that the ferromagnetic phase is such that the polarization vector is in the lattice plane, in a direction that depends on the values of $I_c$ and $I_d$. Applying a magnetic field in the direction of one of the bonds then is equivalent to reinforcing the ferromagnetic Ising terms that act on those bonds. Since these bonds reinforce the tendency to order in plane, then the stabilization of ferromagnetic order in the lattice plane is enhanced when an in plane field is applied, and the system transitions towards and in plane ferromagnet easily. Furthermore, since at these values of $I_c$ and $I_d$ the ferromagnetic phase has a polarization vector parallel to the {\it zz}-bonds (the $[\bar{1}10]$ direction), then the magnetic field in this direction possesses the smallest critical field.  

We will also compare the magnetization in the direction of the field ($OP_{M\cdot H} = \boldmath{M}.\boldmath{H}$) with the ferromagnetic order parameter ($OP_{FM}$) and analyse the susceptibility, $\chi$, of  $OP_{FM}$, to determine the critical field at which ferromagnetic order is realized. The ferromagnetic order parameter will probe the system for the presence of a ferromagnetic state, while $OP_{M\cdot H} $ will probe whether the polarization vector is in the direction of the field or not. 

\begin{figure*}
\begin{minipage}{0.3\textwidth}
\includegraphics[scale=0.65]{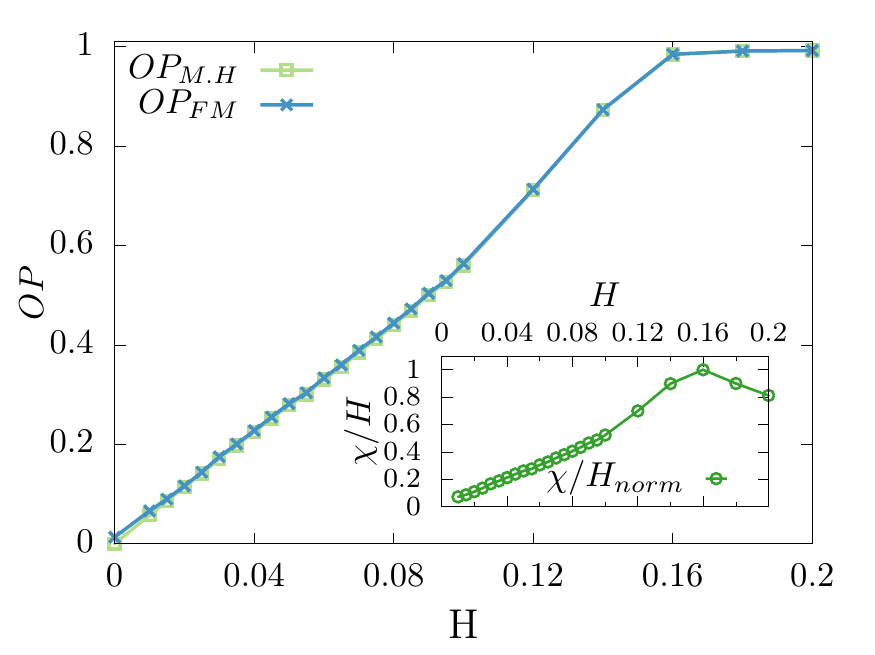}
\end{minipage}
\begin{minipage}{0.3\textwidth}
\includegraphics[scale=0.65]{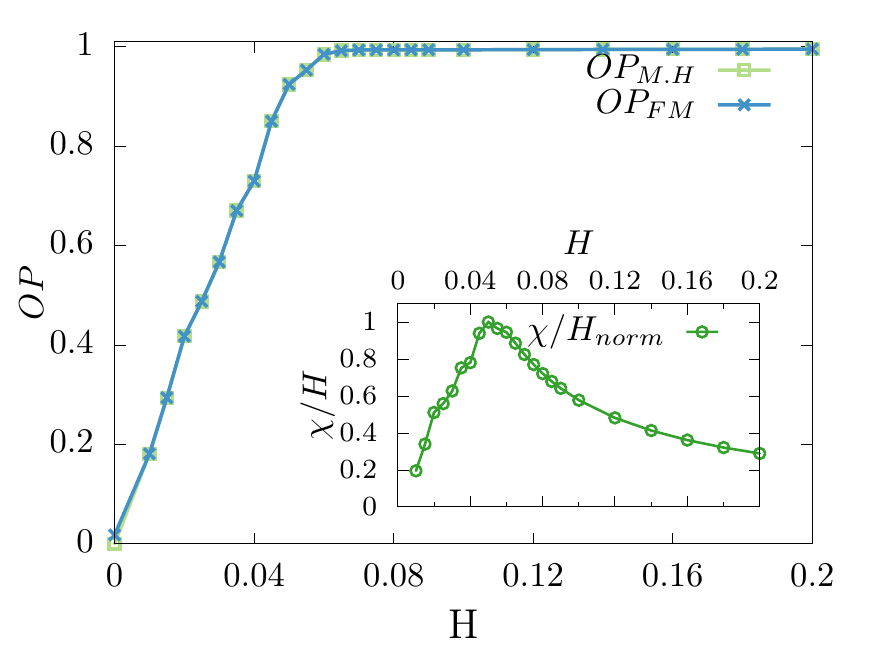}
\end{minipage}
\begin{minipage}{0.3\textwidth}
\includegraphics[scale=0.65]{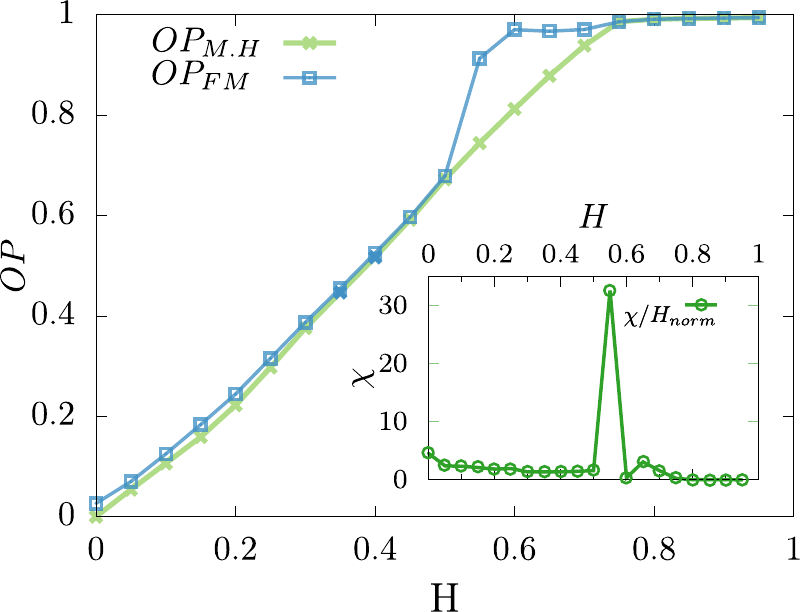}
\end{minipage}
\caption{Magnetization in the global $[11\bar{2}]$ (left), $[\bar{1}10]$ (center), and $[111]$(right) directions (green) and ferromagnetic order parameter (blue) as a function field intensity, for the $I_cI_d$-model {\it Insets} susceptibility of $OP_{FM}$ as a function of field.}
\label{magsic}
\end{figure*}

For both in plane direction,  $[11\bar{2}]$ and  $[\bar{1}10]$ (Figs.~\ref{magsic}(left) and \ref{magsic}(center)), the behaviour in $OP_{FM}$ and $OP_{M\cdot H} $ is the same. The magnetization monotonically increases until saturation is reached at rather small critical fields (in comparison with the relevant couplings of the model). The susceptibility curves show a peak that indicates a phase transition towards the fully polarized state. 

In the case of the field in the $[111]$ direction (Fig~\ref{magsic}(right)) the behaviour of $OP_{M\cdot H} $ and $OP_{FM}$ is rather different. Both curves increase monotonically up to $H = 0.5$ (unless stated otherwise, the magnetic fields are given in units of $|K|$), at this point they separate into two different behaviours. While $\boldmath{M}.\boldmath{H}$ keeps slowly growing until saturation at approximately $H=0.8$, the ferromagnetic order parameter suddenly reaches saturation at $H=0.6$. This indicates a stable intermediate state, a ferromagnetically ordered phase, in which the ferromagnetic state is realized in a direction not parallel to the applied field. The resulting ferromagnet is a state where the spins are out of plane. The polarization vector of this state can be partitioned into two components. A component parallel to the external field, and one perpendicular to it. In the intermediate state, the perpendicular component is non-zero and aligned along the {\it zz}-bonds.

We determine the critical fields by analysing the susceptibility response. We obtain a critical field  (in units of $|K|$) $H_{crit} = 0.05$ for a field direction $[\bar{1}10]$, $H_{crit} = 0.1$ for the magnetic field in the $[11\bar{2}]$ direction, and $H_{crit} = 0.8$ for the field in the $[111]$ direction. 

\subsection{Anisotropic third neighbour model, $\mathcal{H}_A(J_{1,3}, K_{1,2}, \Gamma_{1,2})$}

\begin{figure}[h]
\includegraphics[scale=0.78]{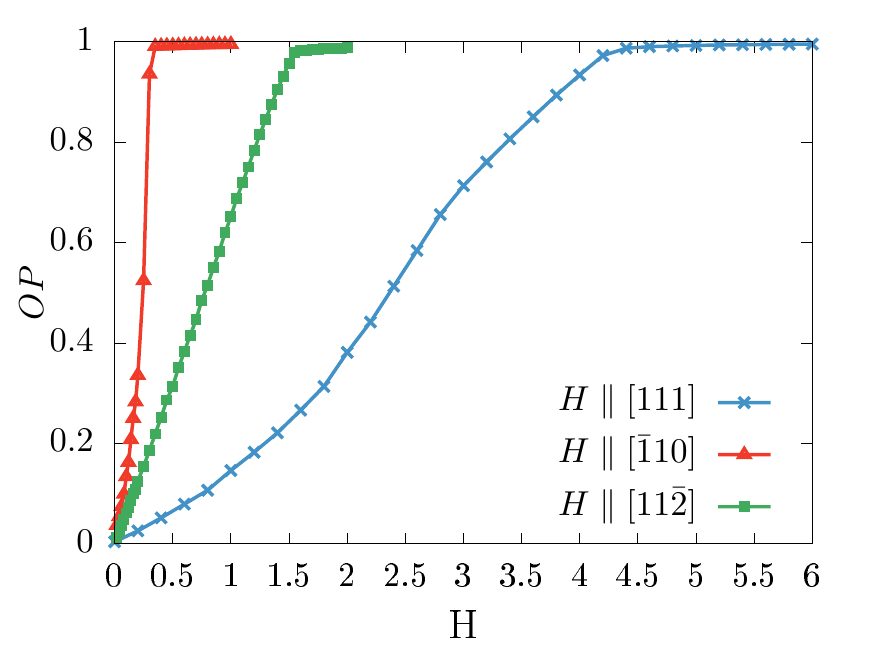}
\caption{Magnetization in the direction of the field, for the model shown in Eq.~\ref{Hamiltonian_W2}, as a function of field intensity for three different field directions: $H \parallel [111]$ (blue curve), $H \parallel [\bar{1}10]$ (red curve), and  $H \parallel [11\bar{2}]$ (green curve). The system exhibits a an easy axis anisotropy in the $[\bar{1}10]$ direction.}\label{W_easy_axis}
\end{figure}

In this section we will study the magnetization processes for the bond anisotropic version of the model shown in Eq.~\ref{Hamiltonian_W2}.  We will employ the same values of the exchange couplings as those in table \ref{Winter_bonds} and the same field directions as for the nearest neighbour model, $H \parallel [111]$, $H \parallel [\bar{1}10]$, and $H \parallel [11\bar{2}]$. As we shall see, the magnetization processes change drastically from those in the nearest neighbour model, given the strong off-diagonal and further neighbour interactions present in the third neighbour model. 
In Fig.~\ref{W_easy_axis} we show the magnetization in the direction of the field for the three field directions studied. We observe that the polarized state is reached at $H \sim 0.25$ (in units of $|K_1|$) in the $[\bar{1}10]$ direction, while for the magnetic field in the $[11\bar{2}]$ direction the critical field is $H \sim 1.5$ and $H \sim 4$ for the field in the $[111]$ direction. This behaviour points towards the existence of an easy axis anisotropy in the $[\bar{1}10]$ direction (please remember that the $[\bar{1}10]$ direction is the direction parallel to the $zz$-bonds).  Note that we have dedicated a big part of the numerical effort to fields $H \lesssim 1$, which correspond to experimentally realizable fields. 

\begin{figure}[h!]
\includegraphics[scale=0.7]{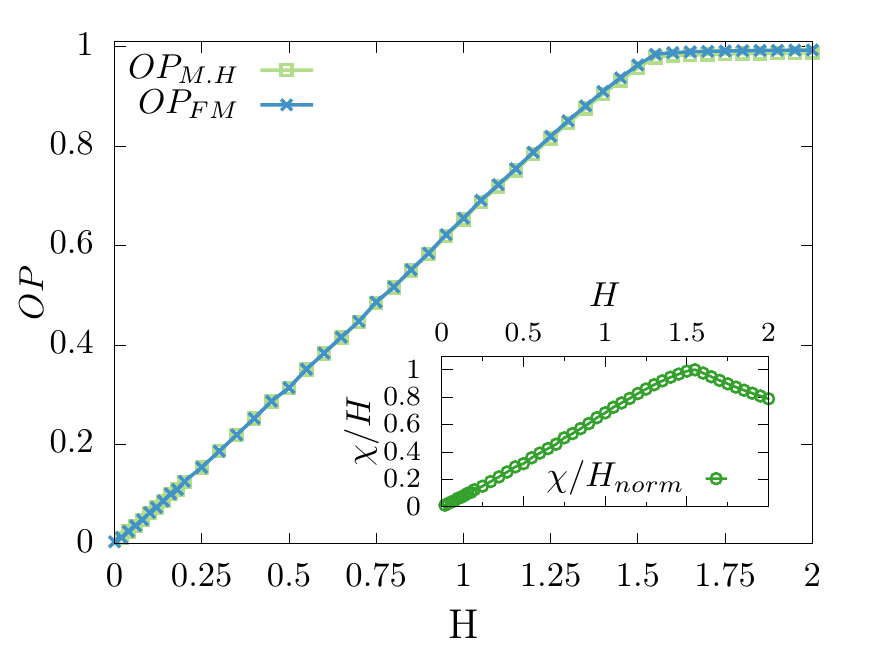}
\includegraphics[scale=0.35]{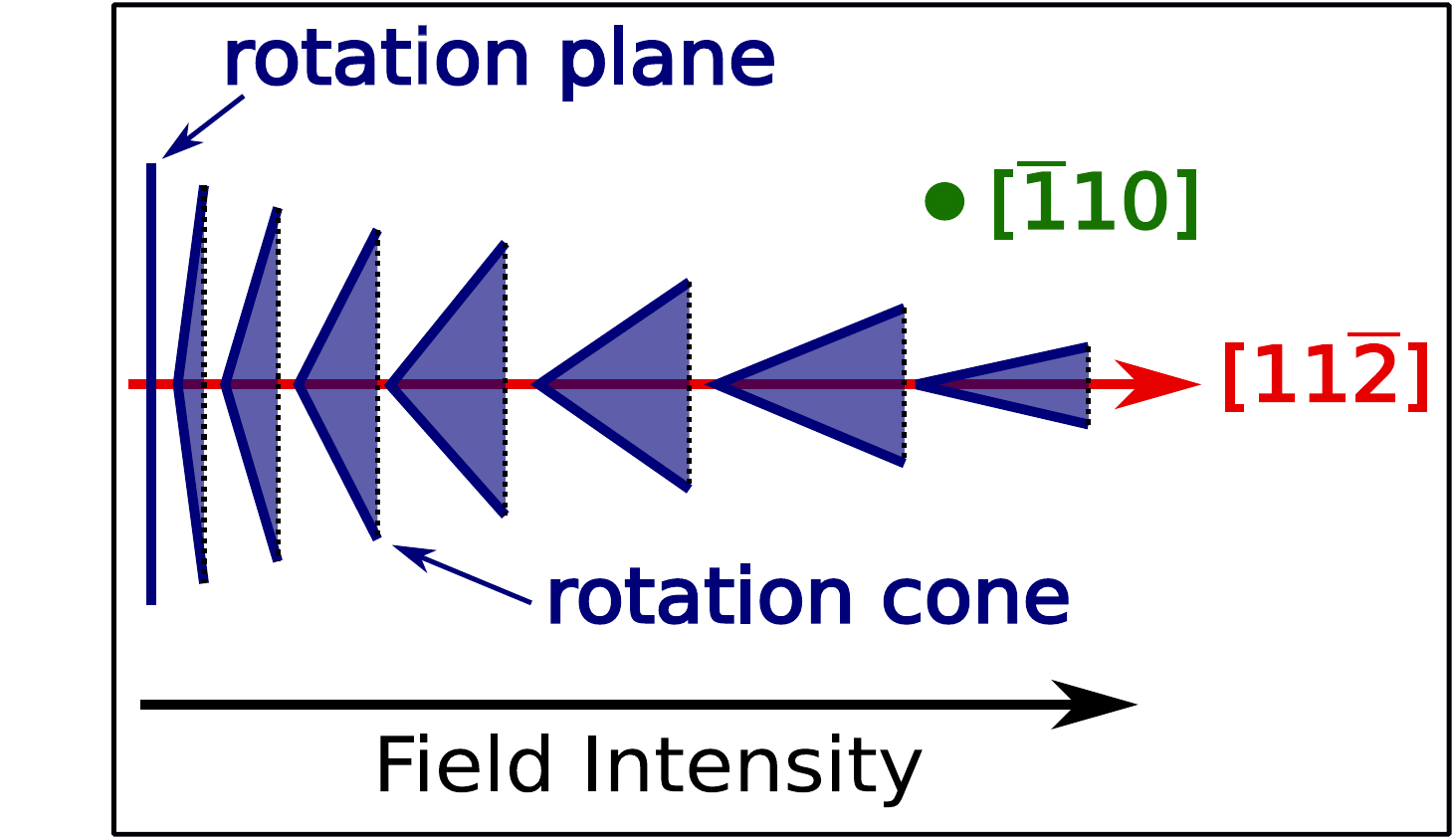}
\caption{{\it Top}: Magnetization in the global $[11\bar{2}]$ direction (green curve) and ferromagnetic order parameter (blue curve) as a function field intensity. {\it Inset:} susceptibility of the ferromagnetic order parameter as a function of field. {\it Bottom:} scheme showing the change from a rotation plane towards a rotation cone (see text).}\label{W_d112}
\end{figure} 

We show in Fig.~\ref{W_d112}(top) that the magnetization for the field in the direction $H\parallel[11\bar{2}]$ grows monotonically with a constant slope, reaching saturation at $H\sim 1.5$.  At $H = 0$ the state is a planar counterrotating spiral as shown in the previous section, but for $H > 0$ a continuous transition between a counterotating planar spiral and a counterotating conical spiral is realized.  As the name indicates, a conic spiral is a helimagnetic state in which the spiral does not rotate in a plane but in a cone around a certain common direction. Since the magnetic field is applied in the direction parallel to the propagation direction, at moderate fields the spins cant in that direction, which transform the plane of rotation into a cone. We show in Fig.~\ref{W_d112}(bottom) a scheme of this process. We indicate the direction of propagation (which coincides with the direction of the applied field, $[11\bar{2}]$) as well as the direction perpendicular to it ($[\bar{1}10]$). At low fields the canting in the spins induces the transition from a rotation plane to a rotation cone. As the field increases the cone gets narrower, until at high fields the spins point in the direction of the field, thus reaching saturation.

\begin{figure}[h!]
\includegraphics[scale=0.7]{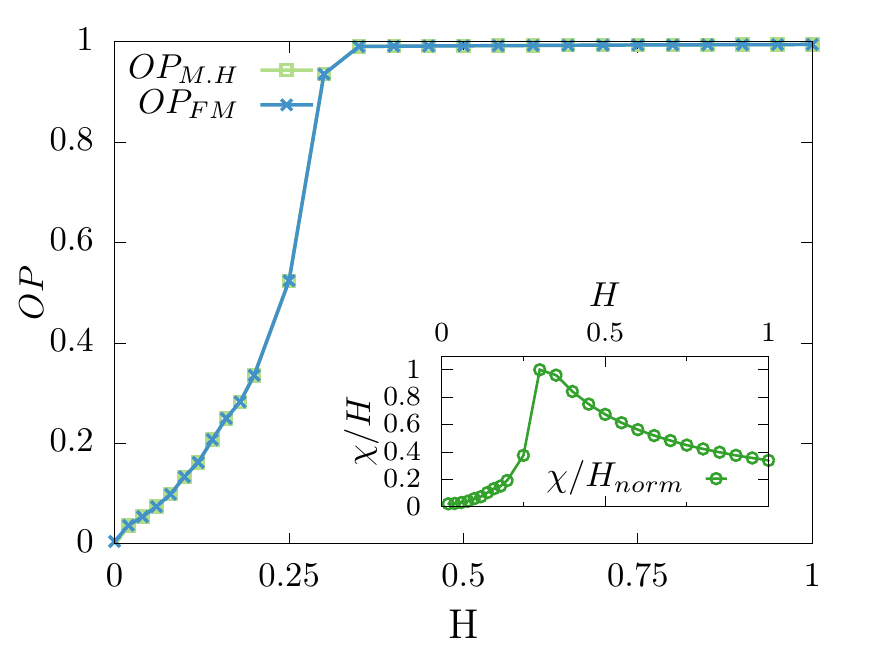}
\includegraphics[scale=0.7]{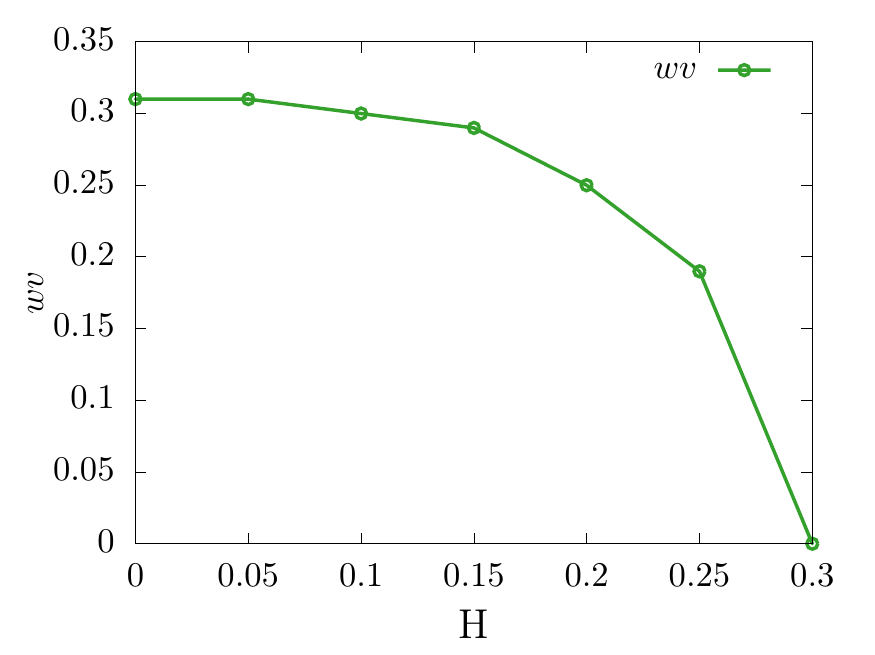}
\caption{{\it Top, main figure}: Magnetization in the global $[\bar{1}10]$ (green curve) and ferromagnetic order parameter (blue curve). {\it Top, inset:} susceptibility of the ferromagnetic order parameter as a function of field. {\it Bottom}: change in the associated spiral wavevector ($q$) as a function field intensity.}\label{W_mag_d-110}
\end{figure}

For the other in plane direction, $H \parallel[\bar{110}]$ (the direction parallel to the $zz$-bonds), the behaviour at small fields is different. We show in Fig.~\ref{W_mag_d-110}(top) the magnetization curve for an applied magnetic field in the $[\bar{1}10]$ direction. We observe that the magnetization monotonically increases with an increasing slope at small fields up to $H\sim 0.3$, at which point it suddenly reaches saturation. This transition towards saturation follows the same mechanism as that developed in section ~\ref{W_ani_pHd} for the commensurate-incommensurate transition.

The spiral phase in the $H=0$ case can be seen, projected over the $[111]$ plane, as that of alternating ferromagnetic and antiferromagnetic domains (as in Fig.~\ref{phd_W_aniso}), where the ferromagnetic domains are oriented in the direction parallel to the $zz$-bonds (the $[\bar{1}10]$ direction). If we introduce a small magnetic field in this direction, the wavevector of the magnetic spiral decreases, which is evidenced in the projection on the $[111]$ plane as the increase of the ferromagnetic domains' size, until they overpower the system at a critical value of the field.  In Fig.~\ref{W_mag_d-110}(bottom) we show the change in the spiral wavevector as a function of field. It can be seen that at small fields, $H < 0.15$ the change in the wavevector is not pronounced, ranging from $q = 0.31$ to $q = 0.25$ (in units of $2\pi$). On the other hand, for $H > 0.15$, the wavevector decreases rapidly, reaching $q=0$  at $H\sim 0.3$. In this case, a vanishing wavevector indicates the onset of a ferromagnetic state, which is confirmed by the real space pattern and the correlation functions obtained from the simulations. It is also at $H \sim 0.3$ that we observe the maxima in the susceptibility of the ferromagnetic parameter (inset of Fig.~\ref{W_mag_d-110}(top)) indicating that the critical field for this field direction is $H_{\text{crit}}\sim 0.3$.
			
\begin{figure*}
\includegraphics[scale=0.7]{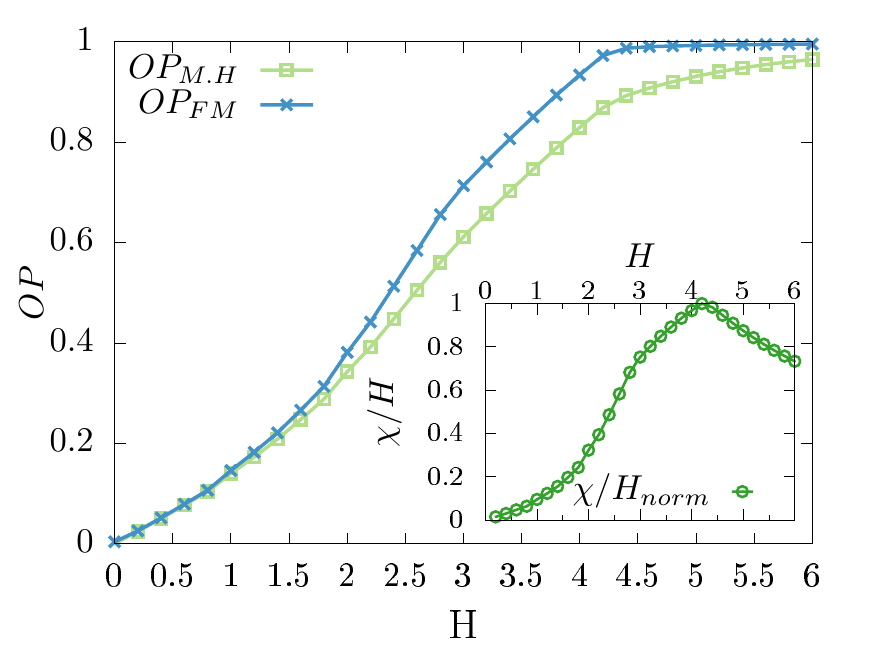}
\includegraphics[scale=0.25]{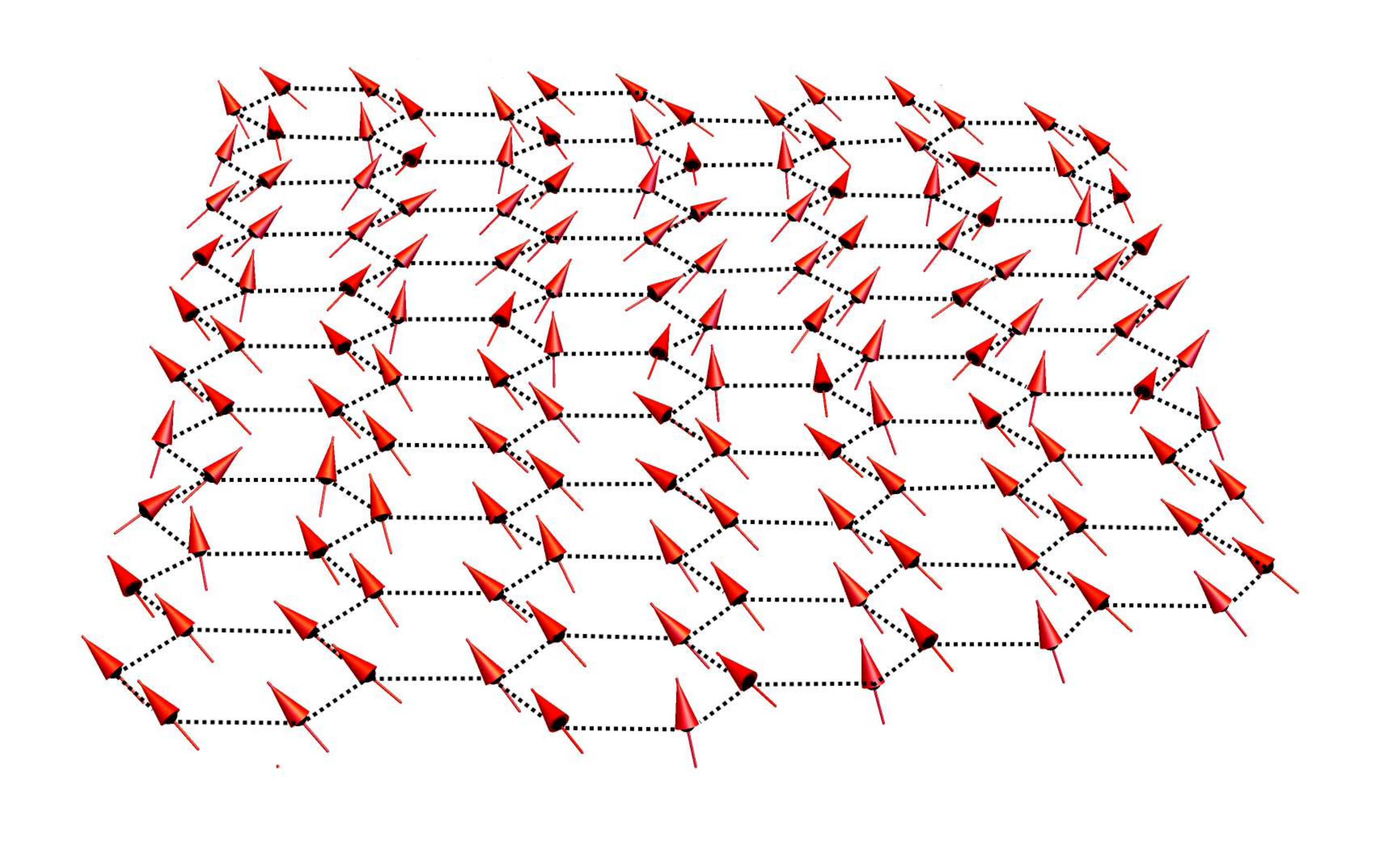}
\caption{{\it Left}: Magnetization in the global $[111]$ direction (green curve) and ferromagnetic order parameter (blue curve) as a function field intensity. {\it Right:} real space spin pattern at $H = 0.325$.}\label{W_d111}
\end{figure*}

Since the model contains strong in plane interactions, saturation is only achieved at strong magnetic fields for out of plane directions. In the case of a magnetic field in the $[111]$ direction (direction perpendicular to the lattice plane) the saturation is reached at $H \sim 4$ as can be observed in Fig.~\ref{W_d111}(left). In this case, as well as in the nearest neighbour model, both order parameters saturate at different field intensities. Both the ferromagnetic order parameter and the magnetization in the direction of the field have to saturate at high enough fields, but while the ferromagnetic order parameter saturates at $H \sim 4$, the magnetization in the direction of the field does not completely saturate up to the biggest calculated fields. 

A magnetic field in the $[111]$ direction tilts the spins in this direction, producing a net magnetization, up to the biggest calculated field, which is deviated from the magnetic field direction. The magnetization process can be understood recalling Fig.~\ref{phd_W_aniso} bottom. Here we mentioned that the spin spiral state can be considered as arrangements of ferromagnetic/antiferromagnetic domains of spins in the $[111]$ plane. When a magnetic field is applied perpendicular to the direction of the ferromagnetic domain's polarization (as is the case here), the spins of the ferromagnetic domains will be tilted in the direction of the field. The same will happen to the spins in the antiferromagnetic domains which are not already pointing parallel to the field, but a different tilt angle. In Fig.~\ref{W_d111}(right) we show the spin pattern at $H = 0.35$, where the tilt in the direction of the field can be observed.

\section{Discussion}
\label{DandC}

We have studied a variety of extended Kitaev Hamiltonians as minimal models of $\alpha$-Li$_2$IrO$_3$. The experimental studies of this material indicate a strong bond anisotropy, which led us to study models which contain strong anisotropic bond dependent interactions beyond Kitaev interactions. Furthermore we have considered interactions ranging from nearest to third neighbours. While all models share similarities with each other, we have shown that the range of the interactions can produce radical differences in the behaviour of the system.

Our nearest neighbour models (those given by $\mathcal{H}(J_1, K_1, I_c \neq 0, I_d = 0)$ and $\mathcal{H}(J_1, K_1, I_c \neq 0, I_d\neq 0)$ Eq.~\ref{Hamiltonian}) exhibit a phase diagram where a big part of it presents an incommensurate counterrotating coplanar spiral state. For the model with non zero $I_c$ and $I_d$ we encounter a spiral state whose wavevector coincides with that of $\alpha$-Li$_2$IrO$_3$ on a section of the phase diagram. However, while in the Iridate material the plane of rotation is tilted away from the lattice plane by $\sim 80^o$, in this model the plane of rotation is parallel to the $XY$ Cartesian plane ($54^o$ tilt with respect to the lattice plane). This can be understood realizing that a Hamiltonian only containing the $I_c$ term is equivalent to an XY model, where in this case the  model posses a U(1) symmetry around the Cartesian $z$-axis. The same is true for the $I_d$ terms when only one of them is present.  When the Kitaev and Heisenberg terms are taken into account together with the $I_c$ term, the U(1) symmetry is broken, but there is a residue of this symmetry, which is exhibited in the  strong anisotropy of the state, where the rotation plane of the spin spiral state is perpendicular to the Cartesian $z$-axis. The introduction of a small term of the same nature as the $I_c$ but over the $xx$- and $yy$-bonds (that is, the $I_d$ term in the nearest neighbour model) will induce a tilt of the rotation plane, induced by the symmetries of these terms. The tilt angle will thus be a function of the ratio $I_c/I_d$, allowing us to reproduce not only the wavevector but also the tilt of the rotation plane characteristic of $\alpha$-Li$_2$IrO$_3$. 

While the introduction of second neighbor interactions do not reproduce the experimental results, as shown in Appendices \ref{2N} and \ref{2NI_c}, recent DFT calculations \cite{Winter2016} have put forward a third neighbour model which indicate that further neighbour interactions are needed. We tested this model, in both its bond isotropic and anisotropic forms. Given that the experiments performed on $\alpha$-Li$_2$IrO$_3$ indicate strong bond anisotropy, is no surprise that the model which reproduced the experimental features of the material is the bond anisotropic third neighbour model, $\mathcal{H}_I(J_{1,3}, K_{1,2}, \Gamma_{1,2})$ (Eq.~\ref{Hamiltonian_W2}).

In the third neighbour models, the interactions are slightly different than in the nearest neighbour models previously studied. We modelled the bond dependent interactions in the nearest and second neighbour models as $I_{c/d}\sum_{<ij>}S^{r_{ij}}_iS^{r_{ij}}_j$, with $S^{r_{ij}}_i = \mathbf{S}_i\cdot\mathbf{r}_{ij}$, where $r_{ij}$ is the bond between sites $i$ and $j$. This in turn takes the form, for the $zz$-bonds (the expression for the remaining bonds is analogous)

\begin{align}
I_{c}\sum_{<ij>}S^{r_{ij}}_iS^{r_{ij}}_j & =  I_c (S^x_iS^x_j + S^y_iS^y_j + S^x_iS^y_j + S^x_jS^y_i)  \notag\\
&  = I_c(S^x_iS^x_j + S^y_iS^y_j) + I_c \Gamma^{xy}\,,
\end{align}

which has the form of the $\Gamma$ terms included in the third neighbour model, supplemented by two Heisenberg terms. This expression indeed shows a clear connection between the nearest neighbour model and the nearest neighbour terms of the third neighbour model, where those terms can be deformed into one another by finely tuning the bond anisotropy parameters. While the nearest neighbour terms are connected on both models, the presence of further neighbour exchanges stabilize the properties of the magnetic order.  For the nearest neighbour model, the parameters which reproduce the experimental signatures of  $\alpha$-Li$_2$IrO$_3$ are constrained to a small part of the phase diagram, close to a phase boundary, which makes the behaviour of the spin spirals highly susceptible to small parameter variation and magnetic fields. On the other hand, the third neighbour model reproduces these signatures over a broader region, making the spiral state robust to parameter variations and applied magnetic fields. Turning to the parameters studied here, it is worth pointing out that we have only mapped the phase diagram as a function of a two parameters per model, the $I_c$ and $I_d$ terms in the nearest neighbour model, and $J_1$ and $J_3$ in the third neighbour model, since we have tried to remain as faithful as possible to the exchange couplings proposed in Refs.~\cite{Winter2016, Williams-2016} However, while it is expected that the dominant exchange interaction in these materials is of Kitaev type and ferromagnetic, it cannot be discarded that different combination of parameters (in particular regarding the Heisenberg exchange in the nearest neighbour model, or the second neighbour Kitaev and $\Gamma$ exchanges in the third neighbour model) could also reproduce the experimental results. 

Since both  nearest neighbour and anisotropic third neighbour models reproduce the experimental features of the specific heat, we proceed to calculate the Curie-Weiss temperature within a mean field approach. For the  nearest neighbour model we find an anisotropic susceptibility $\mathbf{\chi} = (\chi_{xx},\chi_{yy},\chi_{zz})$ arising from the $I_c$ and $I_d$ terms, where the associated Curie-Weiss temperatures are given by

\begin{equation}
\Theta_{xx} =\Theta_{yy} = \frac{S(S+1)}{3K_b}(-3J-I_c-I_d-K)\,,
\end{equation}

\begin{equation}
\Theta_{zz} = \frac{S(S+1)}{3K_b}(-3J-2I_d-K)\,.
\end{equation}

At the point of interest ($I_c=-0.5$ and $I_d = -0.35$), where experimental results are reproduced, we obtain positive Curie-Weiss temperatures: $\Theta_{xx} = \Theta_{yy} = 16.35K$ and $\Theta_{zz} = 14.38K$ in units of the Kitaev coupling, eploying the values of the couplings in meV from Ref.~\cite{Williams-2016}.

In the case of the third neighbour model we also find an anisotropic susceptibility, in this case arising from the anisotropy of the nearest neighbour Heisenberg and Kitaev interactions

\begin{equation}
\Theta_{xx} =\Theta_{yy} = \frac{S(S+1)}{3K_b}(-3J^{XY}_1-K^{XY}_1-2K_2-3J_3)\,,
\end{equation}

\begin{equation}
\Theta_{zz} = \frac{S(S+1)}{3K_b}(-3J^{Z}_1-K^{Z}_1-2K_2-3J_3)\,,
\end{equation}

Evaluated at the experimentally relevant point shown in Table.~\ref{Winter_bonds} we obtain $\Theta_{xx} = \Theta_{yy} = 20.15K$ and $\Theta_{zz} = -17.17K$.

The experimentally measured Curie-Weiss temperature \cite{Singh2012} was obtained from a fit to the high temperature magnetic susceptibility for polycrystalline samples, $\Theta_{\text{exp}} = -33(3)K$. Since the experiments have been performed on a polycrystalline sample, the Curie-Weiss temperature obtained is the average of the anisotropic Curie-Weiss temperatures. In the case of the  nearest neighbour and anisotropic third neighbour models we obtain an averaged Curie-Weiss temperature $\Theta^{I_cI_d}_{av} = 15.7K$ and $\Theta^{W}_{av} = 7.58K$. As previously shown, a considerable part of the phase diagram for the anisotropic third neighbour model reproduces the experimental results. This would indicate that a different selection of coupling strengths could modify the averaged Curie-Weiss temperature of the model as to obtain the experimental value, while maintaining the overall behaviour of the system to be the same as in the neutron diffraction experiments. While we could do a similar analysis for the  nearest neighbour model, the region of the phase diagram that reproduces the results is much more reduced, and as such the variation of the exchange parameters is not enough to change the sign of the Curie-Weiss temperatures. Extra thermodynamic studies on single crystals would prove useful at determining the anisotropic susceptibilities and the respective signs of the associated Curie-Weiss temperatures. Given that the experiments where performed at high temperature, the comparison between our ground state calculations and the experimental results are not sufficient to determine if the anisotropic third neighbour model is the minimal model of $\alpha$-Li$_2$IrO$_3$. 

We studied the magnetization processes for both nearest neighbour and third neighbour models. In both cases we find a strong anisotropy in the magnetic response, with different behaviors for different directions of the applied field.  For the nearest neighbour model we find a strong easy plane anisotropy, arising from the off diagonal terms of the Hamiltonian. Since these terms favor in plane orderings, the tendency to order ferromagnetically when an in-plane magnetic field is applied is strong, which presents in our study as a lower critical field for in-plane rather than out of plane directions  of the external field. The tendency to in-plane ordering is also observed in the magnetization behavior when a field in the $[111]$ plane is applied, as there exist an intermediate state where the magnetization has a net component in the direction of the field, but also a component in the lattice plane. At intermediate fields, the three field directions studied show similar magnetic behavior, which resembles that of a ferromagnetic/paramagnetic transition, where ferromagnetic domains with polarization parallel to the direction of the external field are induced, and which grow in size as the field is increased. We mention that, since the magnetic fields are expressed in terms of $K$, we can extract the value in Tesla since $H=1$ in units of $|K|$ is equivalent to $g\mu_BH/|K|=1$. The gyromagnetic factor is believed to be anisotropic, given the deformations of the octahedral cage of Oxygens, but the exact value is not known. Recent calculations \cite{Winter-2017} show that the g factor can drastically change depending on the octahedral deformation, with values ranging from $\sim 1.5$ to $\sim 4$ for the parallel component of the gyromagnetic factor (the component in the lattice plane), and from $\sim 2.5$ to $\sim 0.5$ for the out of plane component. Given the strong variation of this parameter and the lack of experimental data, we have employed an isotropic factor $g = 2$. Finally, if we assume the interaction strength of the Kitaev interaction to be $-4.5$meV (as proposed in Ref.~\cite{Williams-2016}),  we obtain the critical fields of $\sim 31$T for the $[111]$ field direction, $\sim 6$T for $[11\bar{2}]$, and $\sim 2.7$T for $[\bar{1}10]$.

For the case of the anisotropic third neighbour model the behaviours is quite different. While the system also exhibits an anisotropic magnetic response, in this case it possesses an easy axis anisotropy. Additionally, assuming that the overall energy scale of the couplings determined by DFT is correct, we find that the critical field for an external field in the $[111]$ direction results $\sim 200$T, while it is $\sim 111$T for the $[11\bar{2}]$ direction, and $\sim 22$T for $[\bar{1}10]$. While such high fields are hard to achieve in experiments, the low field regime already exhibits a highly anisotropic behaviour which can be probed experimentally.  For fields in the $[\bar{1}10]$ direction, the magnetization seems to suddenly increase at $|H| \sim 0.25$ (in units of the averaged nearest neighbour Kitaev coupling), exhibiting a change in the spiral wavevector. For fields in the $[11\bar{2}]$ and $[111]$ directions the transition is continuous, presenting a continuous transformation of the spin pattern (in the case of the $[11\bar{2}]$ field direction, from a planar spiral towards a conical spiral).

\section{Conclusions} 

The similarities in the ground states found for both systems indicate that one of them could prove to be the minimal model of $\alpha$-Li$_2$IrO$_3$. The differences in specific heat as well as different magnetization behaviors indicate that further experimental studies are needed to decide which one, if any, is the model corresponding to this material. While the Curie-Weiss temperature for the nearest neighbour model does not reproduce the sign found from the calorimetric experiments, the nature of the experiments (the powder average and the high temperature measures) could obscure further details which could clarify the discrepancy. Studies on single crystals would prove valuable as in these cases the different crystallographic directions could be probed to assert the existence of anisotropic susceptibilities. While Ref.~\cite{Winter2016} proposes a set of values for all couplings in the third neighbour model (which were chosen as the starting point for our study), in both isotropic and anisotropic cases these values do not correspond to spiral phases. By changing the couplings of the nearest and third neighbour Heisenberg exchanges we find a range of values where the experimental results are obtained. This indicates that further studies regarding the quantum chemistry as well as the crystal structure of the material are necessary to further determine the strength of the couplings.

To further distinguish the models, magnetization measurements can be fruitful. While the critical fields for the third neighbour model are far beyond current capabilities, interesting studies can be performed at low field. In particular, for the nearest neighbour model the calculated critical fields are within the possibility of experimental realization, while for the third neighbour model, the anisotropy present at low fields can be measured experimentally (we remind the reader that the critical field measured for a field in the $[\bar{1}10]$ direction in third neighbour model is $22$T). In particular, the low field behaviour, which is different for the different field directions, could be observed in single crystals. In this context it is worth pointing out that even if the critical fields obtained experimentally were to results smaller than the ones reported in this work, that would not indicate a failure of the proposed models. It has been shown previously that quantum fluctuations can further reduce the critical fields \cite{winter18} for $\alpha$-RuCl$_3$ with respect to those obtained from classical solutions, but that the overall magnetization behaviour is not necessarily modified by these fluctuations.

We conclude then, pointing out that questions remain open regarding the minimal model of $\alpha$-Li$_2$IrO$_3$. We have reduced the number of possible models and explored the different magnetization behavior of those models which reproduce the experimental signatures of the material. We expect that magnetization measurements can point in the direction of one these models being correct in the low temperature limit. However, we recommend further studies: We expect that electronic structure calculations could clarify the current situation in which the obtained exchange couplings do not lead to the experimentally measured spin pattern. On the other hand, and with respect to the third neighbour model, it was pointed out in Ref.~\cite{Winter2016} that the crystal structure of the material is not well understood. In their work, the authors of Ref.~\cite{Winter2016} also studied a relaxed crystal structure, which coincided with the couplings later obtained in Ref.~\cite{Freund2016}. We do expect that a study of the kind performed in this work employing the relaxed crystal structure will give a better agreement with experiments for this particular model. However, the importance of further neighbour interactions and anisotropies to the stabilization of the spin spirals will not be modified by this study.   

The recent growth of single crystals could help refine the crystal structure. Furthermore, our study of the third neighbour model suggest that modest long-range interactions can stabilize the counterotating spirals found in $\alpha$-Li$_2$IrO$_3$, but that anisotropy is crucial to reproduce the experimental results. Since materials realizing Kitaev interactions show strong bond anisotropy, it cannot be discarded that perhaps a different combination of interactions with a different combination of anisotropies could also reproduce the experimental results and be relevant in real materials. We further mention that also more Monte Carlo studies can be beneficial. By studying the finite temperature magnetization curves, we might encounter lower critical magnetizations and interesting intermediate states which could be easier to realize experimentally (as was shown to be the case for $\beta$-Li$_2$IrO$_3$ \cite{Ruiz2017}). Furthermore, a study of the full phase diagram in the presence of different external magnetic fields could help model future materials which could be realized. Finally, we mentioned that both models could also be distinguished by their behaviour in the presence of dilution. The study performed in Ref.~\cite{manni14} probe the thermodynamic response of the material when the magnetic ions have been removed, as well as the variation of the spin glass critical temperature as a function of dilution. They conclude that their results indicate that the further neighbour interactions are needed in the minimal model of $\alpha$-Li$_2$IrO$_3$. A detailed study of the models proposed here in the presence of dilution could clarify the picture regarding this material, and it will be left for future studies.

%====================================================================
\section{Acknowledgements}

We would like to thank Roser Valenti, Simon Trebst, Johannes Reuther, and Marek Gluza for fruitful
discussions and help with this manuscript. M. L. B is supported by the Freie Universität Berlin within the Excellence Initiative of the German Research Foundation		.

\bibliography{ref}

\appendix

\section{Benchmark: Heisenberg-Kitaev model}\label{sec:kit-heis}

We will benchmark our code by studying the Heisenberg-Kitaev model on the honeycomb lattice. The Hamiltonian for this models is given by,
\begin{equation}
\mathcal{H}=J\sum_{<ij>}\mathbf{S}_i\cdot\mathbf{S}_j+K\sum_{<ij>}\sum_{\gamma}S^{\gamma}_i S^{\gamma}_j\,,
\label{Hamiltonian2}
\end{equation}

where $\gamma = \{x,y,z\}$  and $\mathbf{S}_i$ are classical spins on sites $i$ of the honeycomb lattice. This model contains exactly solvable points and has been studied in detail by Price {\it et. al} ~\cite{Price-2013} via Monte Carlo simulations employing periodic boundary conditions (PBCs). We reproduce some of the results in Ref.~\cite{Price-2013}, and show how (in the case of FEBs), at big enough system sizes we recover the bulk behaviour expected from a simulation employing PBCs. 

We perform Monte Carlo simulations on the Hamiltonian shown in Eq.~\ref{Hamiltonian2}, for sizes ranging from 24 to 5400 sites. The temperature of the simulations is consistently chosen as $T= 0.001$ in units of $|K|$. We employ $2\times10^5$ Monte Carlo sweeps from which $1\times 10^5$ are used as equilibration steps.  We concentrate here on the behavior of a simple Metropolis-Hastings algorithm, without recurring to parallel tempering or iterative minimization schemes, to test the effectiveness of the Monte Carlo code. For the results in the main text we employ further modifications of the algorithm to reduce the effects of domain walls and of rough energy landscapes as mentioned in section~\ref{method}.

\begin{figure}[h!]
\centering
\includegraphics[scale=1]{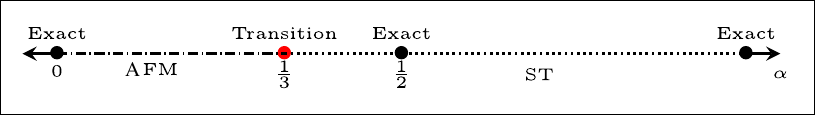}
\caption{Phase diagram for the Heisenberg-Kitaev model. We observe a phase transition from an antiferromagnetic state to a stripy phase occurring at $\alpha = 1/3$. The points $\alpha = 0$, $\alpha = 0.5$, and $\alpha \rightarrow \infty$ are exactly solvable, where $\alpha = 0.5$ presents an emergent SU(2) symmetry. }
\label{HKPhD}
\end{figure}

We have mapped the phase diagram for the Heisenberg-Kitaev model. We parametrized the exchange couplings as $K = 2\alpha$ and $J = 1-\alpha$ and calculated the ground state for values between $\alpha=0$ and $\alpha=1$. We show the phase diagram in Fig.~\ref{HKPhD}. 

The phase diagram exhibits two phases, at $\alpha < 1/3$ we obtain a ferromagnetic state, while for $\alpha > 1/3$ the state is a triple degenerated stripy phase, due to the symmetries of the Kitaev interaction, allowing for the three possible stripy phases, st-X, st-Y, and st-Z.

\subsection{Space pattern and correlation functions}

For this study we will select the lattice plane as the Cartesian XY-plane, and observe the real space spin  pattern on the YZ plane where both N\'eel and stripy phases are easy to distinguish. We will choose to show phases where the spins are oriented maximally in the $z$-direction to ease the comparison, but states where spins are aligned in other directions are also possible, and have been also obtained within our approach.

We study the real space configurations for different values of $\alpha$ to compare with the low temperature results of ~\cite{Price-2013}. We show the results for $\alpha = 0$, $\alpha=0.5$, and $\alpha = 0.75$ obtained from a calculation employing 216 sites. 

\begin{figure}[h!]
\centering
\includegraphics[scale=0.2]{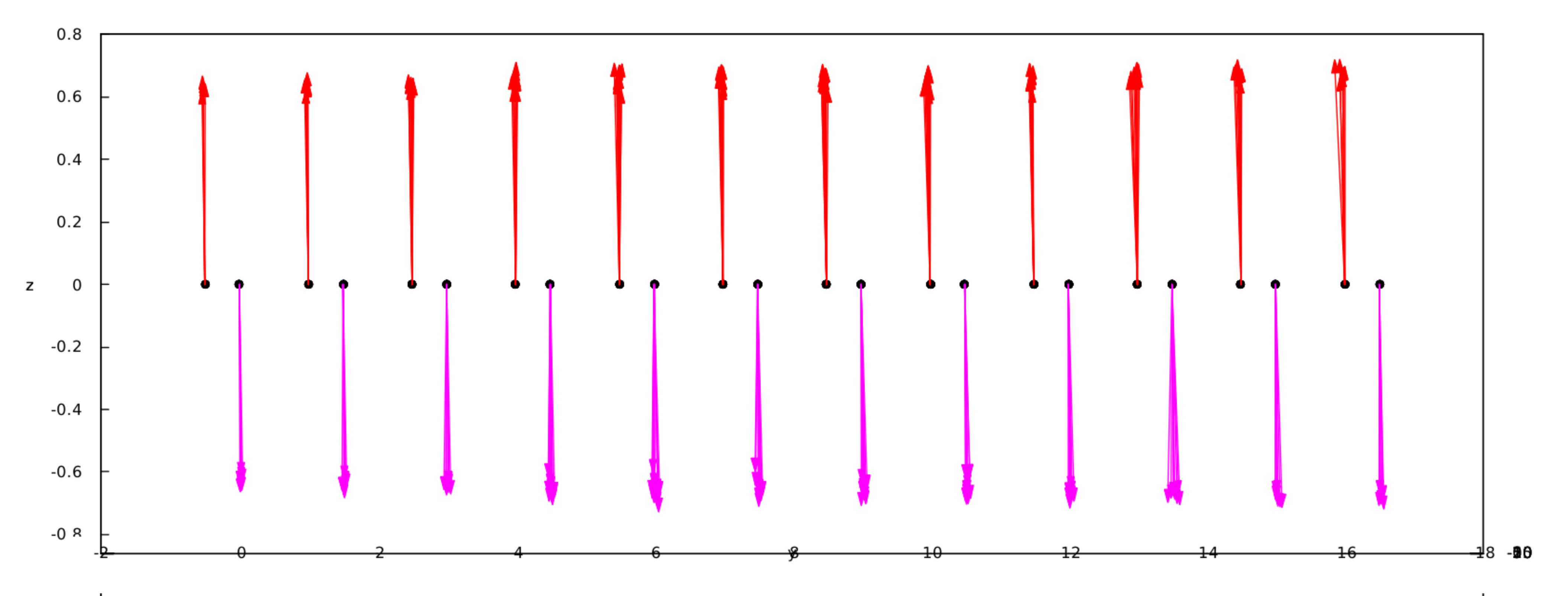}
\caption{YZ plane of the real space configuration for the $\alpha =0$ case. }
\label{AFM_yz_plane}
\end{figure}

For the case $\alpha=0$ we recover the antiferromagnetic Heisenberg model. In Fig~\ref{AFM_yz_plane} we show one of the simulations performed, in which the system exhibits a Neel order, where the spins are ordered along the Cartesian $z$-direction. 

\begin{figure}[h!]
\centering
\includegraphics[scale=0.2]{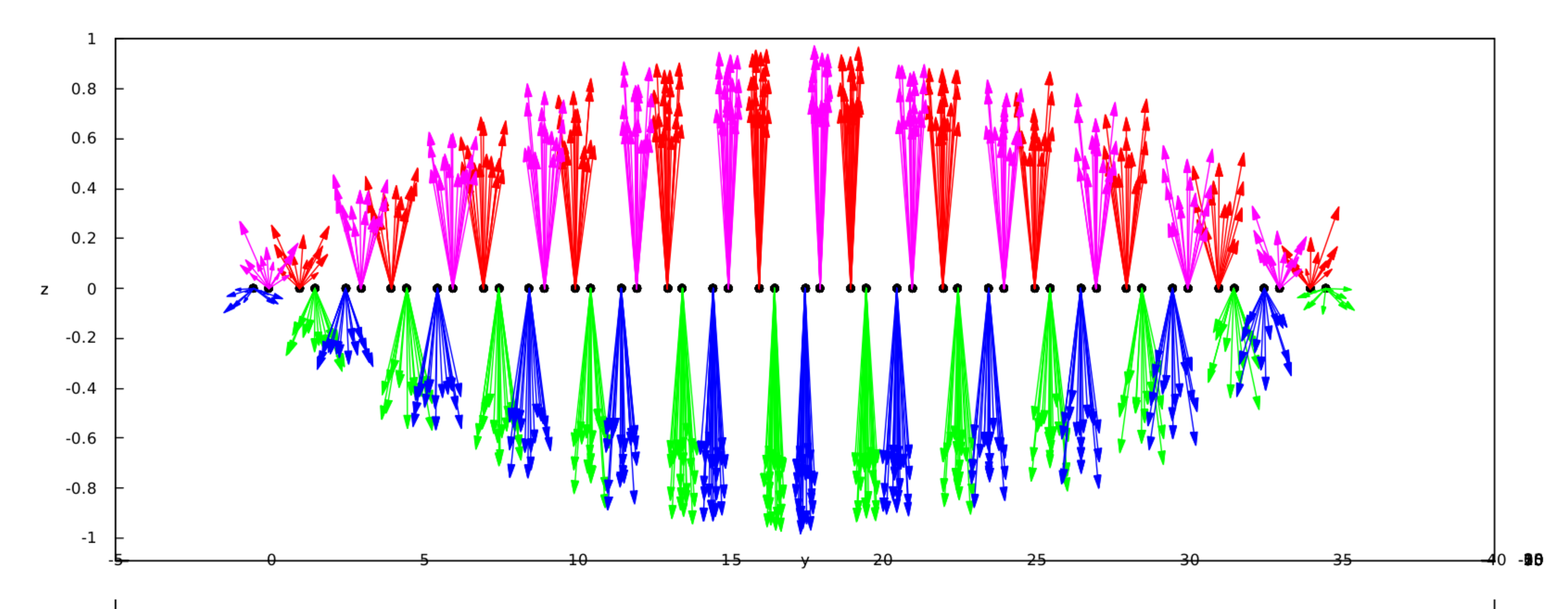}
\caption{YZ plane of the real space configuration for the $\alpha = 0.75$ case. }
\label{ST2_yz_plane}
\end{figure}

For the stripy phases, we show the spin pattern for $\alpha=0.75$ (Fig.~\ref{ST2_yz_plane}). In this case we observe that in the center of the system, the spins point in two directions forming stripes that span the system. In this case we separate the system in four sublattices, two with spins pointing in the $+z$ directions (red and pink spins), and two in the $-z$ (blue and green). The finite size effects are noticeable here, where the spins deviate from the $\pm z$ orientation, and where this deviation is more pronounced at the boundaries of the system.  Even for this small system size, where the finite size effects are considerable, the simulation already reproduces the known results for the Heisenberg-Kitaev model for these parameters.

\begin{figure}[h!]
\centering
\includegraphics[scale=0.2]{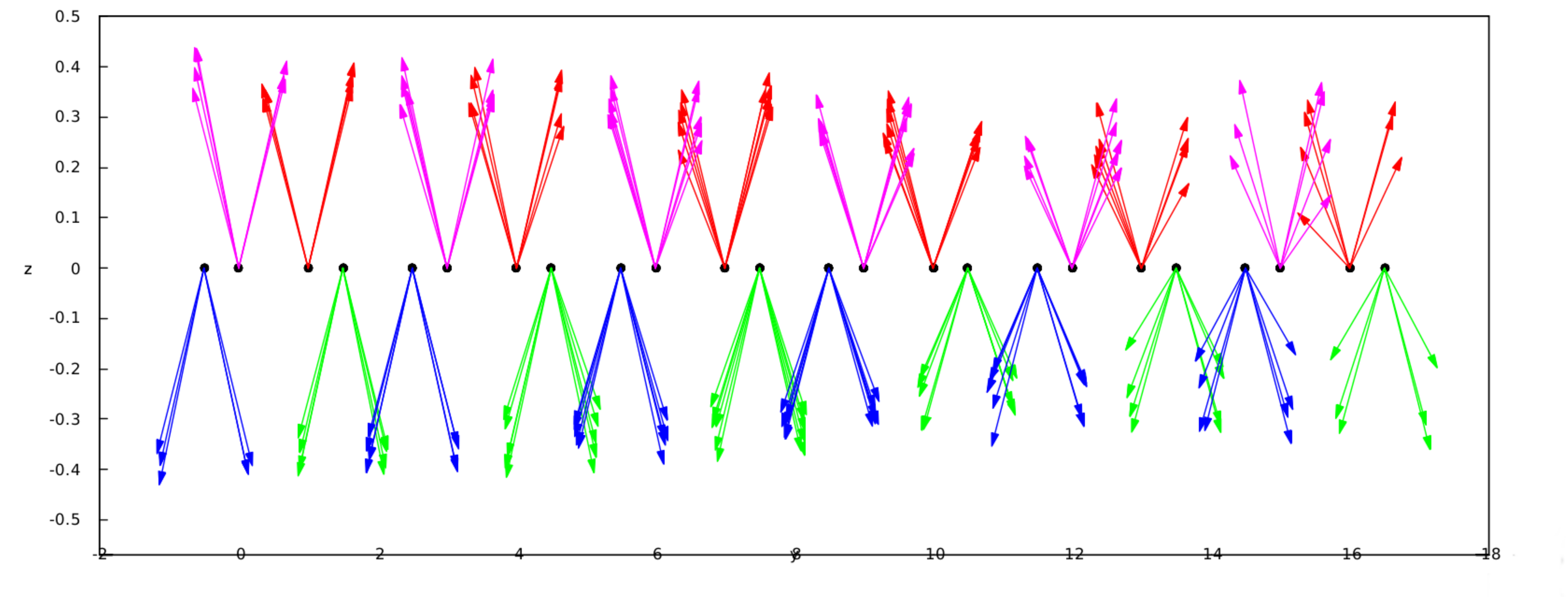}
\caption{YZ plane of the real space configuration for the case $\alpha =0.5$ which presents an emergent SU(2) symmetry.}
\label{ST1_yz_plane}
\end{figure}

For the stripy phase, an important point needs to be mentioned. The point $\alpha=0.5$ is special, since at this point the obtained stripy phase becomes and exact ground state of the system, which can be seen as a ferromagnetic state in a rotated basis. It can be proven that in this case the system exhibits a SU(2) symmetry \cite{Price-2013}. In Fig.\ref{ST1_yz_plane} we show a spin pattern for this case. The ground state realizes a stripy phase, but in this particular case the spins are not arranged along only one direction, due to the emergent symmetry of the state. While in the $\alpha = 0.5$ case some disorder in the state can be seen, these finite size effects are not as pronounced as in the rest of the stripy state.

These findings coincide with what was reported in Ref.~\cite{Price-2013} for the full phase diagram. We have purposefully showed results for a relatively small system size (216 sites), to exemplify the effect of FEBs. 

\begin{figure}[h!]
\centering
\includegraphics[scale=0.6]{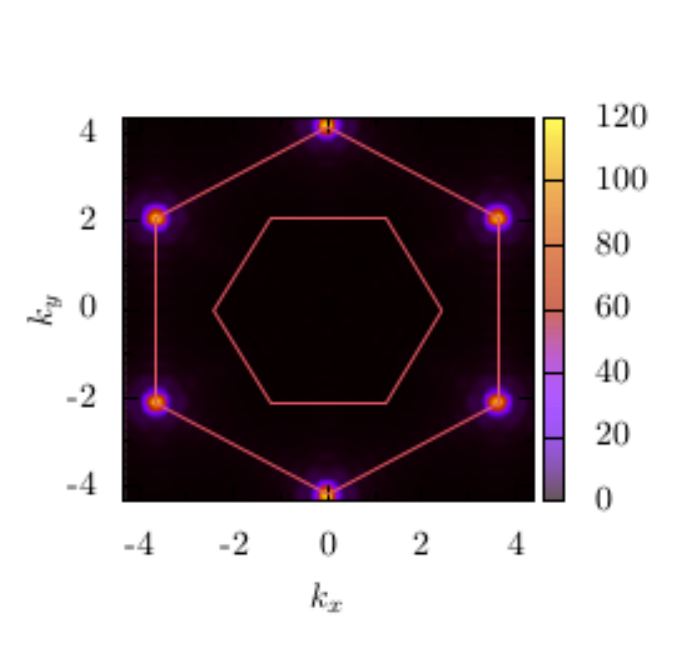}
\includegraphics[scale=0.6]{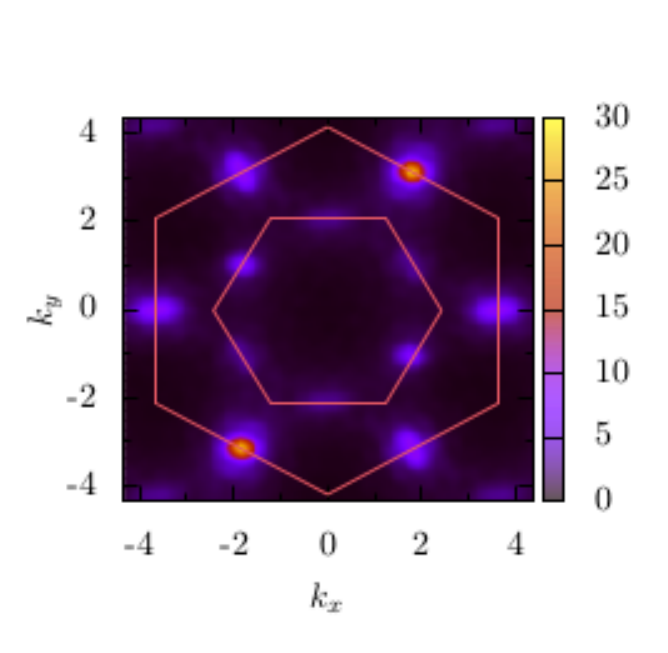}
\caption{Fourier transform of the correlation function. {\it Left}: AF state for $\alpha=0$. {\it Right}: Stripy phase for $\alpha=0.75$}
\label{Sq}
\end{figure}

We proceed to show in Fig.~\ref{Sq} the Fourier transform of the correlation function for both the antiferromagnetic and stripy phases, obtained from simulations on 2400 sites.

Fig.~\ref{Sq}(left) shows the Fourier transform for the antiferromagnetic state. We observe the distinct features of the N\'eel phase, presenting maxima at the corners of the extended Brillouin zone. On the other hand, the stripy phase (Fig.~\ref{Sq}, right) presents two maxima as well as secondary maxima on the sides of the extended Brillouin zone. This maxima indicate that, in this particular simulation, the state is dominated by a st-Y phase. The secondary maxima on the sides of the extended Brillouin are effects arising from the FEBs, where small regions of the system realize the other two degenerate stripy phase.

\subsection{FEBs effect on big system sizes}\label{sec:loc_min}

Even though the effect of FEBs is still noticeable in system sizes as big as 2400 sites, the behavior of the system is not radically affected by them. While the presence of FEBs has the tendency to generate domain walls when the low temperature state might be degenerate, the  simulations consistently converge to states that correspond the expected behavior of the system. 

\begin{figure}[h!]
\centering
\includegraphics[scale=0.2]{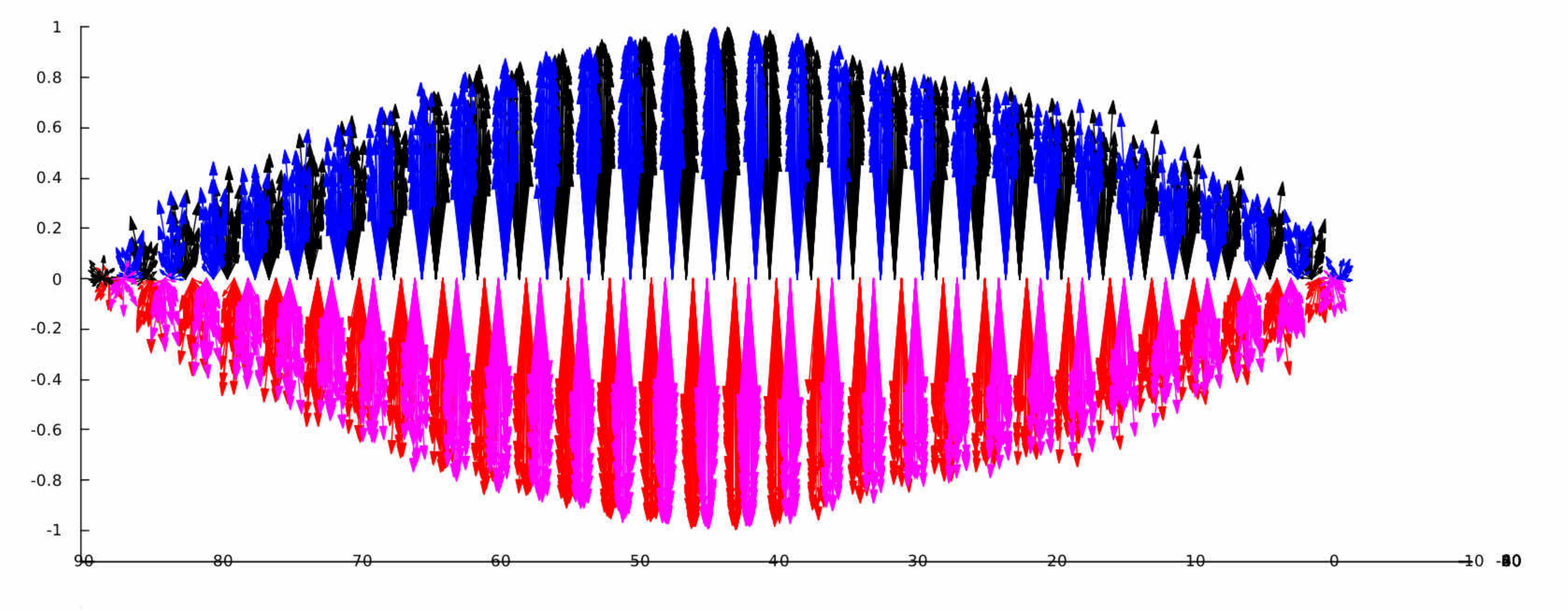}
\caption{Real space snapshot of the XZ plane for the ground state of the case $\alpha = 0.75$ case using 5400 sites.}
\label{stripy075_yz_plane}
\end{figure}

In the following, we show the results of two identical simulations, run with different seeds, for the case $\alpha = 0.75$ and 5400 sites. In this case, both simulations converge to the stripy phase on the $z$-direction. Fig.~\ref{stripy075_yz_plane} shows the stripy phase for a simulation in which no domain walls have been generated. In this case, away from the bulk the spins deviate from their $\pm z$ orientation, but it is only close to the edges of the system where the effects of FEBs destroy the stripy order. Furthermore, these deviations do not take an arbitrary form. When the spins deviate from the $\pm z$ direction, they do it by inducing a non-zero $x$- and/or $y$-spin components arranged according to the st-X and st-Y phases.

\begin{figure}[h!]
\centering
\includegraphics[scale=0.2]{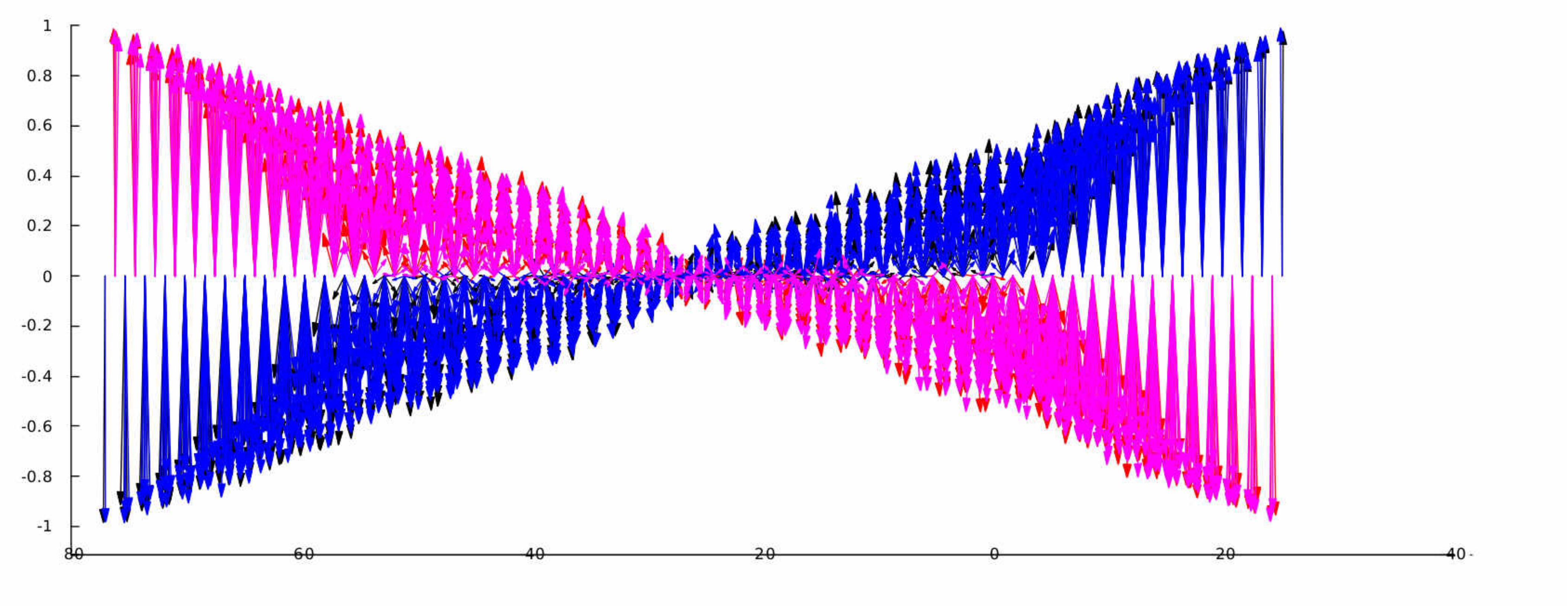}
\caption{Real space snapshot of the XZ plane for a local minima spin configuration for the $\alpha = 0.75$ case using 5400 sites.}
\label{local_minima}
\end{figure}

On the other hand, Fig.~\ref{local_minima} shows a spin pattern where a domain wall has been generated in the middle of the system.  In this case, both sides of the domain wall present a stripy order, but the orientation of the sublattices is flipped.  In this case, the domain wall also generate non-zero x- y-spin components which are arranged accordingly to the st-X and st-Y phases.

While in Fig~\ref{local_minima} the effects of FEBs seem to be more noticeable than in Fig.\ref{stripy075_yz_plane}, the correlations for both states are the same, i.e, the Fourier transform of the correlation function presents the same characteristics as Fig.~\ref{Sq}.

In this work we implement further algorithms to minimize the effect of FEBs, but while their effect can be minimized, they cannot be eliminated entirely. 
With this in mind, and the comparison of our results to the work in Ref.~\cite{Price-2013}, we are ensured that while FEBs will induce domains (when the ground state is degenerated) and edge defects in the systems, these effects will not modify the nature of the studied system nor their correlations. If the simulations is able to converge to a low temperature state (please note that here we are not considering other effects such as local minima or critical slowing down) then we are confident that this state will be a low temperature state of the studied Hamiltonian.

\section{ $\mathcal{H}(J_1, K_1, I_c \neq 0, I_d = 0)$ model}
\label{I_c}

Kimchi {\it et. al.} proposed \cite{Kimchi-2015} a nearest neighbour Hamiltonian that is capable of capturing the counterotating features of the ground state properties common to these three compounds. The Hamiltonian takes the form of Eq.~\ref{Hamiltonian} with $I_d = 0$, explicitly

\begin{align}
\mathcal{H}(J_1, K_1, I_c \neq 0, I_d = 0)&=J\sum_{<ij>}\mathbf{S}_i\cdot\mathbf{S}_j+K\sum_{<ij>_\gamma}\sum_{\gamma}S^{\gamma}_i S^{\gamma}_j\notag\\
&+I_c\sum_{<ij>_{zz}}S^{r_{ij}}_iS^{r_{ij}}_j\,,
\label{Hamiltonian3}
\end{align}

where $<ij>_\gamma$ indicates the bonds in which the interactions acts. This Hamiltonian contains terms that are symmetry allowed by the microscopic structure of the materials, and in their paper\cite{Kimchi-2015} Kimchi and collaborators studied it via a reduction to a 1D chain model and a subsequent solution employing an LT approximation. The 1D toy model relies on the reduction of the Hamiltonian in Eq.~\ref{Hamiltonian3} to that of decoupled zig-zag chains by taking the $I_c$ term to zero and introducing a second neighbour Heisenberg interactions. This Hamiltonian is solved by proposing an ansatz for the spin components and then the interchain coupling arising from the $I_c$ terms is slowly introduced. They subsequently solve the perturbed Hamiltonian by an LT approximation. They find that even when the second neighbour interactions is zero and the $I_c$ term is completely introduced, the spin spirals survive in the phase diagram. We study the model in Eq.~\ref{Hamiltonian3} exactly, via Monte Carlo simulations to test the accuracy of our methodology at detecting incommensurate states. 

\subsubsection{Phase diagram}
 
 \begin{figure}[h!]
\centering
\includegraphics[scale=0.45]{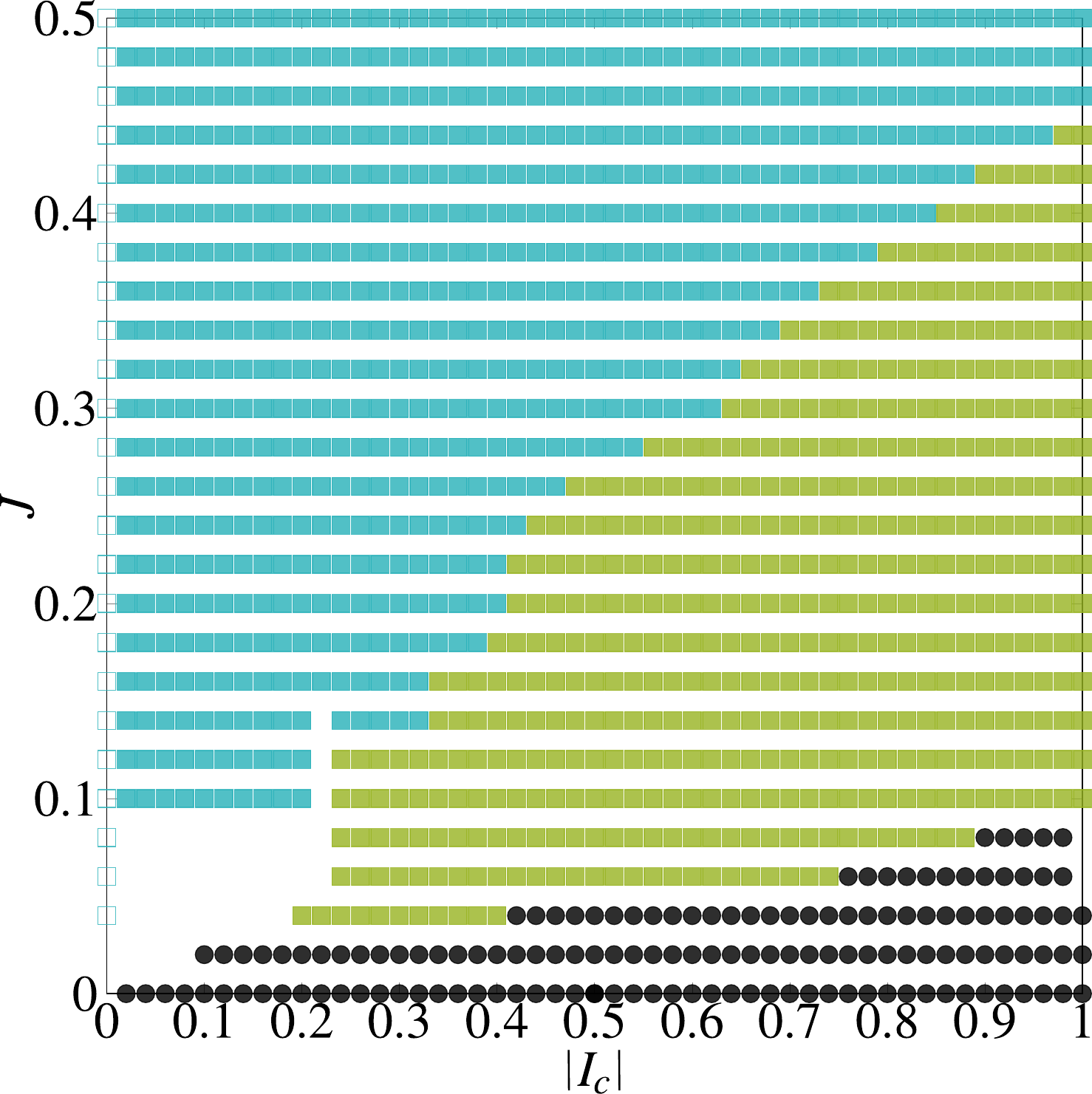}
\caption[Phase diagram for the $JKI_c$ mode;]{\small{ Phase diagram for model shown in Eq.~\ref{Hamiltonian3}. Blue squares represent the st-Z order (see text), green squares the spiral phase, and black dots correspond to ferromagnetic order }}\label{phdJKIc}
\end{figure}

Employing large scale Monte Carlo simulations as detailed in section \ref{MandM} ,we are able to map the phase diagram of this model ($\mathcal{H}(J_1, K_1, I_c \neq 0, I_d = 0)$) in the phase space of the $J$ and $I_c$ couplings, where the Kitaev exchange is consistently set to $-1$. The obtained phase diagram can be seen in Fig.~\ref{phdJKIc}. This diagram shows evidence of the strong Kitaev interactions present, displaying a broad region where the stripy phase lives. At the boundary $I_c = 0$ ($J\neq 0$) we recover the Kitaev-Heisenberg model, which presents a stripy phase for these values of $J$ and $K$. Furthermore, and as expected, this state is affected by a spin locking effect.  As soon as we set $I_c \neq 0$ the degeneracy of the stripy phase is broken, and a stripy in the $z$ direction is chosen (we will name this phase st-Z), which we show in the phase diagram by blue squares. 

On the other limit, when $J=0$ and $I_c \neq 0$ the model in Eq.~\ref{Hamiltonian3} reduces to two ferromagnetic couplings, which produces a large ferromagnetic phase which survives up to finite $J$ (black dots in Fig.~\ref{phdJKIc}). The case $I_c = 0$ and $J= 0$ is special, as this point reduces to the Kitaev model, which is a macroscopically degenerate state without LRO. For finite $J$ and $I_c$ we find an incommensurate spiral phase (green squares in Fig.~\ref{phdJKIc}). It is worth noting that Kimchi {\it et.al} report the existence of a small area where the st-X and st-Y phases should be present, and they locate it at the intersection of the spiral and st-Z phase. 

The incommensurate order present in the phase diagram (green squares in Fig.~\ref{phdJKIc}) exhibits signatures of incommensurate counterotating spirals. This spiral phase propagates in the horizontal direction according to Fig.~\ref{lattice} (the direction perpendicular to the {\it zz}-bonds), with a wavevector that varies from $0.5$ to $0.30$ in units of $2\pi$. The phase of rotation tilted with respect to the lattice plane remains constant throughout the phase diagram at $\theta = 54^o$, i.e, the rotation plane is oriented parallel to the XY-Cartesian plane. This phase mostly reproduces the experimental structure determined from neutron diffraction except for the tilt of the rotation plane (we find a tilt of $54^o$ and the expected value for the tilt angle is $80^o$).

\section{ Isotropic third neighbour model, $\mathcal{H}_I(J_{1,3}, K_{1,2}, \Gamma_{1,2})$ }
 \label{ITM}
\subsubsection{Phase diagram}

We will start by mapping part of the phase diagram of the isotropic case for the model shown in Eq.~\ref{Hamiltonian_W2}. We will implement a dominant nearest neighbour Kitaev coupling $K_1 = -1$ supported by (here and in the following, all exchange couplings are given in units of $|K_1|$) $K_2 = -0.275$, $\Gamma_1 = 1$, $\Gamma_2 = 0.275$, $J_1 \in (-0.1,..., 0.4)$, and $J_3 \in (0,..., 0.4)$.

\begin{figure}[h!]
\centering
\includegraphics[scale=0.45]{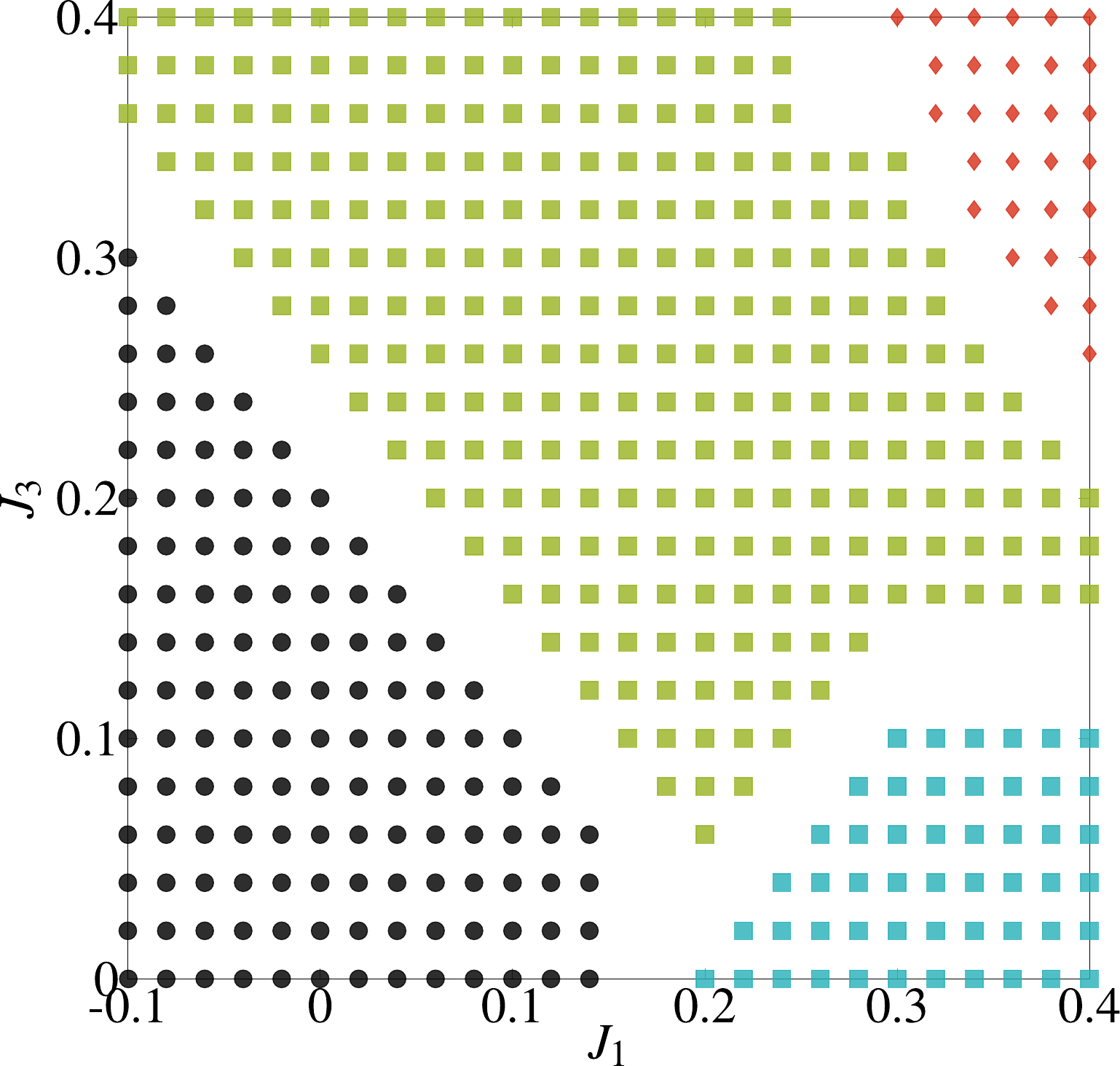}
\caption{Phase diagram for the isotropic case of the Hamiltonian shown in Eq.~\ref{Hamiltonian_W2}. Black dots correspond to ferromagnetic order, light blue squares to stripy order. The incommensurate states are represented by green squares, and the Neel state by red diamonds. }\label{phd_W_iso}
\end{figure}

The phase diagram is shown in Fig.~\ref{phd_W_iso}. In this model, the boundary between the different phases was not clearly distinguished from the Monte Carlo simulations, and future work will be needed to obtain reliable bounds. However, as we are interested in the nature of the phases arising in these models, we leave the study of the precise location of the boundaries for future work. In the phase diagram shown in Fig.~\ref{phd_W_iso} the phase boundaries will be located over the white spaces. 

We observe three commensurate phases, and an incommensurate one. In the limit $J_3 = 0$ a transition from a ferromagnetic to a stripy phase at $J_1 \sim 0.15$ is realized. At a critical value of the third neighbour coupling, $J_3 \sim 0.5$, a commensurate-incommensurate transition takes place and an intermediate incommensurate phase emerges. At a critical value of $J_1$ (which depends on the value of $J_3 \gtrsim 0.05 $) the system enters into a spiral phase (green squares in Fig.~\ref{phd_W_iso}). If $J_1$ is further increased the system can enter a stripy phase ($J_3 \lesssim 0.1$), remains in the spiral phase, or enter an antiferromagnetic state ($J_3 \gtrsim 0.25$). Both antiferromagnetic and stripy phases have spins oriented perpendicular to the lattice plane, which allows us to classify the stripy phase as st-Z order. Both phases are two fold degenerated, and this arises in the spin pattern as domains separating these degenerate states. We have confirmed the existence of these phases via a study of the Fourier transform of the correlation functions obtained from Monte Carlo, as well as with a LT minimization. These observations confirm the results of Ref.~\cite{Winter2016}.

\subsubsection{Spiral properties}

\begin{figure}[h!]
\centering
\includegraphics[scale=1]{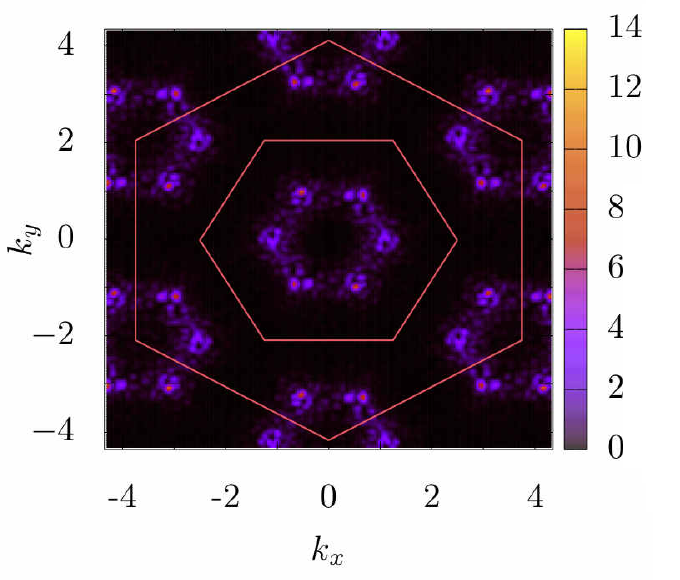}
\caption{Correlation function for the spiral phase found in the bond isotropic version of the model shown in Eq.~\ref{Hamiltonian_W2}. We observe a degenerate set of maxima inside the first Brillouin zone. }\label{Sq_W_iso}
\end{figure} 

The incommensurate order present in this model exhibits the signatures of a co-planar counterrotating spin spiral. The fundamental difference between these spirals and the ones appearing in $\alpha$-Li$_2$IrO$_3$ is that in this case, the spirals are degenerate. Since the model is bond isotropic, the spin spirals can propagate in three symmetry allowed directions, i.e. the propagation direction can be perpendicular to the $x$-, $y$- or $z$ bonds. In our simulations this implies that our state can contain domain walls separating domains where the spin spirals propagate in different directions. This can be seen in Fig.~\ref{Sq_W_iso} the degeneracy is evident in the Fourier transform of the correlation function. Here we see that indeed, we obtain maxima inside the first Brillouin zone, and that these are six fold degenerated. The study of the wavevector and nature of the spirals confirm that these spirals are of the same type as those obtained in the nearest neighbour model shown in the main text, with the caveat that the state in these models is degenerated.

\section{$\mathcal{H}(J_{1,2}, K_{1,2}, I_c = 0, I_d = 0)$ model}
\label{2N}

We proceed now to study the first of the second neighbour models, which consists of first and second Heisenberg and Kitaev interactions. We show in Fig.~\ref{lattice} the exchange couplings in the honeycomb lattice for this model. The Hamiltonian corresponding to this model  is given by Eq.~\ref{Hamiltonian4}. 

\begin{align}
\mathcal{H}(J_{1,2}, &K_{1,2}, I_c = 0, I_d = 0)=J_1\sum_{<ij>}\mathbf{S}_i\cdot\mathbf{S}_j\notag\\
&+K_1\sum_{<ij>}\sum_{\gamma}S^{\gamma}_i S^{\gamma}_j +J_2\sum_{<<ij>>}\mathbf{S}_i\cdot\mathbf{S}_j\notag\\
&+K_2\sum_{<<ij>>}\sum_{\gamma}S^{\gamma}_i S^{\gamma}_j
\label{Hamiltonian4}
\end{align}

This model was previously studied in the quantum limit, and in the context of the $\mathrm{Li}_2\mathrm{IrO}_3$ family, via the pseudofermionic functional renormalization group method (PFFRG) \cite{Reuther-2014}. In their paper, Reuther {\it et. al} parametrize the different couplings via two angles, $P_1$ and $P_2$, as $J_1 = \mathrm{cos}(\pi P_1/2)$, $K_1 = -\mathrm{sin}(\pi P_1/2)$, $J_2 = -g\mathrm{cos}(\pi P_2/2)$, $K_2 = g\mathrm{sin}(\pi P_2/2)$, and map the phase diagram of \ref{Hamiltonian4} for $P_1 \in (0, 1)$ and $P_2 \in (0, 1)$. 

They find two incommensurate spiral phases for $P_2 \gtrsim 0.5$, the spirals SP1 and SP2, while for values below $0.5$ they find a ferro and antiferromagnet. Studying the maximum of the susceptibility they find that the state SP1 corresponds to maxima outside of the first Brillouin zone, while the SP2 has maxima inside the first zone.  We will employ Monte Carlo simulations to study the classical equivalent of this model, and we will restrict ourselves to the value $g=0.8$, as according to the evidence in Ref.\cite{Reuther-2014} the phase diagram does not change drastically for different values of $g$. 

\subsubsection{Phase diagram}

\begin{figure}[h!]
\centering
\includegraphics[scale=0.5]{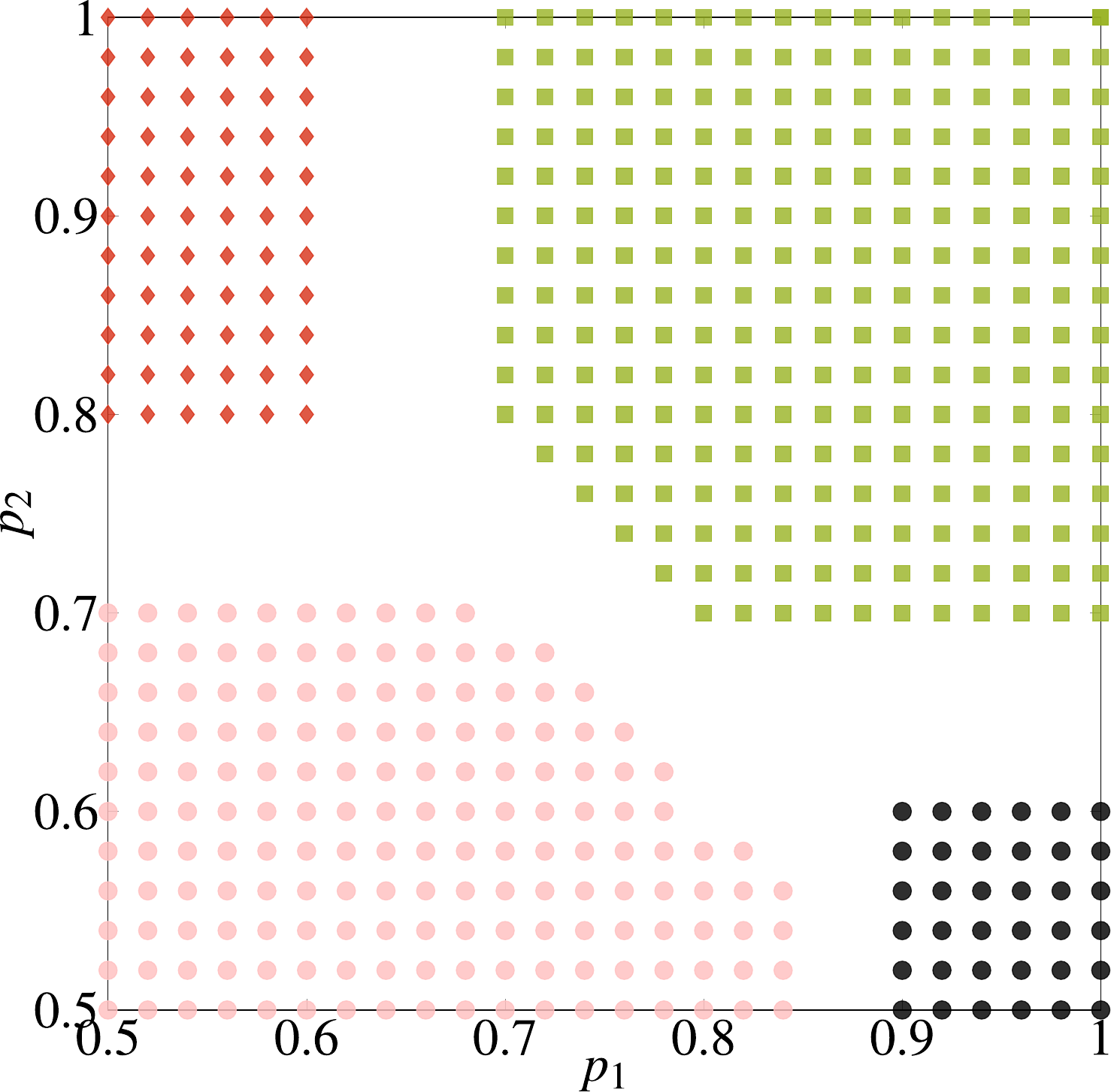}
\caption[Phase diagram for the $\mathcal{H}(J_{1,2}, K_{1,2}, I_c = 0, I_d = 0)$ model]{\small{Phase diagram for the $2N$-model. Pink dots represent antiferromagnetic order, black dots correspond to ferromagnetic order, red diamonds to a spiral phase SP1, and green squares to another spiral phase SP2.}}\label{phdJKJK}
\end{figure}

The phase diagram for the this model, as obtained from our Monte Carlo simulations, is shown in Fig.~\ref{phdJKJK}. Please note that since the aim of this study is to identify the nature of the spiral phases, we have not gone to great lengths mapping the boundary in between the phases in this model. In the phase diagram shown in Fig.~\ref{phdJKJK} the phase boundaries will be located somewhere over the white spaces. 

The phase diagram presents two commensurate and two incommensurate phases. Pink dots represent antiferromagnetic order, black ones indicate the onset of ferromagnetic order. We indicate by red diamonds the incommensurate spiral phase SP1, and by green squares the phase SP2. We will study these phases in detail in the next section.

The ferro and antiferromagnetic phases exhibit clear features in the correlation function which confirms their nature. However, when we study the real space pattern of the spins we find that these phases come together with domain walls as well as with vortex-like defects.  

\subsubsection{Spiral properties}

\begin{figure}[h!]
\centering
\includegraphics[scale=0.6]{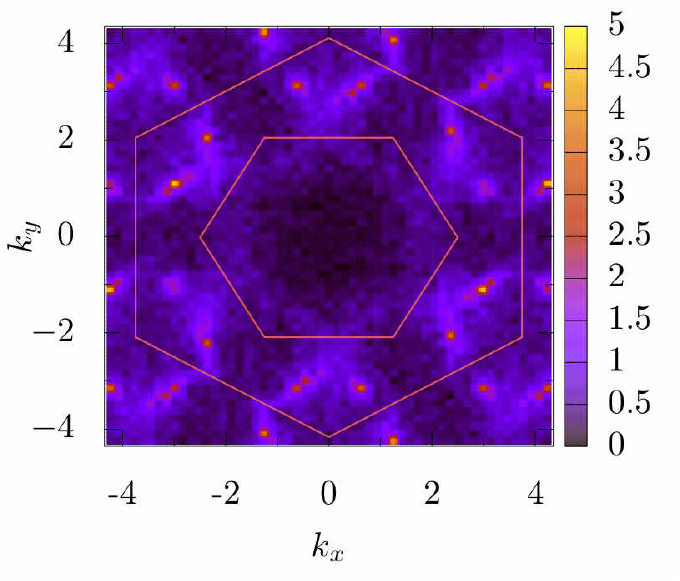}
\includegraphics[scale=0.6]{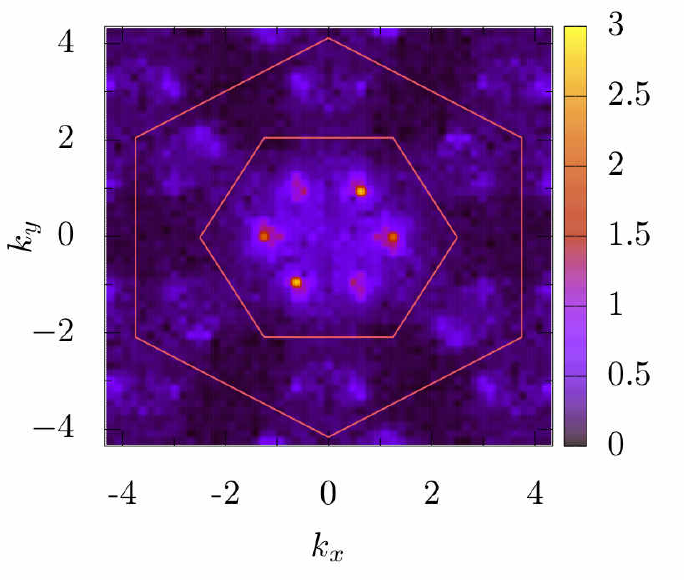}
\caption[Correlation functions for the SP1 and SP2 phase]{\small{Fourier transform of the correlation function for the $\mathcal{H}(J_{1,2}, K_{1,2}, I_c = 0, I_d = 0)$ model shown in Eq.~\ref{Hamiltonian4} with parameters $P_1 = 0.6$ and $P_2=0.8$ (SP1 phase, {\it left}) and $P_1 = 1$ and $P_2=0.7$ (SP2 phase, {\it right})}.}\label{SqJKJK}
\end{figure}

The two incommensurate spin spiral phases we find correspond to the red diamonds in Fig.~\ref{phdJKJK} (SP1), and to the  green squares (SP2). Upon examination of the correlation function (Fig.~\ref{SqJKJK}) we see that the SP1 phase present maxima outside of the first Brillouin zone, while the maxima corresponding to the SP2 phase are contained within the first Brillouin zone.

The Fourier transform of the correlation function for the SP2 phase presents maxima as satellites around the $\Gamma$ point, but no secondary maxima around the $K$ points are observed, unlike the expected signal for $\alpha$-Li$_2$IrO$_3$. Furthermore a degeneracy is present, in which the maxima appear in the three symmetry related positions, which would indicate that the spin spirals propagate in the three directions allowed by the Kitaev symmetry. On the other hand, the spiral phase SP1 presents only maxima as satellite peaks around the $K$ points. 

\begin{figure}[h!]
\centering
\includegraphics[scale=0.305]{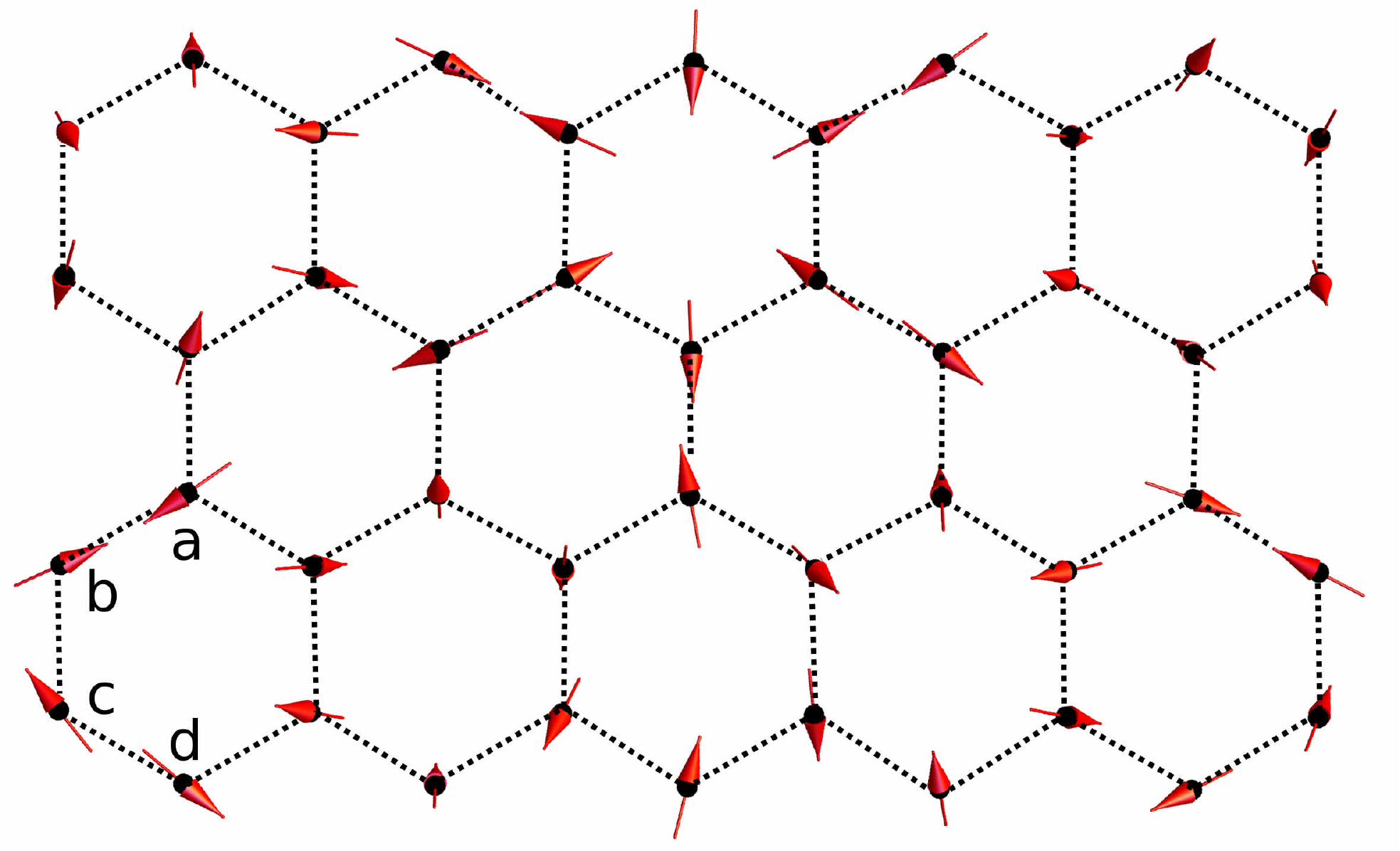}
\includegraphics[scale=0.25]{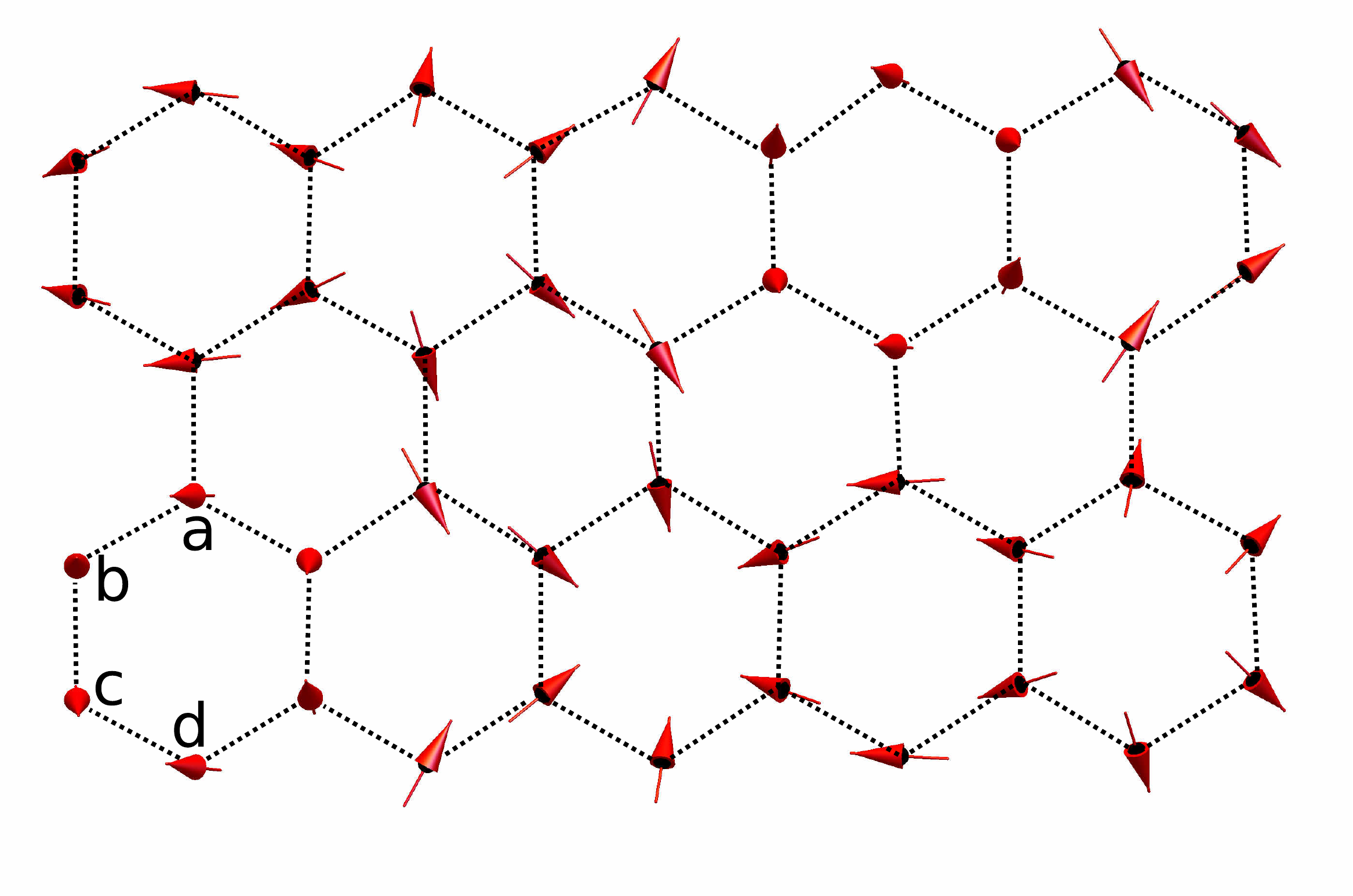}
\caption[Spiral structures SP1 and SP2 found in the $2N$ model]{\small{Spiral structure obtained for the model shown in Eq.~\ref{Hamiltonian4} within the phase SP1, i.e. parameters $P_1 = 0.6$ and $P_2=0.8$. ({\it top}) and within phase SP2, $P_1 = 1$ and $P_2=0.t$. ({\it bottom}).}}
\label{spiralsJKJK}
\end{figure}

Upon a closer inspection of the real space spin pattern we see that for both the SP1 and SP2 phases, the actual spiral present in this model does not coincide with that of $\alpha-\mathrm{Li}_2\mathrm{IrO}_3$. In the experimental case one observes co-planar spirals, where the plane of rotation can be defined by $(\mathbf{S}_1 \times \mathbf{S}_2)\cdot \mathbf{S}_3$, where $S_x$ (with $x= \{1,2,3\}$) are consecutive spins on a single spiral. However, this model exhibits a non-coplanar helimagnet, where a plane of rotation cannot be defined. 

\begin{figure}[h!]
\centering
\includegraphics[scale=0.6]{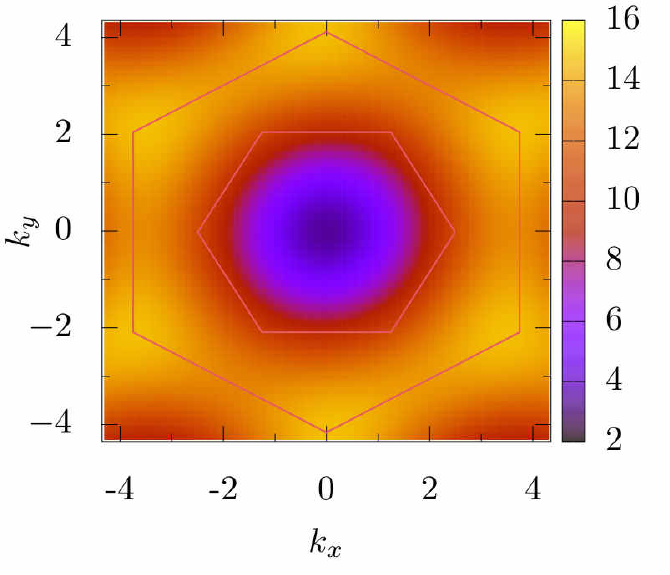}
\includegraphics[scale=0.6]{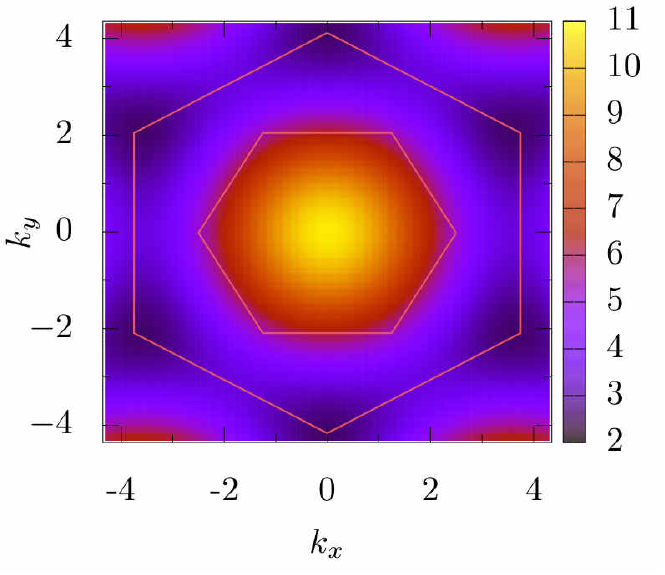}
\caption[Nearest neighbor correlation functions for the SP1 and SP2 phase]{\small{Fourier transform of the nearest neighbor correlation function for the model shown in Eq.~\ref{Hamiltonian4} with parameters $P_1 = 0.6$ and $P_2=0.8$ (SP1 phase, {\it left}) and $P_1 = 1$ and $P_2=0.7$ (SP2 phase, {\it right})}.}
\label{SqJKJKNN}
\end{figure}

While both spirals are non-coplanar, we can further distinguish them by studying their nearest neighbor correlations. Calculating the correlation function up to nearest neighbours, and Fourier transforming it, we obtain Fig.~\ref{SqJKJKNN}. From here we see that the main distinguishing feature between both non-coplanar spirals resides in the nature of the nearest neighbor correlations. While the maxima are broadened, we see that for phase SP1 we find maxima in the corners of the extended Brillouin, which indicates antiferromagnetic nearest neighbor correlations. On the other hand, SP2 has a maximum in the $\Gamma$ point, which coincides with ferromagnetic correlations. 

\section{$\mathcal{H}(J_{1,2}, K_{1,2}, I_c \neq 0, I_d = 0)$ model}
\label{2NI_c}
Since $\alpha-\mathrm{Li}_2\mathrm{IrO}_3$ presents bond dependent interactions, a question that arises after studying the $2N$-model is what the effect of other symmetry allowed interactions are in this particular model. It is not absurd to think that an $I_c$-like term could break the degeneracy and perhaps induce a coplanar spiral. For this purpose we will now study a modification of the $2N$ model, where we introduce an $I_c$ term of the same form as used in the nearest neighbour models. The Hamiltonian results. 

\begin{align}
\mathcal{H}=&J_1\sum_{<ij>}\mathbf{S}_i\cdot\mathbf{S}_j+K_1\sum_{<ij>}\sum_{\gamma}S^{\gamma}_i S^{\gamma}_j+J_2\sum_{<<ij>>}\mathbf{S}_i\cdot\mathbf{S}_j\notag\\
&+K_2\sum_{<<ij>>}\sum_{\gamma}S^{\gamma}_i S^{\gamma}_j+I_c\sum_{<ij>}S^{r_{ij}}_iS^{r_{ij}}_j
\label{Hamiltonian5}
\end{align}

As before we parametrize the different couplings via two angles, $P_1$ and $P_2$, as $J_1 = \mathrm{cos}(\pi P_1/2)$, $K_1 = -\mathrm{sin}(\pi P_1/2)$, $J_2 = -g\mathrm{cos}(\pi P_2/2)$, $K_2 = g\mathrm{sin}(\pi P_2/2)$, and set $g=0.8$. The coupling $I_c$ will take values in the range  $\{-1, -0.9, ..., 0\}$. We run Monte Carlo simulations for the pairs $\{P_1 = 0.5, P_2 =0.5\}$, $\{P_1 = 0.5, P_2 =1\}$, $\{P_1 = 1, P_2 =0.5\}$, $\{P_1 = 1, P_2 =1\}$, and $\{P_1 = 0.9, P_2 =0.9\}$ as a way to probe the effect of the $I_c$ term at different points in the phase diagram.

\subsubsection{$p_1 = 0.5$, $p_2 =0.5$}

\begin{figure}[h!]
\centering
\includegraphics[scale=0.45]{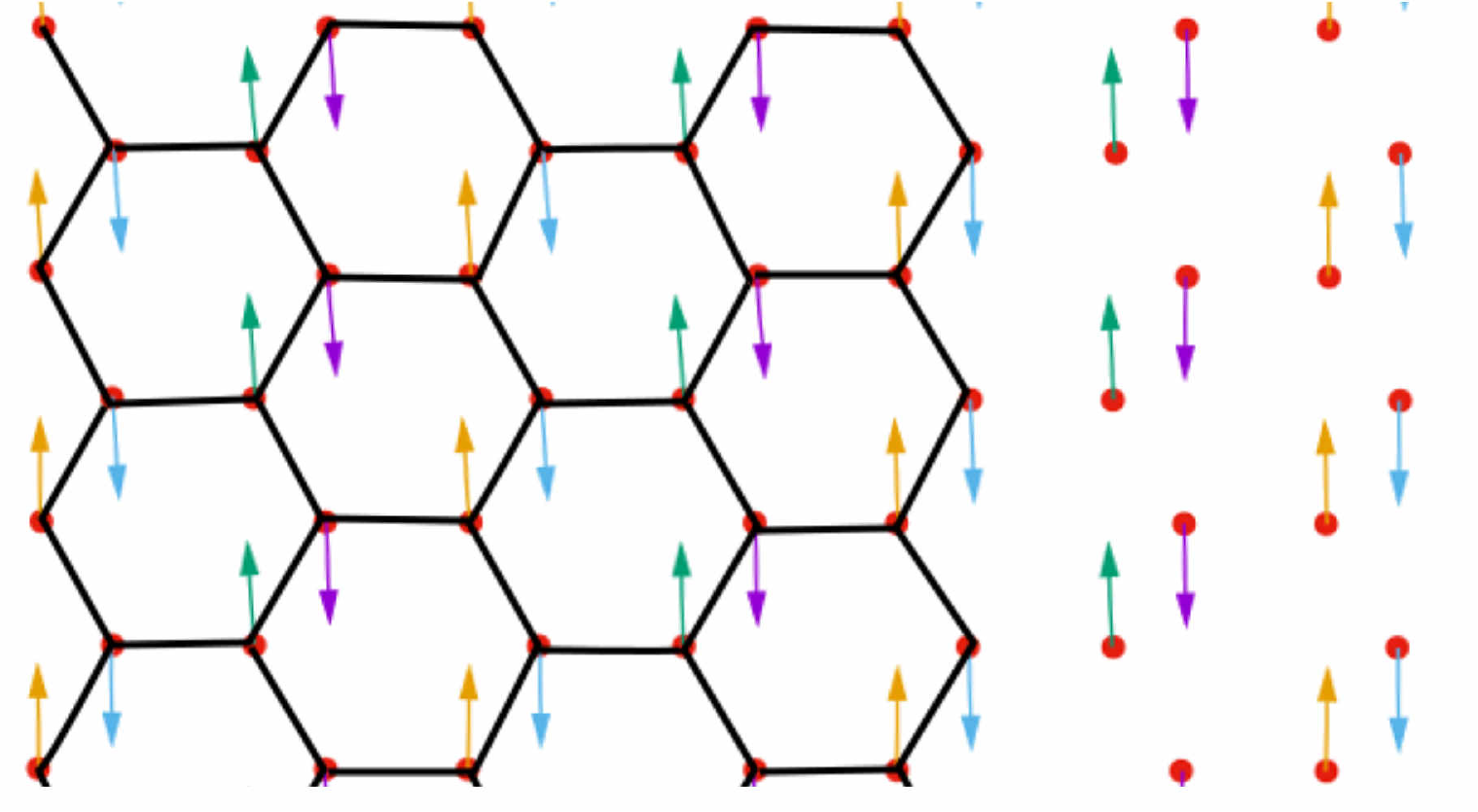}
\caption[Antiferromagnetic phase in the $2NI_c$-model, for $I_c=1$]{\small{Stripy phase in the $2NI_c$-model, for $I_c=1$.}}
\label{afm_JKJKIc}
\end{figure}

For the limit $I_c=0$ an antiferromagnetic state is realized, as shown in Fig.~\ref{phdJKJK}. This state is degenerate, with spins able to point in different symmetry allowed directions. For $I_c>-0.5$ the state remains an antiferromagnet, but now the degeneracy disappears  and the ground state selects the state with spins pointing in plane and perpendicular to the $zz$-bonds, as shown in Fig.~\ref{afm_JKJKIc}. For $I_c < -0.5$ the state changes slightly, maintaining its antiferromagnetic nature, but with the staggered magnetization in a direction perpendicular to the lattice plane.

\subsubsection{$p_1 = 1$, $p_2 =0.5$}

For these set of parameters, a ferromagnetic state is realized, where vortex-like defects appear. When the value of $I_c$ is non zero, the vortex defects disappear, and we find a ferromagnetic state with a net magnetization in the direction of the {\it zz}-bonds. This configuration is reached for the smallest values of $I_c$ studied and remains unchanged through the whole range. Here, two possible orientations of the net magnetization are possible, and they appear through the simulation separated by extended domain walls spanning through the system. 

\subsubsection{$p_1 = 0.5$, $p_2 =1$}

\begin{figure}[h!]
\centering
\includegraphics[scale=0.3]{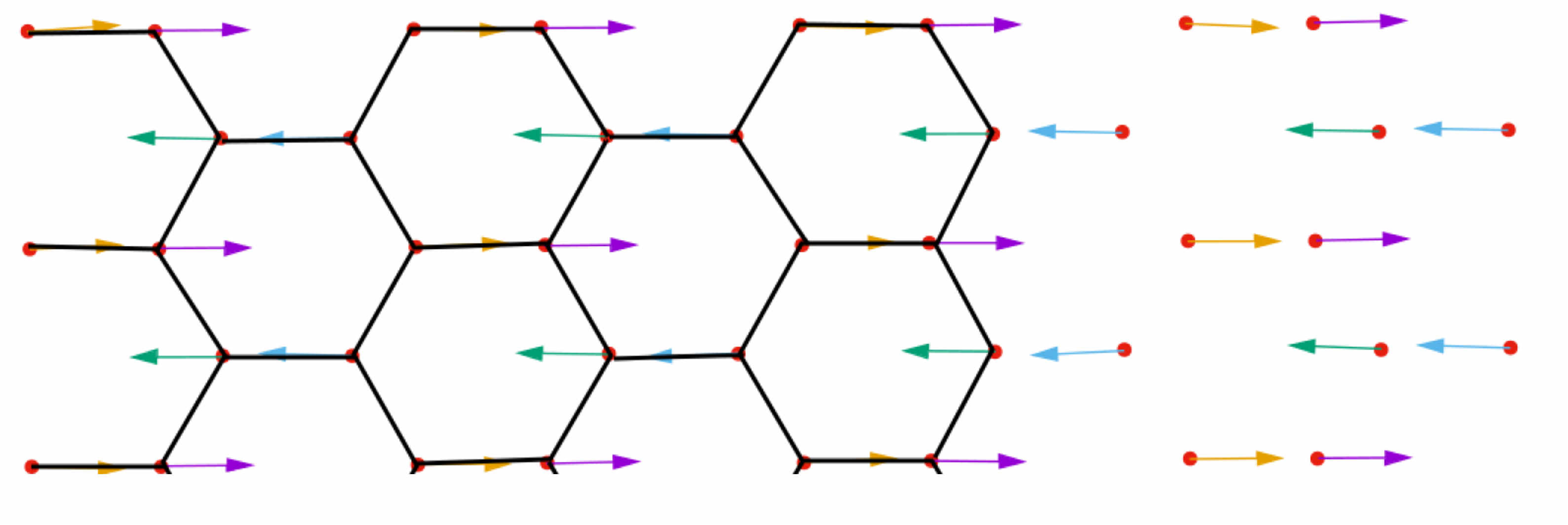}
\caption[Stripy phase in the $2NI_c$-model, for $I_c=1$]{\small{Stripy phase in the $2NI_c$-model, for $I_c=1$. The spins are coloured according to the conventions in Fig.~\ref{afm_JKJKIc}.}}
\label{stripy_JKJKIc}
\end{figure}

The case for $I_c=0$ realized a helimagnetic state for these parameter values.The inclusion of a finite $ I_c > -0.2 $ breaks the degeneracy of the spin spiral states, and spirals only propagate in the direction perpendicular to the $zz$-bonds. The nature of the spin spiral also changes with respect to the case $I_c = 0$, the spirals are still non-coplanar, but the nearest neighbor correlations are not purely antiferromagnetic, as the $I_c$ term introduces a ferromagnetic binding in the $zz$- bonds. For $I_c = -0.3$ a stripy phase dominates the phase diagram, with domain walls separating different orientations of the spins. In this stripy phase, the spins are in plane, aligned in the direction of the {\it zz}-bonds. We show in Fig.~\ref{stripy_JKJKIc} the resulting stripy phase for the case $I_c=1$.

\subsubsection{$p_1 = 1$, $p_2 =1$, and $p_1=0.9$, $p_2 =0.9$}

For the limit $I_c=0$ we observed a spin spiral state which we analysed in the previous section. When $I_c>0$ the non-coplanar nature of the spin spiral disappears, the anisotropy along the $zz$-bonds introduced by the ferromagnetic $I_c$ term enforces a ferromagnetic in plane alignment of the spins in this bond, which transforms the spin spiral into an antiferromagnetic state. As $I_c$ decreases the state remains unchanged.

Since one would expect that the introduction of an $I_d$ term would modify the plane of rotation, but not the wavevector, as we have shown in the nearest neighbor model that only an $I_c$ term is enough to generate incommensurate counterotating spirals, a further modification of the model shown in this appendix was not attempted. We can conclude then, that the second neighbour models are not minimal models for $\alpha$-Li$_2$IrO$_3$.

\end{document}